\documentclass[prd,showpacs,preprintnumbers,amsmath,amssymb,nofootinbib]{revtex4}

\usepackage{graphicx}% Include figure files
\usepackage{dcolumn}% Align table columns on decimal point
\usepackage{bm}% bold math

\newcommand{\PL}[1]{Phys.\ Lett.\ {\bf #1}}

\newcommand{\PR}[1]{Phys.\ Rev.\ {\bf #1}}
\newcommand{\PRL}[1]{Phys.\ Rev.\ Lett.\ {\bf #1}}

\newcommand{\Od}{{\cal O}}

\newcommand{\Tr}{\mbox{Tr}}

\newcommand{\im}{\mbox{Im}\,}
\newcommand{\re}{\mbox{Re}\,}
\newcommand{\sgn}{\mbox{sgn}}

\newcommand{\intcc}{\int_C \! d^4x}
\newcommand{\intccp}{\int_C \! d^4x'}

\newcommand{\modp}{\vert \vec{p} \vert}

\newcommand{\betatp}{\tilde\beta_p}

\newcommand{\be}{\begin{equation}}
\newcommand{\ee}{\end{equation}}
\newcommand{\ba}{\begin{eqnarray}}
\newcommand{\ea}{\end{eqnarray}}

\newcommand{\IR}{{\Bbb R}}

\newcommand{\gsim}{\raise.3ex\hbox{$>$\kern-.75em\lower1ex\hbox{$\sim$}}}
\newcommand{\lsim}{\raise.3ex\hbox{$<$\kern-.75em\lower1ex\hbox{$\sim$}}}

\begin{document}
%\baselineskip=20pt
% declarations for front matter

\title{Chemical nonequilibrium for interacting bosons: applications to the pion gas}

\author{D. Fern\'andez-Fraile}
\email{danfer@fis.ucm.es} \affiliation{Departamento de F\'{\i}sica
Te\'orica  II. Univ. Complutense. 28040 Madrid. Spain.}
\author{A. G\'omez Nicola}
\email{gomez@fis.ucm.es} \affiliation{Departamento de F\'{\i}sica
Te\'orica II. Univ. Complutense. 28040 Madrid. Spain.}

\begin{abstract}
We consider an interacting pion gas in a stage of the system evolution where
thermal but not chemical equilibrium has  been reached, i.e., for temperatures between
thermal and chemical freeze-out $T_{ther}<T<T_{chem}$ reached in Relativistic
 Heavy ion Collisions. Approximate
 particle number conservation is implemented by a nonvanishing pion
number chemical potential $\mu_\pi$ within a diagrammatic thermal field
 theory approach, valid in principle for any bosonic field theory in this regime.
  The resulting Feynman rules are derived here
 and applied within the context of Chiral Perturbation Theory
 to discuss thermodynamical quantities of  interest
 for the pion gas such as the
 free energy, the quark condensate and thermal self-energy. In particular, we derive the $\mu_\pi\neq 0$ generalization of Luscher and Gell-Mann-Oakes-Renner type relations. We pay special
 attention to the comparison with the conventional kinetic theory
 approach in the dilute regime, which allows for a check of consistency of our approach.    Several  phenomenological applications are discussed, concerning chiral symmetry restoration, freeze-out conditions and Bose-Einstein  pion condensation.
\end{abstract}


%\vspace{-.5cm}
%\rule{\textwidth}{.1mm}

\maketitle

\section{Introduction}

One of the ongoing research lines in  Heavy Ion Physics is  the
thermal and chemical evolution of the expanding hadronic gas.
Roughly speaking, the accepted picture is that  the
evolution of the cooling system reaches chemical freeze-out
before the thermal one, so that when hadrons fully decouple the
chemical potentials associated to particle number conservation are
not zero. The chemical composition of the gas can be determined
experimentally by looking at the relative abundances of the
different hadron species and their spectra
\cite{Kataja:1990tp,Bebie:1991ij,Hung:1997du,BraunMunzinger:2003zd,Kolb:2002ve}.
The presence of such a chemically not-equilibrated phase is more
likely to exist for higher collision energies such as those in
RHIC or LHC than for  SPS or AGS experiments
\cite{BraunMunzinger:2003zd}. For the pion component, different
estimates based on local thermal equilibrium and particle spectra
analyses predict $\mu_\pi\sim $ 50-100 MeV at a thermal freeze-out
temperature $T_{ther}\sim$ 100-120 MeV, with chemical freeze-out
taking place at about $T_{chem}\sim$ 180 MeV
\cite{Bebie:1991ij,Song:1996ik,Hung:1997du,Kolb:2002ve,Torrieri:2005va,Letessier:2005qe}. On the other hand, the
 plasma is almost electrically neutral, so that it is a good approximation to keep vanishing
  charge or isospin chemical pion chemical potentials.

For low and moderate temperatures, the dominant component is the
pionic one. In that phase, the mean free path of pions is small
compared to the system size, so that local thermal equilibrium
prevails \cite{Goity:1989gs,Bebie:1991ij,Schenk:1993ru}. On the
other hand, the chemical relaxation rate through inelastic $\pi\pi\leftrightarrows \pi\pi\pi\pi$ processes is very small
\cite{Goity:1993ik,Song:1996ik} due to a strong phase space
suppression. Therefore, in the range of temperatures
$T_{ther}<T<T_{chem}\lsim T_c$, with $T_c$ the chiral restoration
critical temperature, the system is in thermal equilibrium and
dominated  by elastic collisions so that $\mu_\pi\neq 0$. In that
temperature range, it is valid to use the theoretical framework of
Chiral Perturbation Theory (ChPT) and  it is also reasonable to
adopt a dilute gas description, since the mean particle density is
small. In addition,  neglecting  dissipative effects such as
viscosities, entropy is conserved in the evolution.

The system described above, i.e., a pion gas with $\mu_\pi\neq 0$
is the one we will consider here. Clearly, it is an oversimplified
version of the real hadron gas, but we will take it as a
physically relevant working example for our present analysis. So
far, pion number chemical potential effects in such a system have
been incorporated basically in two ways. One of them   is the
limit of free particles (where one has actually exact particle
conservation) used  for the evaluation of the partition function,
including resonances explicitly  \cite{Bebie:1991ij,Song:1996ik}.
This allows, via   entropy conservation requirements, to describe
rather accurately the isentropic dependence $\mu_\pi (T)$ in the
range of temperatures of phenomenological relevance indicated
above. The other one is to use kinetic theory arguments to include
the $\mu_\pi\neq 0$  dependence directly in the distribution
function. The latter has been followed for instance in the
calculation of the thermal width \cite{Goity:1989gs}, in the
evaluation of transport coefficients
\cite{Prakash:1993bt,Dobado:2003wr} or in the virial approach for
low densities \cite{Dobado:1998tv}. Finally, it is worth
mentioning that there are phenomenological analyses, like that in
\cite{Baier:1996if} for the dilepton rate, where the same
prescription is followed, i.e., replacing the distribution
function, but for propagators at the diagrammatic level, inspired
on the nonequilibrium formulation of thermal field theory
\cite{Chou:1984es}.

 We will be interested  in a
diagrammatic formulation of this system, i.e, we will derive the
 Feynman rules to be used when approximate particle number conservation is
 valid. The Feynman rules of thermal field theory with {\em exact}
 conserved charges can be obtained  straightforwardly
 \cite{Landsman:1986uw} but this is a completely different
 situation, since particle number is not exactly conserved in an
 interacting
 bosonic field theory (is only conserved in the free case) and
 therefore there is not a local number operator to be added to the
 lagrangian in the usual way. This will also be related to  the
 impossibility to define a proper Matsubara imaginary-time
 formalism.   The motivation for our field-theory description is twofold:
first, it will provide
 a formal proof of the consistency
and validity of the different prescriptions used in the literature and mentioned in the previous paragraph.  Second, it will allow to deal in  a natural  way with
 pion interactions when $\mu_\pi\neq 0$, which is
 particularly interesting in order to describe corrections
to dynamical quantities such as the thermal pion self-energy, but also to  evaluate the effect of interactions in  thermodynamical observables.

The paper is  organized as follows: in the first part   we will describe our formalism, based on
holomorphic path integrals, which naturally leads to the relevant
Feynman rules. Although the results in that part are actually
valid for any real scalar field theory provided one neglects the
contributions of number-changing scattering processes or, in other
words, if the gas is dilute enough, we will be primarily
interested in the pion gas, where chemical nonequilibrium is
actually reached during the expansion. We explain more clearly
this physical motivation in section \ref{sec:motivation}, where we discuss
 the relevant distribution function to describe the system.   The
Feynman rules we derive (section \ref{sec:formal}) had not been considered before in the
interacting case and,  as we will see, they are really meaningful
only in the real-time formalism of thermal field theory. The
second part (section \ref{sec:apppion} ) deals with the
application of our formalism to the pion gas. We will analyze
corrections  both in thermodynamical (free energy, entropy,
particle number and quark condensate) and dynamical (thermal mass
and width) observables,  the former being understood as a
 generalization of the usual thermodynamical
variables during the chemical nonequilibrium phase.  We also
compare to previous works in the literature and we discuss several
phenomenological consequences regarding chiral symmetry
restoration, Bose-Einstein
 condensation of neutral and charged pions, as well as  thermal and chemical
 freeze-out.
Appendices \ref{app:holo} and \ref{app:therprop} contain detailed
results used in the main text  about holomorphic path integrals
and thermal propagators respectively. In particular, in Appendix
\ref{app:therprop} we discuss some relevant differences between
the case of particle number chemical potential considered here and
the more usual one associated to the electric charge exact
conservation, concerning especially the way in which KMS boundary
conditions are broken.

%%%%%%%%%%%%%%%%%%%%%%%%%%%%%%%%%%%%%%%%%%
\section{Physical motivation}
\label{sec:motivation}
%%%%%%%%%%%%%%%%%%%%%%%%%%%%%%%%%%%%%%%%%%
As stated in the introduction, we are interested in describing the physical system constituted by a pion gas in expansion, during the time when the number of pions is approximately conserved. This is the case of  the pionic component of the hadronic gas produced after a relativistic heavy-ion collision. The pionic component is the
dominant one in the hadron gas around thermal freeze out \cite{Kataja:1990tp}, although considering additional degrees of freedom in the gas (kaons, etas, nucleons and resonances) and the interactions among them is relevant at temperatures close to the chiral phase transition. We will not consider those extra components here, although their inclusion in the chiral framework, together with the corresponding additional chemical potentials (strangeness and baryon number) is a feasible extension of this work. Unlike other treatments \cite{Bebie:1991ij}, where it was shown that it is a reasonable approximation to introduce in the partition function all the states (asymptotic states as well as resonances) up to a given energy as \emph{free} degrees of freedom, in our approach the resonances present in the pion gas, the $f_0(600)/\sigma$ and the $\rho(770)$, are generated dynamically by means of unitarization methods so the actual degrees of freedom in the lagrangian are only pions. It is however important to mention that, even when introduced as explicit degrees of freedom, the processes $\rho\leftrightarrows \pi\pi$ and $\sigma\leftrightarrows\pi\pi$ do not restore chemical equilibrium in the pionic component, because $\mu_\rho=\mu_\sigma=2\mu_\pi$, the truly relevant particle-changing process being  $\pi\pi\leftrightarrows \pi\pi\pi\pi$.  When pions and resonances are in chemical equilibrium with respect to each other we talk about a \emph{relative} chemical equilibrium, since it is possible to choose their chemical potentials to maintain it, whereas \emph{absolute} chemical equilibrium is only possible for $\mu_\pi=0$ \cite{Song:1996ik}.

The evolution of the pionic fireball can be divided into three stages as it cools down according to the corresponding temperature ranges \cite{Bebie:1991ij}: (I) $T_{chem}\leq T\leq T_c$, the pion gas is produced after hadronization from a quark-gluon plasma phase and it is in full statistical equilibrium (thermal and chemical). (II) $T_{ther}\leq T\leq T_{chem}$, the mean-free path of elastic collisions, $\lambda_{el}$ is smaller than the typical size of the fireball, $R\sim 5-10\ \mathrm{fm}$ \cite{Adler:2001zd}, so that thermal equilibrium is maintained, whereas the mean-free path of inelastic collisions $\lambda_{in}$ is larger than $R$ so the total number of pions $N_\pi\equiv N_{\pi^0}+N_{\pi^+}+N_{\pi^-}$ remains approximately constant\footnote{For instance, at $T=150\ \mathrm{MeV}$ the relaxation time of elastic $\pi\pi$ collision is $\tau_{el}\sim 2\ \mathrm{fm}$, whereas the relaxation time of the process $\pi\pi\leftrightarrows \pi\pi\pi\pi$ is $\tau_{in}\sim 200\ \mathrm{fm}$ \cite{Song:1996ik}, the typical mean velocities at those temperatures being $\bar v\sim c$. See also our comments in section \ref{sec:pimasswidth}.} and a finite chemical potential associated to $N_\pi$ builds up, so that the system is out of chemical equilibrium. (III) $T\leq T_{ther}$, $\lambda_{el}$ is larger than $R$ so the pions stop interacting and their momentum distributions become frozen, so that thermal equilibrium is lost.

We shall analyze the phase II of the evolution, where the total number of pions is approximately conserved (and the gas remains dilute enough) so that the introduction of a finite pion chemical potential $\mu_\pi$ associated to $N_\pi$ is necessary. The existence of such a chemically not equilibrated phase during the fireball space-time evolution is supported by several phenomenological results. For instance when analyzing experimental data from the NA44 collaboration,  it was shown in \cite{Hung:1997du} that in order to fit properly the pion spectrum at low transverse momentum in PbPb reactions, one needs to introduce a finite chemical potential of order $\mu_\pi\sim 60-80\ \mathrm{MeV}$ at thermal freeze-out. A similar conclusion is reached in \cite{Kolb:2002ve} for RHIC AuAu collisions. In addition, the analysis of total particle yields and yield ratios for SPS and RHIC energies are fitted with values of fugacities compatible  with the pionic component being significatively out of chemical equilibrium  \cite{Torrieri:2005va,Letessier:2005qe}.

Neglecting electromagnetic interactions, the pions are described by neutral scalar fields. For a neutral boson field theory, particle number is conserved only in the free case. Our aim here is to provide a field-theory description of the non-equilibrium state corresponding to  phase II. In this respect, it is important to remark that there are fundamental differences between total particle number and charges which are exactly conserved by the dynamics, such as the net electric charge or baryon number in QCD. In a pion gas,   the total number of pions is expressed in terms of individual number operators as $\hat{N}={\hat N}_{\pi^0}+{\hat N}_{\pi^+}+{\hat N}_{\pi^-}$ whereas electric charge, or the third isospin component  (they are equivalent in the pion $SU(2)$ case) is measured by $N_{\pi^+}-N_{\pi^-}$ \cite{SonLoewe,Eletsky:1993hv}. The main difference is of course that charge is exactly conserved in the second case, which implies several important consequences: first, from the field-theoretical point of view, in the charge case there is a local charge operator $\hat Q$ written in terms of the field and its derivatives, which allows for a straightforward derivation of the corresponding Feynman rules, adding the usual $\mu_Q \hat Q$ term to the lagrangian \cite{Landsman:1986uw,KapGale06}. However, that is not the case for particle number, which instead has a natural formulation in terms of canonical creation and annihilation operators. That is the main reason why we will develop our holomorphic representation of the path integral in section \ref{sec:formal}. Second, in the $\mu_\pi$ case we are really facing a nonequilibrium description, which is only consistent if $\mu_\pi$ and $T$ are not independent parameters, the function $\mu_\pi(T)$ parameterizing the deviations from chemical equilibrium and vanishing at $T=T_{chem}$. This  signals the end of phase II, or its beginning if we think in terms of proper time, as inverse of temperature evolution in a hydrodynamical description (remember that in phase II local thermal equilibrium is assumed). The form of $\mu_\pi(T)$ has to be fixed by additional physical assumptions. We will rely here (see section \ref{sec:thermo}) on the isentropic condition stating that the ratio of entropy density to pion density $s/n$ remains constant along the chemical evolution, which has phenomenological support \cite{Bebie:1991ij,Hung:1997du}. Finally, we remark that these differences between the charge $\mu_Q\neq 0$ case and the pion number $\mu_\pi\neq 0$ one translate into a  different way in which the KMS boundary conditions characteristic of equilibrium are broken. We discuss this issue in detail in Appendix \ref{app:therprop}. Throughout this work we will take $\mu_Q=0$, which corresponds to an electrically neutral pion gas, which seems to be well supported by the phenomenological values of the fugacities \cite{Letessier:2005qe}.

Since the system in phase II is in thermal equilibrium and there is an approximate conserved  operator $\hat{N}$ with chemical potential $\mu_\pi$ associated, the appropriate non-equilibrium partition function is
\begin{equation}\label{partitionf}
\tilde{Z}_\beta(t)\equiv\Tr\left\{\mathrm{e}^{-\beta(\hat{H}-\mu_\pi\hat{N})}\right\}\ ,
\end{equation}
where quantities with a tilde will refer always to the nonequilibrium $\mu_\pi\neq0$ case throughout this paper. By including source terms we can then derive thermal correlation functions. Note that $\tilde{Z}_\beta$ is independent of the position in space (we consider an homogeneous system), but it actually depends on (proper) time $t$ during  the gas expansion  through temperature $\beta(t)\equiv 1/T(t)$ and the chemical potential $\mu(T(t))$, along the lines  discussed in the previous paragraph. The validity of the out-of-(chemical) equilibrium distribution function in (\ref{partitionf}) will be subject to times $t<t_\mathrm{II}$, where $t_\mathrm{II}$ is the duration of phase II. It is in this nonequilibrium effectively time-dependent situation that our results for the partition function and thermodynamical observables (see section \ref{sec:evalz}) have to be understood.

For  $t<t_\mathrm{II}$ inelastic processes are scarce at temperatures $T_{ther}\leq T\leq T_{chem}$, therefore if at some time $t_1$ the system is in a state with well defined number of particles equal to $N$, $\hat{N}|n(t_1)\rangle=N|n(t_1)\rangle$, then at another time $t_2$ with $t_2-t_1<t_\mathrm{II}$, $\hat{N}|n(t_2)\rangle\simeq N|n(t_2)\rangle$, and thus in Heisenberg's picture $\hat{N}(t_1)\simeq \hat{N}(t_2)$ and  from Heisenberg's equation
$\mathrm{i}d{\hat{N}}(t)/dt=[\hat{N}(t),\hat{H}]$
we infer:
\begin{equation}
0\simeq\mathrm{i}(\hat{N}(t_2)-\hat{N}(t_1))=\int\limits_{t_1}^{t_2}[\hat{N}(t),\hat{H}]\, \mathrm{d}t\quad\Rightarrow\quad[\hat{N}(t),\hat{H}]\simeq 0\ ,\quad\mathrm{for}\ t_1\leq t\leq t_\mathrm{II}\ .
\end{equation}
In the following section, the condition $[\hat{N}(t),\hat{H}]\simeq 0$ (valid for times in phase II) will be used to derive a field-theory description of the system based on the holomorphic path-integral representation of the partition function provided by (\ref{partitionf}).

\section{Formalism:  chemical potentials for neutral bosons}
\label{sec:formal}

As we mentioned in the introduction, the operator $\hat{N}$ has a non-local representation in terms of the field operator  \cite{Itzykson:1980rh}, so that the appropriate representation for this operator is instead in terms of creation and annihilation operators. The holomorphic path-integral representation uses these convenient operators, and its main ideas can be found in \cite{ZJ02}. We will give the essential steps of the derivation for our system in this section, more technical aspects being relegated to Appendix \ref{app:holo}. As we saw in the previous section, the generating functional of thermal correlation functions will be constructed from the non-(chemical)equilibrium partition function (\ref{partitionf}). In order to simplify further the discussion, we will consider first a quantum-mechanical gas of Bose particles and next we will extend it straightforwardly to the QFT case. Throughout this section and for simplicity we use the notation $\mu$ for the chemical potential associated to particle number ($\mu_\pi$ in the pion gas case, which we will analyze extensively in section \ref{sec:apppion}).

Let us consider then a single-frequency quantum oscillator (free Hamiltonian) coupled to an external force $j(t)$. The Hamiltonian and number operators are then:

\begin{eqnarray}
\hat{H}&=&\frac{1}{2}\hat{p}^2+\frac{1}{2}\omega^2\hat{q}^2-j(t)\hat{q}\equiv
\hat{H}_0-j(t)\hat{q}
\nonumber\\&=&\frac{\omega}{2}\left(\hat{a}^\dagger\hat{a}+\hat{a}\hat{a}^\dagger\right)
-\frac{1}{\sqrt{2\omega}}\left(\hat{a}^\dagger+\hat{a}\right)j(t)
\label{hamil}\\
\hat{N}&=&\hat{a}^\dagger\hat{a}
\end{eqnarray}
where $\hat{q}$ and $\hat{p}$ are respectively the position and
conjugate momentum operators (whose role will be played by the
field and its conjugate momentum) and the  creation and
annihilation operators are defined in the usual way:

\begin{equation}
\hat{a}=\frac{ i }{\sqrt{2\omega}}(\hat{p}- i \omega\hat{q})
\,\quad\hat{a}^\dagger=-\frac{ i }{\sqrt{2\omega}}(\hat{p}+ i \omega\hat{q})
\end{equation}
satisfying canonical commutation relations
$[\hat{a},\hat{a}^\dagger]=\hat{1}$.

In the holomorphic representation \cite{ZJ02} traces of operators
are evaluated in the space of complex analytic functions of one
complex variable $z$ and creation and annihilation operators act
on this space as:

\begin{equation}
\hat{a}^\dagger\mapsto z,\quad\hat{a}\mapsto\frac{\partial}{\partial
z} \label{aaplusholo}
\end{equation}

We have included in Appendix \ref{app:holo} some of the technical
details to perform the relevant calculations in this formalism. In
particular, the partition function for any Hamiltonian ${\hat H}$
reads, from (\ref{tracehol}):

\begin{equation}\label{approx1}
\tilde{Z}_\beta=\ \int\frac{\mathrm{d}z\
\mathrm{d}\bar{z}}{2\pi i }\ \mathrm{e}^{-\bar{z}z}\
\langle z|\mathrm{e}^{-\beta(\hat{H}-\mu\hat{N})}|\bar{z}\rangle
\end{equation}

Now,   if the number operator is approximately conserved, then
$[\hat{H},\hat{N}]\simeq 0$ and equation (\ref{approx1}) can be
recast, by inserting the identity once, as:

\begin{widetext}
\begin{equation}\label{approx2}
\tilde{Z}_\beta\simeq\ \int\frac{\mathrm{d}z\
\mathrm{d}\bar{z}}{2\pi i }\ \mathrm{e}^{-\bar{z}z}
\int\frac{\mathrm{d}z'\ \mathrm{d}\bar{z}'}{2\pi i }\
\mathrm{e}^{-\bar{z}'z'}\ \langle
z|\mathrm{e}^{\beta\mu\hat{N}}|\bar{z}'\rangle\langle
z'|\mathrm{e}^{-\beta\hat{H}}|\bar{z}\rangle
\end{equation}
\end{widetext}

This is the key step of the derivation, since it contains our main
approximation, which  is equivalent to consider only up to
two-particle states in the trace (\ref{partitionf}). Therefore, it
is physically appropriate to describe a dilute regime where
elastic collisions dominate and particle number is approximately conserved.

Now, the first matrix element in (\ref{approx2}) can be calculated
directly, using (\ref{evol}) with $j=0$, $t_f=t_i-i\beta$ and
$\omega=-\mu$:

\begin{equation}
\langle
z|\mathrm{e}^{\beta\mu\hat{N}}|\bar{z}'\rangle=\exp\left(z\bar{z}'\mathrm{e}^{\beta\mu}\right)
\end{equation}
so that, using (\ref{delta}) we arrive to:

\begin{equation}\label{partint}
\tilde{Z}_\beta=\int\frac{\mathrm{d}z\
\mathrm{d}\bar{z}}{2\pi i }\ \mathrm{e}^{-\bar{z}z}\ \langle
ze^{\beta\mu}|\mathrm{e}^{-\beta\hat{H}}|\bar{z}\rangle\
\end{equation}

From this representation of the partition function we define the
corresponding generating functional (in the QM case):

\begin{equation}\label{genfunint}
\tilde{Z}_\beta[j]\equiv\int\frac{\mathrm{d}z\
\mathrm{d}\bar{z}}{2\pi i }\ \mathrm{e}^{-\bar{z}z}\ \langle
ze^{\beta\mu}|\mathrm{e}^{-\beta(\hat{H}-j\hat{q})}|\bar{z}\rangle\
\end{equation}
so that correlators of any function of the position operator
$\hat{q}$ (the field operator in the QFT
 case) can be expressed in
terms of functional derivatives of $\tilde{Z}_\beta[j]$ with respect
to $j$ at $j=0$ in the usual way.

We will now proceed to the evaluation of $\tilde{Z}_\beta[j]$ when
the Hamiltonian is the free one plus the source term, i.e,
$\hat{H}=\hat{H}_0-j\hat q$ in (\ref{hamil}). Then, as usual, by
functional derivation we will get the generating functional for the
interacting case. We first separate the normal-ordered part as
customarily, i.e, $\hat{H}_0=\omega/2+\omega\hat{a}^\dagger\hat{a}$
where the first term is the vacuum energy. Therefore, we have:

\begin{equation}\label{genfunfree}
\tilde{Z}^0_\beta[j]=e^{-\beta\omega/2}\int\frac{\mathrm{d}z\
\mathrm{d}\bar{z}}{2\pi i }\ \mathrm{e}^{-\bar{z}z}
\mathcal{U}_0(ze^{\beta\mu},\bar{z};-i\beta)
\end{equation}

\begin{figure}[b]
\vspace*{-3cm}
\includegraphics[scale=.4]{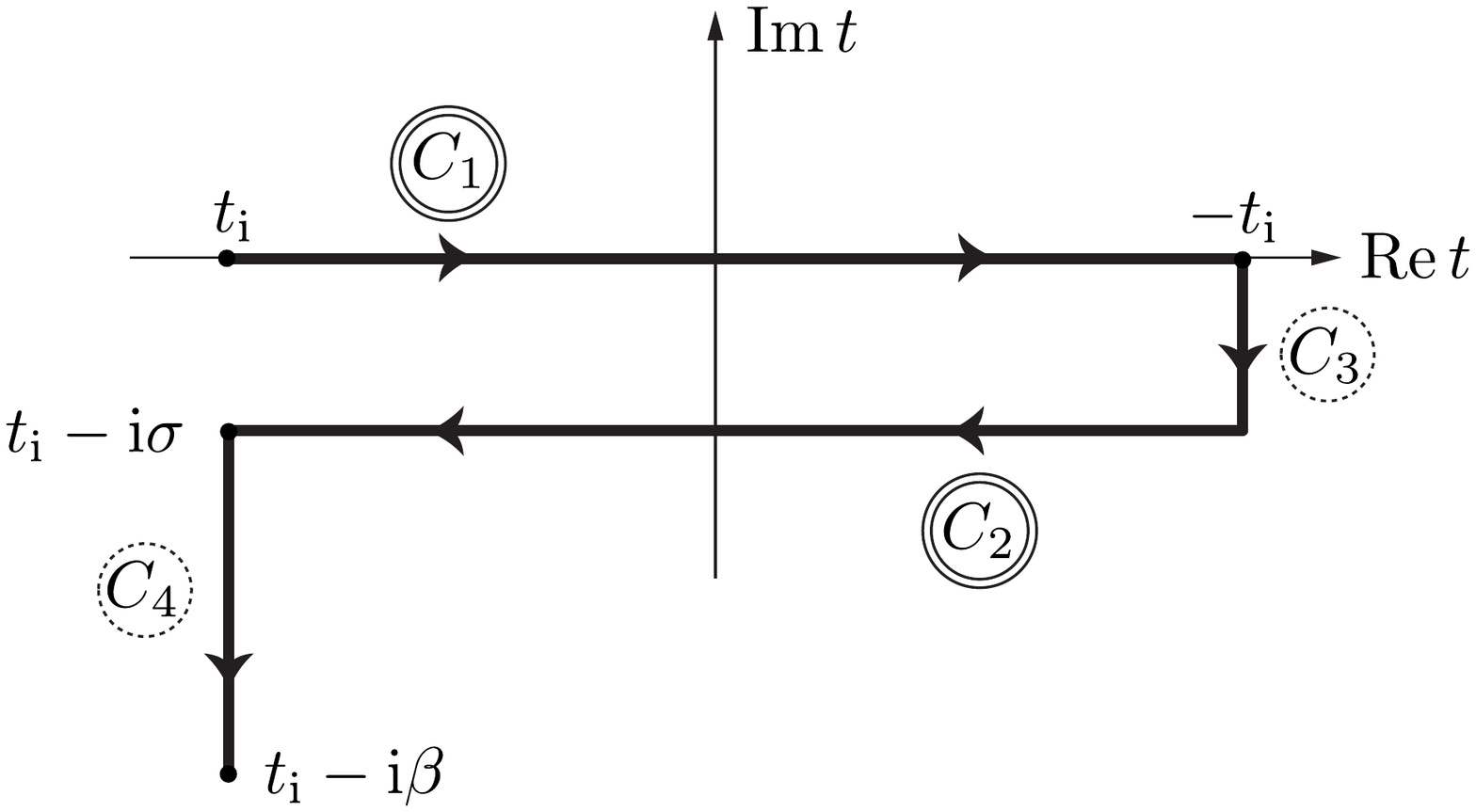}
\vspace*{-4cm}
 \caption{\rm \label{fig:realcontour} Complex time contour including real and imaginary time paths, used in
  the derivation of the $\mu\neq 0$ Feynman rules, where $\sigma\in[0,\beta]$.}
\end{figure}

The function $\mathcal{U}$ is  defined in (\ref{unitop}) and for
the present case, its expression is given in
(\ref{evol})-(\ref{Sigma}). For its evaluation, we have
considered, as detailed in Appendix \ref{app:holo}, the complex
time contour showed in Fig.\ref{fig:realcontour} joining the
points $t_i$ and $t_i-i\beta$ with $\sigma\in[0,\beta]$, which
contains the usual real-time and imaginary-time paths of Thermal
Field Theory and satisfies the usual requirements for the path
integral to be well defined. i.e., $\im t$ is monotonically
decreasing along the contour \cite{Landsman:1986uw}. The
imaginary-time contour  runs in a straight line from $t_i=0$ down
to $-i\beta$ and is denoted as $C_4$, while the $C_1$ and $C_2$
are the paths used in the real-time formulation (see below).

Now, replacing in (\ref{evol})-(\ref{Sigma}) $z\rightarrow
ze^{\beta\mu}$, $\bar{z}'\rightarrow\bar{z}$, $t_f\rightarrow
t_i-i\beta$ is equivalent to replace:

 \begin{equation}\label{betatildedef}
 \beta\rightarrow \tilde{\beta}=\beta\left(1-\frac{\mu}{\omega}\right)
 \end{equation}
except for the $\beta$ appearing in the $C$ contour.  With this replacement one can follow the same steps as in
\cite{ZJ02} for the evaluation of the remaining integral in
(\ref{genfunfree}). Namely, one goes back to the discretized version
of the path-integral (see Appendix \ref{app:holo}), uses again
(\ref{multipleint}) with the modified $A$ matrix and finally takes
again the continuum limit. The final result is:

\begin{equation}\label{pathmec}
\tilde{Z}_\beta^0[j]=\tilde{Z}^0_\beta
\exp\left(-\frac{1}{2}\int\limits_C\mathrm{d}t\ \mathrm{d}t'\
j(t)\tilde{G}_\mathrm{F}(t-t')j(t')\right)\
\end{equation}
with the free partition function
\begin{equation}
\tilde{Z}^0_\beta=\frac{\mathrm{e}^{-\beta\omega/2}}{1-\mathrm{e}^{-\beta(\omega-\mu)}}\
\end{equation}
and the free propagator:
\begin{equation}\label{proplibreqm}
\tilde{G}(t)=\frac{1}{2\omega}
\left[\mathrm{e}^{- i \omega|t|}(1+n(\omega-\mu))
+\mathrm{e}^{ i \omega|t|}n(\omega-\mu)\right]\
\end{equation}
where the  Bose-Einstein  function:

\begin{equation}\label{modbe}
n(x)=\frac{1}{e^{\beta x}-1}
\end{equation}
so that $n(\omega-\mu)$ is the free distribution function at
nonzero $\mu$ in (\ref{bosedisfun}) for a particle of positive energy. Note that we must restrict to $\mu<\omega$ in order that the
previous expressions for the partition function are well defined (see also comments in Appendix \ref{app:therprop}).
The upper limit would correspond to   Bose-Einstein condensation (see below).
 In the above propagator, $\vert
t-t'\vert$ has to be understood in terms of the relative position
of times $t$ and $t'$ with the path routing shown in Fig.\ref{fig:realcontour}.

The result (\ref{pathmec}) for the quantum-mechanical case for $\mu\neq 0$ is one of our main results.
Its importance relies on the fact that we can now easily construct
the generating functional in the interacting case, say
$\hat{H}=\hat{H}_0+V(\hat{q})$ with $V$ the potential, in the usual
way, i.e., by expanding formally in series of $V$ and writing every
term in the expansion in terms of functional derivatives of
$\tilde{Z}_\beta^0[j]$ with respect to $j$. From there, the
extension to a QFT for a real scalar field \footnote{We remark that in the field-theory case, whereas
the Hamiltonian can be expressed as a space integral of a local
field operator,  that is not the case for the number operator when
infinite frequencies appear. This is only possible for exactly
conserved currents.} with lagrangian density:

\begin{equation}
\mathcal{L}=\frac{1}{2}(\partial_\mu\phi)^2-\frac{m^2}{2}\phi^2-\mathcal{V}(\phi)-j\phi
\end{equation}
is given by:

\begin{widetext}
\begin{equation}\label{pathfinal}
\tilde{Z}_\beta[j]=\tilde{Z}_\beta^0\ \exp\left(- i \intcc
\mathcal{V}\left(\frac{\delta}{ i \delta
j(x)}\right)\right)\exp\left(-\frac{1}{2}\intcc\intccp
j(x)\tilde{G}(x-x')j(x')\right)
\end{equation}
\end{widetext}
where $\intcc\equiv\int_C dt \int d^3 \vec{x}$.

 The  generating functional (\ref{pathfinal}) for the interacting case at $\mu\neq 0$  and the corresponding Feynman rules which we discuss below constitute central results of this paper and, to the best of our knowledge, they had not been considered before. It is valid for any
scalar theory, provided one works in the regime where elastic
collisions dominate and particle number is approximately
conserved. The propagator $\tilde G$ appearing in (\ref{pathfinal}) is the
generalization of (\ref{proplibreqm}) to the QFT case when
$\omega\rightarrow E_p$, the particle energy, and therefore we will
restrict in the following to $\mu\leq m$. Recall that the QFT
generalization of $\tilde\beta$ in (\ref{betatildedef}) is  $\betatp$ in (\ref{betatilde}).
 The  explicit
expression of the propagator coincides, as it should, with the
free two-point function (\ref{freeprop})-(\ref{G<}) derived in
Appendix \ref{app:therprop} directly within the canonical
formalism at $\mu\neq 0$ for $t\in\IR$. In this sense, one could somehow expect that the generalization to $\mu\neq 0$ of the generating functional is the one given in (\ref{pathfinal}), although there was no rigorous proof available in the literature. We insist that the usual field-theory derivation for the case  of an exactly conserved charge is not applicable here.

Next, we will discuss the Feynman
rules needed for diagrammatic calculations. The $\mu\neq 0$ case for approximate particle conservation is essentially a nonequilibrium situation, as commented several times before and thus it presents many subtleties to be borne in mind. One of them is the impossibility of defining
properly a Matsubara or imaginary-time formalism, which is related to the way in which the KMS conditions are broken. We  will separate this discussion from the real-time case, where   a suitable formulation is possible, at least to the order we are considering here.

\subsection{Imaginary-time formalism}
\label{sec:itfor}

The Imaginary Time  Formalism (ITF) corresponds to the choice
$t_i=\sigma=0$ for the contour in Fig.\ref{fig:realcontour} so
that one is left only with purely imaginary times $t=-i\tau$ with
$\tau\in[0,\beta]$. At $\mu=0$, this formalism is usually best
suited to deal with thermodynamic quantities such as the free
energy, while retarded Green functions can be derived from it by
analytic continuation in the external frequencies \cite{lebellac}. However, the $\mu\neq 0$
propagator  shows a distinctive feature that complicates
diagrammatic calculations, generating in some cases
ill-defined results. The origin of the problem is the way in which the standard equilibrium KMS periodicity conditions are broken.
 As explained in Appendix \ref{app:therprop}, to which we refer for notation, in our case  (particle number conservation valid both for neutral and charged bosons) we have, in the mixed representation of the propagator,  $\tilde\Delta_T(\tau+\betatp,p)=\tilde\Delta_T(\tau,p)$ instead of the familiar KMS condition $\tilde\Delta_T(\tau+\beta,p)=\tilde\Delta_T(\tau,p)$. This momentum-dependent
periodic condition makes it impossible to define properly a
Matsubara representation in Fourier space, which  can only be done for $\tau\in[-\betatp,\betatp]$,
e.g. eq.(\ref{matsu}), instead of the required $[-\beta,\beta]$
interval where time differences appearing in propagators are
evaluated (note that $\betatp<\beta$). As explained in  Appendix B, this departure of the standard equilibrium KMS condition in the particle number case is crucially different from that of a chemical potential associated to a exact charge conservation, like the electric charge for charged particles. In the charge case, the departure of KMS is given by the constant $\mu_Q$-dependent multiplicative factor in (\ref{kmscharg}) and, as explained in the Appendix, there is no obstruction to define
the Matsubara representation for $\tau\in[-\beta,\beta]$ in that case, e.g. eq.(\ref{chargedmats}), which amounts just to a shift in the Matsubara frequencies.

 Turning again to the case analyzed in this paper, the KMS-breaking mentioned in the above paragraph may
be a problem for instance in diagrams
contributing to the partition function (closed diagrams) whenever there is momentum exchange (time
propagation) inside the diagram, i.e, more than one interaction
vertex, since in that case the imaginary-time variables running in
the internal propagators lie in the interval $[-\beta,\beta]$
while those propagators
  are only $\beta_p$-periodic. When there is just one interaction vertex, time integration factorizes trivially and
  the answer is proportional to powers of the tadpole-like
contribution $\tilde\Delta_T(\tau=0,\vec{x}=0)$ given in Appendix
\ref{app:therprop}. That will be the case for all the
contributions to leading order $\Od(T^6)$ in the calculation of
the ChPT partition function. The diagrams that contribute are
given in Fig.\ref{fig:diagsGL} (see section
\ref{sec:evalz}). However, consider for instance the diagram
labeled 8b in Fig.\ref{fig:diagsGL}, contributing to
    the ChPT free energy density to $\Od(T^8)$.
Taking for simplicity constant vertices, as in the case of
$\mathcal{V}=\lambda\phi^4/4!$, this diagram in the ITF would be
proportional to:

\begin{equation} \label{itz8b}
I=\frac{\tilde
G^2(0)}{\beta}\int\frac{\mathrm{d}^3\vec{p}}{(2\pi)^3}
\int_0^\beta d\tau' \int_0^\beta d\tau \tilde\Delta_T
(\tau-\tau',p) \tilde\Delta_T (\tau'-\tau,p)
\end{equation}
with $\tilde G (0)=\tilde\Delta_T(0)$ given in
eqs.(\ref{tadpole})-(\ref{g1tilde}). Now, as commented above, we
cannot just replace the Fourier representation for $\tilde
\Delta_T$ in (\ref{matsu}) since it is only defined for the
$[-\betatp,\betatp]$ interval. This obstruction produces
additional unnatural terms. The appearance of those terms can be
seen by using the mixed representation for $\Delta_T$ in terms of
$\tilde G^>$ and $\tilde G^<$ given in (\ref{G>})-(\ref{G<}) and
performing explicitly the $\tau,\tau'$ integrals in (\ref{itz8b}).
We get:

\begin{eqnarray} \label{iprob}
I&=&-\tilde G^2(0)\frac{\partial}{\partial m^2}\tilde
G(0)\nonumber\\&+& \frac{\tilde
G(0)}{\beta}\int\frac{\mathrm{d}^3\vec{p}}{(2\pi)^3}
\frac{1}{8E_p^4}\left\{
\left[1+\tilde{n}_p(E_p)\right]^2\left[e^{2\beta\mu}-1\right]
%\right.\nonumber\\&+&\left.
\left[\tilde{n}_p(E_p)\right]^2\left[e^{-2\beta\mu}-1\right]\right\}
\end{eqnarray}
with $E_p^2=\vert\vec{p}\vert^2+m^2$.
The first term above gives the standard result for $\mu=0$
 with the replacement of the distribution function
$n\rightarrow \tilde n$, as one would expect from  kinetic theory
arguments, while this property does not hold for the additional
terms. The remaining contributions vanish for $\mu=0$ but they do
not do so in the $T\rightarrow 0^+$ limit where they diverge. This
contradicts the natural physical expectation that in the
$T\rightarrow 0^+$ limit and for $\mu<m$, the free energy should
reduce to the vacuum contribution.

A related conflict arises when trying to calculate correlation functions in the ITF. The loss of KMS $\beta$-periodicity implies that the dependence on external times is not only through time-differences. In particular, this means that correlators depend on $t_i$.  Consider for instance the tadpole-like contribution (we omit the spatial dependence for simplicity):

\begin{equation}
\int_{it_i}^{it_i+\beta} d\tau \tilde\Delta_T (\tau_1-\tau) \tilde\Delta_T (\tau-\tau_2)=\int_{\tilde T}^{\tilde T+\beta} d\tau \tilde\Delta_T (\tau_1-\tau_2-\tau) \tilde\Delta_T (\tau)
\label{ittadpole}
\end{equation}
where $\tilde T=it_i-\tau_2$. Now, if we make $\int_{\tilde T}^{\tilde T+\beta}=\int_0^\beta-\int_0^{\tilde T}+\int_\beta^{\tilde T + \beta}$, the change of variable $\tau\rightarrow \tau+\beta$ in the third integral does not cancel the second one due to the loss of $\beta$-periodicity of $\tilde\Delta_T$. Therefore, the result does not depend only on $\tau_1-\tau_2$ but on $\tilde T$, i.e., depends on $\tau_1$ and $\tau_2$ independently and, as a consequence, the dependence on $t_i$ does not vanish.

As we will see in sections
  \ref{sec:evalz} and \ref{sec:pimasswidth}, terms of the type showed above  appear in the self-energy to leading order and in the partition function  at  order $\Od(T^8)$. In the latter case,  our approximation reaches its   validity limit, since particle-changing
  processes start playing an important role. However, precisely for that reason, at the temperatures where  the $\Od(T^8)$ needs to be included we
  may consider in practice $\mu\ll
  T,m$ for these contributions. Recall that in fact, the
  conflictive terms in (\ref{iprob}) are $\Od(\mu/T)$  so that we will be introducing only small corrections by
  neglecting them. The presence of those unnatural terms in the ITF may be also understood  if we note that
  we are facing a nonequilibrium situation,  where
  the ITF is not appropriate and which must be formulated using a  contour including real times
  \cite{Chou:1984es,Altherr:1994jc}. We will indeed see next that one can define a suitable Real-Time Formalism (RTF) so that
  these problems are not present, at least to the order we consider here, and one can calculate properly not only thermal
  correlators but also  vacuum diagrams contributing to the free energy.

\subsection{Real-time formalism}
\label{sec:rtfor}

We consider now the full contour in Fig.\ref{fig:realcontour}
and, following the standard notation, we denote by $\tilde
D_{ij}=\tilde G (t_i-t_j)$ with $t_i\in C_i, t_j\in C_j$. We then
have for the $C_{1,2}$ parts of the contour (we omit the spatial
dependence for simplicity):

\begin{eqnarray}
\tilde D_{11}(t-t')&=&\tilde G^>(t-t')\theta(t-t')+\tilde G^<(t-t')\theta(t'-t)\nonumber\\
\tilde D_{22}(t-t')&=&\tilde G^<(t-t')\theta(t-t')+\tilde G^>(t-t')\theta(t'-t)\nonumber\\
\tilde D_{12}(t-t')&=&\tilde G^<(t-t'+i\sigma)=\tilde D_{21}(t'-t)
\label{rtpropprev}
\end{eqnarray}
where $t,t'\in\IR$ and $\tilde G^>,\tilde G^<$ given in
(\ref{G>})-(\ref{G<}) and so on for the remaining components.

In order to formulate properly the RTF at $\mu\neq 0$ we take first, as customary, $t_i\rightarrow-\infty$. This is necessary if we want to calculate Green functions with arbitrary real time arguments. In principle, this choice implies also that, imposing vanishing asymptotic conditions for the $j$ currents and for the spectral function, which hold also in our case, the  generating
functional for $\mathcal{V}=0$ can  be factorized as \cite{Landsman:1986uw}:

\begin{equation}
\tilde{Z}_{\beta,C}^{\mathcal{V}= 0}[j]=\mathcal{N}\
\tilde{Z}_{\beta,C_{12}}^{\mathcal{V}= 0}[j]\
\tilde{Z}_{\beta,C_{34}}^{\mathcal{V}= 0}[j]\ ,
\end{equation}
so that one could calculate real-time correlation functions
without worrying about the imaginary-leg contributions. However,
as it  was pointed out  in \cite{Niegawa:1989dr,Gelis:1999nx},
there are imaginary-time contributions that still survive in
particular diagrams, for instance self-energy insertions, which
indeed we will calculate here. Nevertheless, there is a standard
rule for collecting all the relevant contributions but using only
the propagators in $C_{1,2}$, the so called $\vert p_0
\vert$-prescription \cite{Niegawa:1989dr,Gelis:1999nx,lebellac}.
This prescription amounts to use in Fourier space $n(\vert p_0
\vert)$ instead of the seemingly equivalent $n(E_p)$ when
multiplied by the on-shell  $\delta$-function, as in
(\ref{tildeGp0}). For instance, with this prescription one obtains
that a simple constant tadpole-like insertion in the self energy
such as the diagram showed in Fig.\ref{fig:selfenergy}a with a
constant vertex, amounts  to a redefinition of the mass, as
expected. It also guarantees that there are no ill-defined
contributions, such as products of $\delta$ distributions at the
same point which in principle could appear when multiplying the
RTF propagators. What we will show here is that  for  $\mu\neq 0$
there is also a natural  prescription which works, now in terms of
$n\rightarrow \tilde n_p (p_0)$, leading to the same
properties at the order considered here. However, it must be pointed out that to higher orders, there may be additional ill-defined terms arising from a  nonequilibrium distribution \cite{Altherr:1994jc}.  Our RTF avoids the main obstruction that we faced in the
ITF,  since the length $\beta$ of the imaginary leg disappears
from the integration limits in momentum space, whose  Fourier
representation is well defined now. Moreover, we also choose
$\sigma\rightarrow 0^+$. Therefore, for Green functions with
real-time arguments for which we neglect (with the above
prescription) the $C_{3,4}$ parts, we end up with a Keldysh-like
contour characteristic of nonequilibrium Thermal Field Theory
\cite{Chou:1984es}. With this procedure we will see that an
additional property holds:  most  results can be written as
functionals of $\tilde n$, which encodes all the $T,\mu$
dependence. This is also a expected property from kinetic theory
arguments, at least for the leading order corrections in $\tilde
n$ (dilute gas regime).

This allows then to calculate properly any real-time correlation
function directly, i.e., without appealing to the analytic
continuation from the ITF, which is cumbersome for $\mu\neq 0$. In
addition, as we will see below, one can also obtain information about the free energy density
without using the ITF. Let us then write  the propagators
(\ref{rtpropprev}) in momentum space for our choice of contour
(note that the $\tilde D_{11}$ component corresponds to the free
propagator $\tilde G$ in (\ref{tildeGp0})):

\begin{eqnarray}
\tilde
D_{11}(p_0,p)&=&\frac{i}{p_0^2-E_p^2+i\epsilon}+2\pi\delta(p_0^2-E_p^2)
n(\vert
p_0 \vert-\mu)
\nonumber\\
\tilde D_{22}(p_0,p)&=&
\frac{-i}{p_0^2-E_p^2+i\epsilon}+2\pi\delta(p_0^2-E_p^2)n(\vert
p_0 \vert-\mu)\nonumber\\
\tilde D_{12}(p_0,p)
&=&
2\pi\delta(p_0^2-E_p^2)\left[\theta(-p_0)+n(\vert
p_0 \vert-\mu)\right]\nonumber\\
\tilde
D_{21}(p_0,p)&=&
2\pi\delta(p_0^2-E_p^2)\left[\theta(p_0)+n(\vert
p_0 \vert-\mu)\right] \label{rtpropmom}
\end{eqnarray}

In the above propagators,  we have  chosen, as discussed above,  the  $\vert p_0
\vert$ prescription ensuring that the distribution function does
not depend explicitly on $E_p$, as in the $\mu=0$ case. We will
see below that this yields the same expected properties as for
$\mu=0$.  It can also be readily checked that our $\mu\neq 0$ RTF
propagators above coincide with those given in \cite{Baier:1996if},  obtained assuming
a direct replacement of the distribution function by the $\mu\neq
0$ nonequilibrium one. In fact, the free propagators (\ref{rtpropmom}) can be readily recast into the general
nonequilibrium Keldysh form, given for instance in \cite{Chou:1984es,Altherr:1994jc} by taking for the nonequilibrium distribution function our $\tilde n_p(k_0)$ given in (\ref{ntildedef})-(\ref{betatilde})\footnote{The convention in \cite{Altherr:1994jc} is such that the $D_{12}$ and $D_{21}$ components are reversed with respect to ours.}, which satisfies the property (\ref{disfunprop}), as required for general nonequilibrium derivations \cite{Altherr:1994jc}.

To provide a particularly relevant example of our previous
statements, let us consider the tadpole-like correction to the
self-energy given by the diagram in Fig.\ref{fig:selfenergy}a
with a constant vertex (the generalization to derivative vertices
appearing in  ChPT calculations will be straightforward).  The
external leg is fixed to be of ``type 1", since we are calculating
the two-point function with real arguments, i.e, the first-order
correction to  $D_{11}$. Then, if we consider only the $C_{1,2}$
contributions,  this diagram gives in position space:

\begin{equation}\label{defF}
F(x-y)=i\sum_{j=1,2}\int_{C_j} \tilde D_{1j}(x-z) \tilde D_{jj} (0) \tilde D_{j1} (z-y)
\end{equation}

Note that correlators depend only on space and time differences in the RTF, so that the problems discussed in the previous section, related to the ITF version of the tadpole in eq.(\ref{ittadpole}) are not present now.

Now, we take into account that $\tilde D_{11}(0)=\tilde
D_{22}(0)=\tilde G(0)$ in (\ref{tadpole})-(\ref{g1tilde}). Then,
the Fourier transform of $F$ is:

\begin{eqnarray}
F(p_0,p)=i\tilde G(0) \left[ \tilde D_{11}^2 (p_0,p)-\tilde D_{12}
(p_0,p) \tilde D_{21}(p_0,p)\right]
\end{eqnarray}

We replace in the above equation the propagators in
(\ref{rtpropmom}) and use
$\frac{\delta(x)}{x+i0^+}=-\frac{\delta'(x)}{2}-i\pi\delta^2(x)$
where as customary we keep the regulator in the definition of
$\delta(x)=\frac{i}{2\pi}\left(\frac{1}{x+i0^+}-\frac{1}{x-i0^+}\right)$.
Thus, we can write:

\begin{widetext}
\begin{eqnarray}
F(p_0,p) &=& i\tilde
G(0)\left\{\left[\frac{i}{p_0^2-E_p^2+i0^+}\right]^2-2\pi i
n(\vert p_0 \vert-\mu)\delta'(p_0^2-E_p^2)\right\}
=-\tilde G(0)\frac{\partial}{\partial m^2} \tilde D_{11}(p_0,p)
       \label{tadpolert}
        \end{eqnarray}
\end{widetext}

Note that it is in the last step in the previous equation where it
is  crucial to use the $\vert p_0 \vert$ prescription chosen above
since $n(\vert p_0 \vert-\mu)$ is independent of $m^2$.
Therefore, the result (\ref{tadpolert}) implies that the only
modification in the $\tilde D_{11}$ propagator is $m^2\rightarrow
m^2-\tilde G(0)$, which is the expected result of mass
renormalization which in addition is obtained from the $\mu=0$
case by replacing $n\rightarrow \tilde n$ in the (finite) thermal
correction to the tadpole diagram given by the function  $\tilde
g_1(m,T,\mu)$ in (\ref{g1tilde}). Note that to this order and with this prescription we have been able to get rid of the ill-defined $\delta^2$ terms. However, this prescription might not be  enough when higher orders are included, since it has been shown in \cite{Altherr:1994jc} for a $\lambda\phi^4$ theory that  additional nonequilibrium ill-defined terms arise, which should be properly regulated with a nonzero particle width. At the order corresponding to our previous result (\ref{tadpolert}) we coincide with \cite{Altherr:1994jc}.  We will comment more about this issue in section \ref{sec:pimasswidth}.

Two more important remarks are in order. The first one is that the
spectral properties of the interacting theory are really defined
from retarded Green functions, not from time-ordered ones. From
the ITF, retarded correlators are defined directly by analytic
continuation. However, we have seen that this is not a well
defined procedure for $\mu\neq 0$. The solution of the problem of
finding retarded Green functions from the RTF time-ordered product
was given in \cite{Kobes:1990kr}. In that work, a set of rules
(so-called circling rules) were provided in order to define a function
that has the required causal retarded properties, namely that
satisfies that one of the outgoing lines of the corresponding
diagram has the largest time component. It was then shown in
several examples that this function coincides with the analytic
continuation of the ITF correlator. Now,  it can be checked that
the same properties of the free propagators used in
\cite{Kobes:1990kr} for the derivation of the circling rules hold
for our $\tilde D_{ij}$ propagators and therefore the same rules
lead to the RTF retarded function at $\mu\neq 0$. The application of those rules
is trivial for the tadpole case discussed above, since there is
only one vertex. However, they will be of use for the case
of higher order contributions to the self energy which we will
consider below, like the thermal width arising from diagram
\ref{fig:selfenergy}b.

The second remark has to do with the calculation of thermodynamic
quantities within the RTF, i.e., the partition function or the
free energy density. In principle, due to the factorization of the
imaginary-leg commented above, the contribution to vacuum graphs
when summing over fields of type 1 and 2 vanishes identically.
However, it was shown in \cite{Fujimoto:1984kh} that fixing one of
the vertices of a vacuum diagram to be ``external" of type 1 and
summing over the remaining internal vertices with an overall
$\beta$ factor reproduces the free-energy result and for $\mu=0$
coincides with the ITF. The functional arguments used in those
papers are also applicable to our $\mu\neq 0$ case and in fact,
the direct use of that prescription leads to the expected answers.
Let us show this for the case of the $\Od(T^8)$ diagram 8b in
Fig.\ref{fig:diagsGL}, analyzed in section \ref{sec:itfor} in the
ITF. Applying the previous prescription and with constant
vertices, we get that this diagram is now proportional to:

\begin{eqnarray}\label{rtz8b}
i\tilde G^2 (0)\sum_{j=1,2}\int_{C_j} \tilde D_{1j}(x-z) \tilde
D_{j1}
(z-x)=\tilde G(0) F(0)
%\nonumber\\
=-\tilde G^2 (0)\frac{\partial}{\partial m^2} \tilde G(0)
\end{eqnarray}
with $F$ in (\ref{defF}). We then see that we arrive to the ITF
result (\ref{iprob}) but without the additional terms discussed in
that section, since the proportionality factors between this
diagram and (\ref{rtz8b}) or (\ref{itz8b}) come only from
combinatorics and are therefore identical. We will use this
real-time prescription to define properly our free energy.

\section{Applications to the pion gas}
\label{sec:apppion}

\subsection{Evaluation of the ChPT free energy.}
\label{sec:evalz}

We apply our previous results to the pion gas, described by ChPT
with two light quark flavors of mass $m_u=m_d\equiv m_q$ \cite{Gasser:1983yg,Gerber:89}. The lagrangian is  constructed as an
expansion in derivatives and pion masses,  generically $\Od(p)$
with $p\ll\Lambda_\chi\sim$ 1 GeV, so that ${\cal L}={\cal
L}_2+{\cal L}_4+\cdots$ with ${\cal L}_{2k}=\Od(p^{2k})$. In the
range of temperatures and chemical potentials we are interested,
both $T,\mu_\pi=\Od(p)$ formally, which corresponds to $T$  below
$T_c\sim$ 200 MeV. The ChPT $\Od(p^D)$ power of a given diagram is given by Weinberg's power counting $D=2(N_L+1)+\sum_k 2 N_k(k-1)$ \cite{we79} where $N_L$ is the number of loops and $N_k$ is the number of vertices coming from ${\cal L}_{2k}$. In our approach, we do not perform any formal
chiral expansion in $\mu_\pi$, except in  higher order
contributions (see our discussion below and in section \ref{sec:formal}) where it
is reasonable to expand in $\mu_\pi/T$. We will follow closely the
notation and conventions in \cite{Gerber:89}, where the explicit
expressions of the ${\cal L}_2$ and ${\cal L}_4$ can be found. The
lagrangian ${\cal L}_2$ is the nonlinear-$\sigma$ model, whose
free parameters are the pion decay constant and mass to leading
order $f=f_\pi(1+\Od(p^2))$ with $f_\pi\simeq$ 93 MeV and
$m=m_\pi(1+\Od(p^2))$, $m_\pi\simeq$ 140 MeV.  To fourth order,
${\cal L}_4$ contains five independent low-energy constants
$l_{1-4}$ and $h_1$ which absorb the divergences of the one-loop
diagrams with only ${\cal L}_2$ vertices. The renormalized $\bar
l_i$ appear in physical processes such as pion scattering and
therefore their values can be fitted experimentally. We will use
the same central values given in \cite{Gasser:1983yg,Gerber:89} in
order to compare more easily with the results in \cite{Gerber:89}
at $\mu_\pi=0$. Those values are $\bar l_1=-6.6$, $\bar l_2=6.2$,
$\bar l_3=2.9$, $\bar l_4=3.5$. The constant $h_1$ multiplies a
contact term and appears in the vacuum free energy and quark condensate. We use also
the estimate in \cite{Gasser:1983yg,Gerber:89} of $\bar h_1\simeq$ 3.4.
The lagrangians of higher orders will only appear through
renormalization either of the vacuum energy or the pion mass and
therefore the  low-energy constants of those orders will not show
up once the results are expressed in terms of the physical pion
mass (see details below).

\begin{figure}[h]
\hspace*{-1.5cm}
\includegraphics[scale=.6]{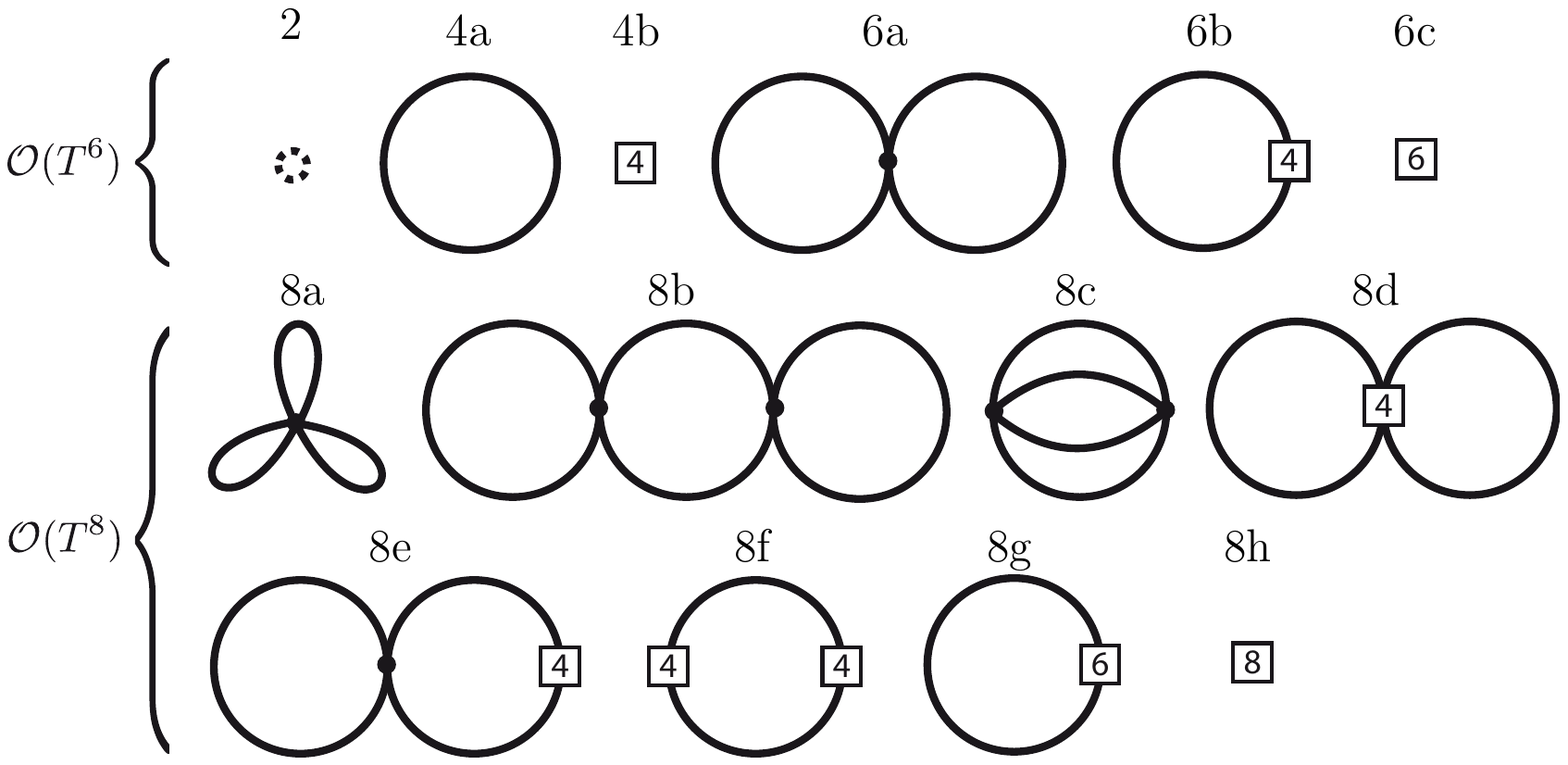}
%\vspace*{-9.5cm}
\vspace*{-11cm}
 \caption{\rm \label{fig:diagsGL} Feynman diagrams contributing to the partition function of the
pion gas up to and including $\Od(T^8)$. The first row includes diagrams up to $\Od(T^6)$ while the second and third rows are the $\Od(T^8)$ contributions. The dots denote interaction vertices coming from ${\cal
L}_2$ while those vertices coming from higher order lagrangians
are indicated by a square box. The notation is the same as in  \cite{Gerber:89}.}
\end{figure}

The free energy density $z$, from which  thermodynamical observables can be obtained,
 is defined as customary:

\begin{eqnarray}
\tilde z(T,\mu_\pi)=-T\lim_{V\rightarrow\infty} \frac{1}{V}\log\tilde Z_\beta (T,\mu_\pi)
\end{eqnarray}

We also define the thermodynamic pressure as in \cite{Gerber:89}, i.e, subtracting its $T=0$ contribution given by the vacuum energy density:

\begin{equation}
\tilde P (T,\mu_\pi) =  \tilde z_0 - \tilde z(T,\mu_\pi) \quad ; \quad \tilde z_0=\lim_{T\rightarrow 0^+}\tilde z
\end{equation}

It is important to emphasize that all our results for the pressure and quantities derived from it have to be understood strictly as time-dependent throughout the plasma expansion, in the sense explained in section \ref{sec:motivation}, the time evolution towards a chemically equilibrated phase being driven by $\mu_\pi(T)$.
The diagrams contributing to the free energy in ChPT are the closed diagrams showed in Fig.\ref{fig:diagsGL}, where we follow the same convention as
\cite{Gerber:89} to name the diagrams. The number assigned to each diagram indicates the order in the chiral expansion and the numbers inside the boxes in the vertices refer to the lagrangian order, the case of ${\cal L}_2$ being indicated  by a dot. Recall that for a given order of the lagrangian, there are vertices with arbitrary number of (even) pions due to the chiral expansion of the $SU(2)$-valued chiral field $U=\exp(i\pi^a\tau_a/f)$ where $\tau^a$ are the Pauli matrices and $\pi^a$ the pion field.

 The leading order $\tilde z_{2}=-f^2m^2$ coming from the contact term (independent of the pion field) in ${\cal L}_2$, is independent of $T$ and $\mu_\pi$ and therefore contributes only to the vacuum energy density $\tilde z_0$. Note that, according to our discussion in the previous sections, we will ensure that all our contributions have a well-defined $T\rightarrow 0^+$ limit for $\mu_\pi<m$, i.e, that the contributions to $\tilde z_0$ to any chiral order are  $\mu_\pi$-independent.  The next order corresponds to diagram 4a and 4b. $\tilde z_{4a}$ corresponds to the quadratic pion field contribution in ${\cal L}_2$ and is therefore
 nothing but the free partition function given in
(\ref{freeparfung0}) multiplied by 3 accounting for the three pion
degrees of freedom. The divergent contribution to $\tilde z_{4a}$ is $T$
and $\mu_\pi$ independent and therefore it merely renormalizes
$\tilde z_0$.

The next order in the chiral expansion is $\Od(T^6)$ and the
diagrams contributing are $\tilde z_{6abc}$ in Fig.\ref{fig:diagsGL}.
It is important to remark that this is the first order where pion
interactions show up. Graph 6c renormalizes $\tilde z_0$, while 6b is of
the same form as 4a and therefore gives rise to the free partition
function contribution but with the mass shifted by its tree level
${\cal L}_4$ renormalization (see section \ref{sec:pimasswidth})
which depends on $l_3$. As for diagram 6a, taking into account
(\ref{secder}), its contribution is proportional to $\tilde
G^2(0)$. As discussed in section \ref{sec:formal}, in this case
the result is trivially identical in both ITF and RTF and
corresponds to the result in \cite{Gerber:89} replacing
$G(0)\rightarrow \tilde G(0)$:

\begin{equation}
\tilde z_{6a}=\frac{3m^2}{8f^2}\tilde G^2(0) \label{z6a}
\end{equation}

The divergent contribution in (\ref{z6a}), according to
(\ref{tadpole})-(\ref{tadpolezero}), contains a contribution to $\tilde z_0$ and
another one which cancels, as it should, with the one in $l_3$ so
that, using (\ref{derfor}), the total finite result for the pressure to $\Od(T^6)$ is:

\begin{equation}
\tilde P =\frac{3}{2}\tilde g_0(m_\pi,T,\mu_\pi)-\frac{3}{8}\frac{m^2}{f^2} \left[\tilde g_1(m,T,\mu_\pi)\right]^2+\Od(T^8)
\label{z6}
\end{equation}
with the functions $\tilde g_1$ and $\tilde g_0$   given in (\ref{g1tilde}) and (\ref{g0tilde}) respectively and
where $m_\pi$ is the physical pion mass at $T=\mu_\pi=0$, related to the bare mass $m$ to this order as \cite{Gasser:1983yg}:

\begin{equation}
m_\pi^2=m^2\left[1-\frac{\bar l_3}{32\pi^2}\frac{m^2}{f^2}+\Od(m^4)\right]
\label{pimasszero}
\end{equation}

Recall that, to this order, the difference between $m_\pi$ and $m$
is only relevant in the $\tilde g_0$ contribution in (\ref{z6}).
The same applies to the distinction between $f$ and $f_\pi$:

\begin{equation}
f_\pi^2=f^2\left[1+\frac{\bar
l_4}{8\pi^2}\frac{m^2}{f^2}+\Od(m^4)\right] \label{fpizero}
\end{equation}

\begin{figure}[ht]
%\vspace*{-2cm}
%\hspace*{-2.5cm}
%\includegraphics[scale=0.55]{2pito4pi.pdf}
\includegraphics[scale=0.65]{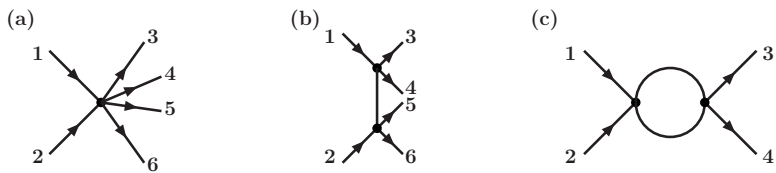}
%\vspace*{-12cm}
\vspace*{-14cm}
 \caption{\rm \label{fig:2pito4pi} (a), (b) Diagrams contributing to leading order (tree level) to $2\pi\rightarrow 4\pi$ processes.  (c)
A one-loop contribution to elastic $\pi\pi$ scattering.}
\end{figure}

We consider now the $\Od(T^8)$ contributions showed in Fig.\ref{fig:diagsGL}. Now, there are several aspects which make the
calculation qualitatively different from the $\Od(T^6)$ one. An
important point is that to this order we may expect that our
approximation of  particle number conservation is less accurate,
since vertices entering number-changing processes   show
up. Consider for instance the  diagrams contributing to
 $2\pi\leftrightarrow 4\pi$ processes in the thermal bath, which to
leading order in ChPT are given by the tree level diagrams showed
in Fig.\ref{fig:2pito4pi}a and \ref{fig:2pito4pi}b. Now, unlike the $\Od(T^6)$
case, one can draw  vacuum diagrams from these processes by
identifying external lines. For instance, joining lines in pairs in the graph in
Fig.\ref{fig:2pito4pi}a
as 1-2, 3-4, 5-6 and equivalent combinations  leads to diagram 8a in
Fig.\ref{fig:diagsGL}. Similarly, joining 1-2, 3-6, 4-5 in
Fig.\ref{fig:2pito4pi}b produces diagram 8c. This is not a
one-to-one correspondence. For instance, joining 1-3 and 2-4 lines
in the elastic one-loop diagram in Fig.\ref{fig:2pito4pi}c yields
also diagram 8c. Diagram 8b can also be obtained from a elastic
process (Fig.\ref{fig:2pito4pi}c joining 1-2, 3-4) or from an
inelastic one (Fig.\ref{fig:2pito4pi}b joining 1-3, 2-6, 4-5). The
crucial point is that none of the $\Od(T^6)$ vacuum closed diagrams in Fig.\ref{fig:diagsGL} can be obtained from the lowest order inelastic diagrams in
Fig.\ref{fig:2pito4pi}a,\ref{fig:2pito4pi}b. This distinctive feature can be interpreted as a way to
identify the validity range of our approximation. However, we should bear
in mind that these $\Od(T^8)$ corrections are meant to be relevant
only very near $T_c$ \cite{Gerber:89} and therefore in the region
where chemical equilibrium is nearly restored and
$\mu_\pi\rightarrow 0$, not surprisingly due to the presence of
the particle-changing processes just discussed \cite{Song:1996ik}.
Precisely for this reason, the $\mu_\pi$-dependence of these
diagrams is suppressed in powers of $\mu_\pi/T$ and
$\mu_\pi/m_\pi$. Therefore, numerically our approach will be still justified to this order. In addition, as we have explained in section
\ref{sec:formal}, taking $\mu_\pi/T$ small justifies in practice
to  get rid of unnatural terms in the ITF formulation.

With the above considerations in mind, we proceed to evaluate the
$\Od(T^8)$ diagrams in Fig.\ref{fig:diagsGL}. Graph $8h$
renormalizes $\tilde z_0$ and  graphs 8f,g renormalize the pion mass
to $\Od(m^6)$.
Graph 8a  is proportional to a third power
of the propagator at the origin, with the same coefficient as in
\cite{Gerber:89}:

\begin{equation}
\tilde z_{8a}=-\frac{25m^2}{48 f^4} \tilde G^3(0)
\end{equation}
which  contains divergent contributions, according to
(\ref{tadpole}).

The graph 8b has been analyzed in section \ref{sec:formal}. The
relevant integral contributing to this graph is (\ref{itz8b}) in
the ITF and (\ref{rtz8b}) in the RTF with the prescription
discussed in that section. The difference between both
formulations is of $\Od(\mu_\pi/T)$ and therefore expected to be
numerically small, for the reasons just discussed. The rest of the
contributions
 to this graph are proportional to $\tilde G^3(0)$ and the proportionality constants
 are the same as in \cite{Gerber:89}.
 Thus, adopting the RTF prescription, we get:

 \begin{equation}
\tilde z_{8b}=\frac{m^2}{16 f^4}\tilde G^2(0)\left(8+3m^2\frac{\partial}{\partial m^2}\right)\tilde G(0)
 \label{z8b}
 \end{equation}
whose divergent contribution can be also separated using (\ref{tadpole}).

Graphs 8d,e have the same form as graph 6a in
(\ref{z6a}), but due to the form of the ${\cal L}_4$ lagrangian and following also our previous RTF prescriptions, we arrive to the same structure as in \cite{Gerber:89}:

 \begin{align}
 &\tilde z_{8d}+\tilde z_{8e}= -\frac{3}{f^4}\left\{(2l_1+4l_2)\left[\tilde G_{\mu\nu}\right]^2\right.\nonumber\\
 &+\left.\tilde G(0)\left[\left(3l_1+l_2+l_3\right)m^4\tilde G(0)-\frac{l_3}{2}m^6\frac{\partial}{\partial m^2}\right]\tilde G(0)\right\}
 \label{z8de}
 \end{align}
 where $\tilde G_{\mu\nu}=\partial_\mu\partial_\nu \tilde G (0)$, which has the same properties as in \cite{Gerber:89}, namely, its divergent contribution is the same ($T$ and $\mu_\pi$ independent) while its finite part can be written in the same way in terms  of $\tilde g_0$ and $\tilde g_1$ .

The remaining graph is 8c. Following again the RTF prescription,
this contribution is:

\begin{equation}
\tilde z_{8c}=\frac{1}{48 f^4}\left[3m^4\tilde J_1-72\tilde J_2+16m^2\left(\tilde G(0)\right)^3\right]
\label{z8c}
\end{equation}
where:

\begin{eqnarray}
\tilde J_1&=&i\int d^4 x \left[\tilde D_{11}^4(x)-\tilde
D_{12}^4(x)\right]
%\nonumber\\&=&
=\int d^3\vec{x}\int_0^\infty dt
\left\{\left[\tilde G^>(t,\vec{x})\right]^4-\left[\tilde
G^<(t,\vec{x})\right]^4\right\} \nonumber\\
\tilde J_2&=&i\int d^4 x \left[\left(\partial_\mu\tilde D_{11}(x)\partial^\mu\tilde D_{11}(x)\right)^2
%\right.\nonumber\\&-&\left.
-\left(\partial_\mu\tilde D_{12}(x)\partial^\mu\tilde D_{12}(x)\right)^2\right]
\label{tildej12}
\end{eqnarray}

Written in the above form,
it is not difficult to show that for $\mu_\pi=0$, when the propagators
are $\beta$-periodic, i.e. $G^<(t+i\beta)=G^>(t)$, one has for instance
$J_1=\int d^3\vec{x} \int_0^\beta d\tau \Delta_T^4(\tau,\vec{x})$ and similarly for $J_2$ \cite{Gerber:89}. As we have seen, for $\mu_\pi\neq 0$ the
periodicity condition does not hold. However, for this diagram,
instead of working directly with the RTF expressions
(\ref{tildej12}), we will make use of the fact that $\tilde
G^<(t+i\beta)=\tilde G^>(t)+\Od(\beta\mu_\pi)$ and neglect the
non-periodic terms, so that we end up with $\tilde J_1\simeq\int
d^3\vec{x} \int_0^\beta d\tau \tilde \Delta_T^4(\tau,\vec{x})$ and $\tilde J_2\simeq\int
d^3\vec{x} \int_0^\beta d\tau \left(\partial_\mu\tilde \Delta_T(\tau,\vec{x})\partial^\mu\tilde \Delta_T(\tau,\vec{x})\right)^2$. This approximation simplifies considerably the renormalization of this graph, since now we
can follow the same steps as in \cite{Gerber:89}. First we
separate $\tilde\Delta (\tau,\vec{x})=\tilde G^>(-i\tau,\vec{x})=\tilde\Delta (\tau,\vec{x})^{T=\mu_\pi=0}+\delta\tilde\Delta (\tau,\vec{x})$ using the representation (\ref{G>})-(\ref{G<}). The divergent contributions in the integrals  (\ref{tildej12}) are then contained in the $(\delta\tilde\Delta\tilde\Delta^0)^2$, $\delta\tilde\Delta(\tilde\Delta^0)^3$ and $(\tilde\Delta^0)^4$ terms and can be  renormalized with the same  counterterms as in \cite{Gerber:89}  replacing the $g_{0,1}$ by  $\tilde g_{0,1}$.  The finite part of the $\tilde J_{1,2}$ integrals can be evaluated numerically. A crucial point is that  this approximation is consistent, as far as renormalization is concerned, with our previous evaluation of the $\tilde z_{8abde}$ diagrams since the divergent parts of the terms proportional to $\tilde g_0^2,\tilde g_0\tilde g_1$ and $\tilde g_1^2$ arising from the $\tilde J_{1,2}$ integrals cancel exactly with those coming from the other four diagrams, while the terms proportional to $\tilde g_1$ add together to renormalize the physical pion mass according to the definition

\begin{equation}
m_\pi=-\lim_{T\rightarrow 0^+} T\log\tilde P (T,\mu_\pi=0)
\end{equation}

In addition, as it happens for $\mu_\pi=0$, this ensures that neither the tree level constants from ${\cal L}_6$ nor the $T,\mu_\pi$ independent renormalization constants needed to render $\tilde J_{1,2}$ finite appear in the final expression for the free energy once it is expressed in terms of $m_\pi$. We remark that with our representation, not only the renormalization procedure is consistent, but the final answer for
the full $\Od(T^8)$ contribution amounts  to replace
$n(E_p)\rightarrow n(E_p-\mu_\pi)$ in all the spatial momentum
integrals,  without dealing with unnatural terms,
like those discussed in section \ref{sec:formal}.

After the previous detailed evaluation, we arrive finally to a finite expression for the free energy, suitable for numerical evaluation, with the  approximations discussed above implying that the $\Od(T^8)$ corrections are reliable only for small $\mu_\pi$. From this expression we proceed to present our results for the $\mu_\pi$ dependence of several relevant observables.

\subsection{Results for thermodynamical observables}
\label{sec:thermo}

From the energy density, we  obtain the quark condensate (the order parameter of the chiral transition), the entropy density and the pion number density in the
standard way:

\begin{eqnarray}
\langle \bar q q \rangle(T,\mu_\pi)&=&\langle \bar q q \rangle(0,0)\left[1+\frac{c}{f^2} \frac{\partial \tilde P (T,\mu_\pi)}{\partial m_\pi^2}\right]\label{conddef}\\
\tilde s(T,\mu_\pi)&=&\frac{\partial \tilde P (T,\mu_\pi)}{\partial T}\label{sdef}\\
\tilde n(T,\mu_\pi)&=&\frac{\partial \tilde P (T,\mu_\pi)}{\partial \mu_\pi}
\label{ndef}
\end{eqnarray}
where $c=1-m^2 (4\bar h_1+\bar l_3-1)/(32\pi^2 f^2)+\Od(m^4)$.

\begin{figure*}
%\vspace*{-2cm}
%\hspace*{-2cm}
\centerline{\includegraphics[width=6cm,height=4.2cm]{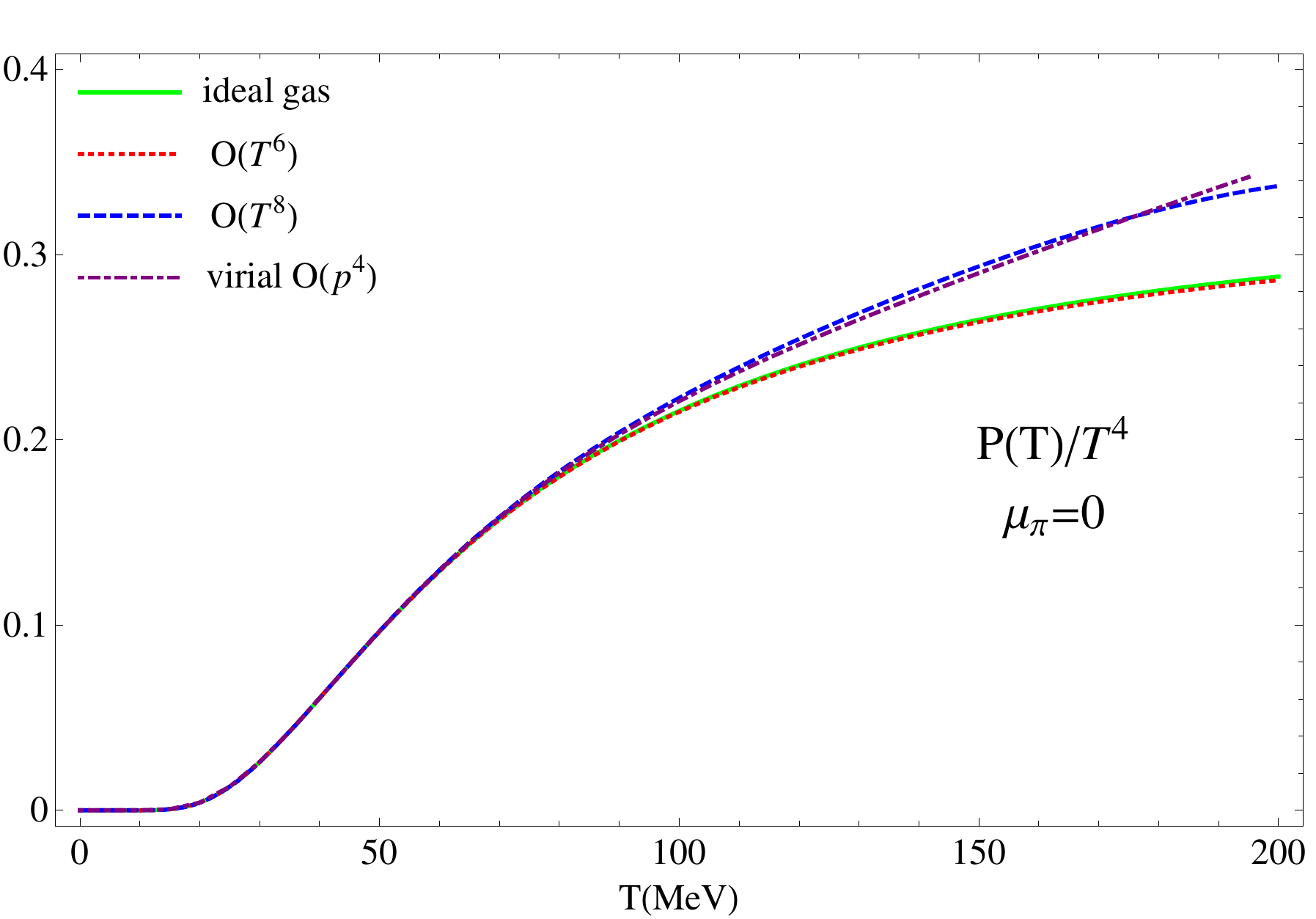}\includegraphics[width=6cm,height=4.2cm]{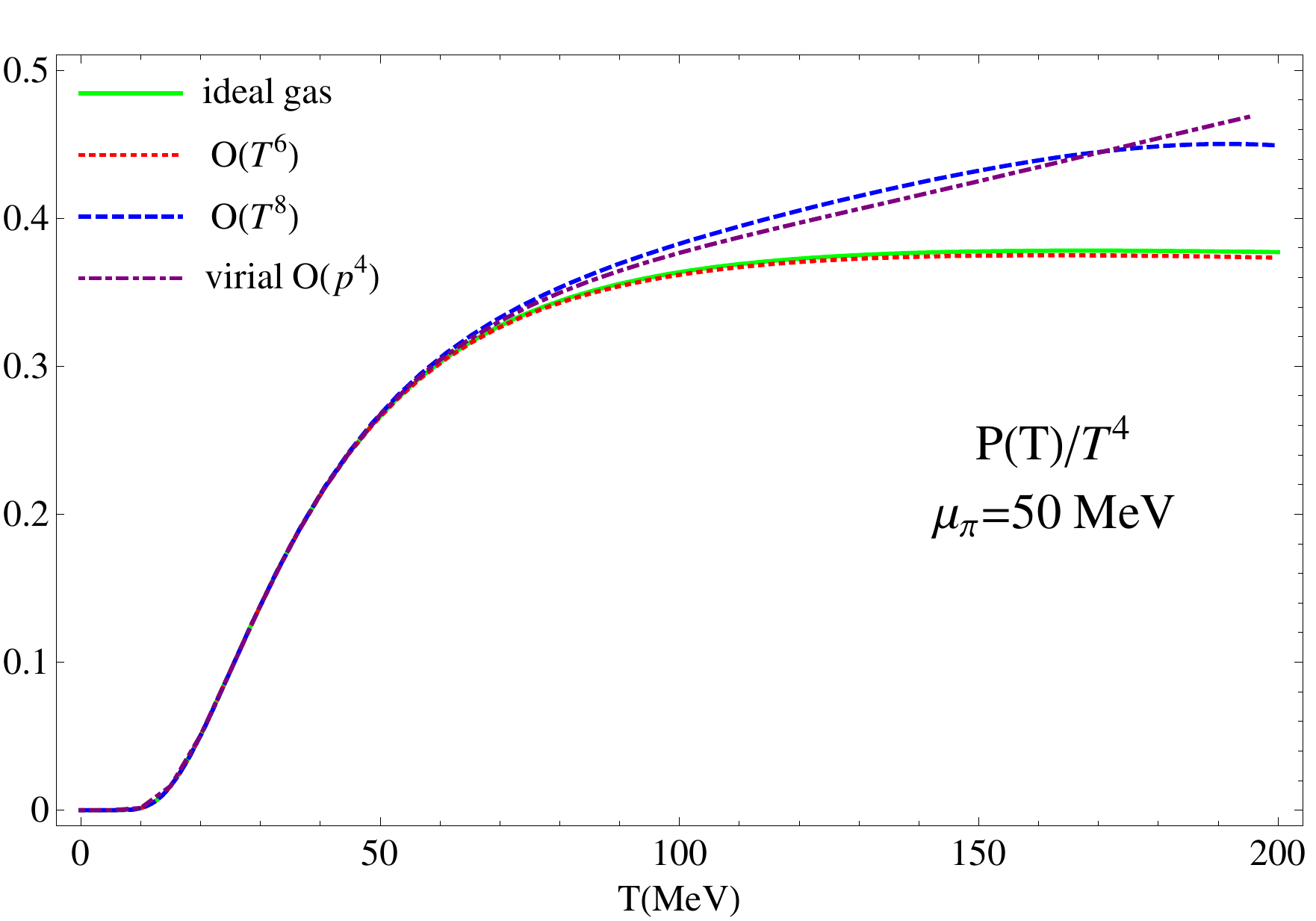}\includegraphics[width=6cm,height=4.2cm]{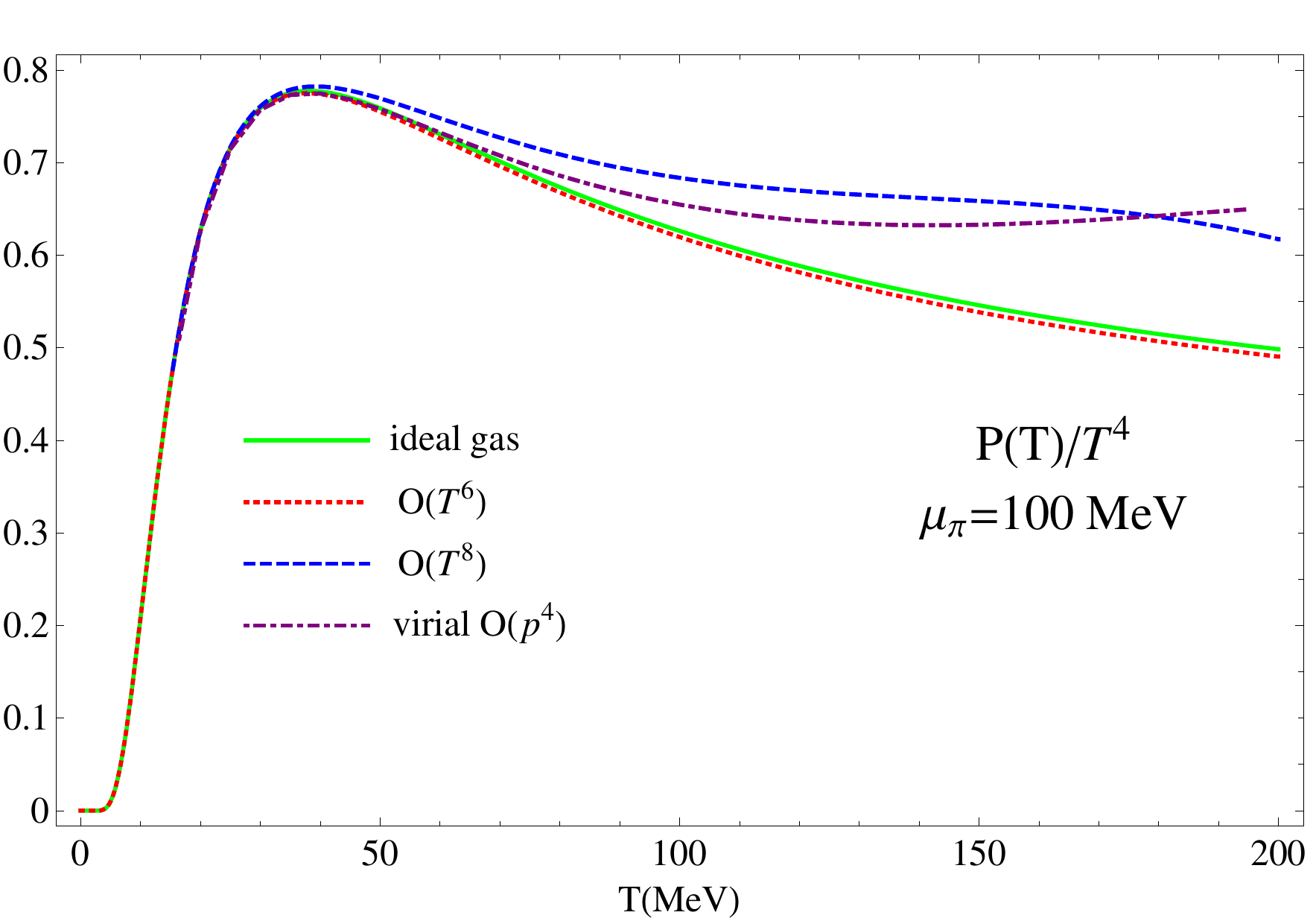}}
\centerline{\includegraphics[width=6cm,height=4.2cm]{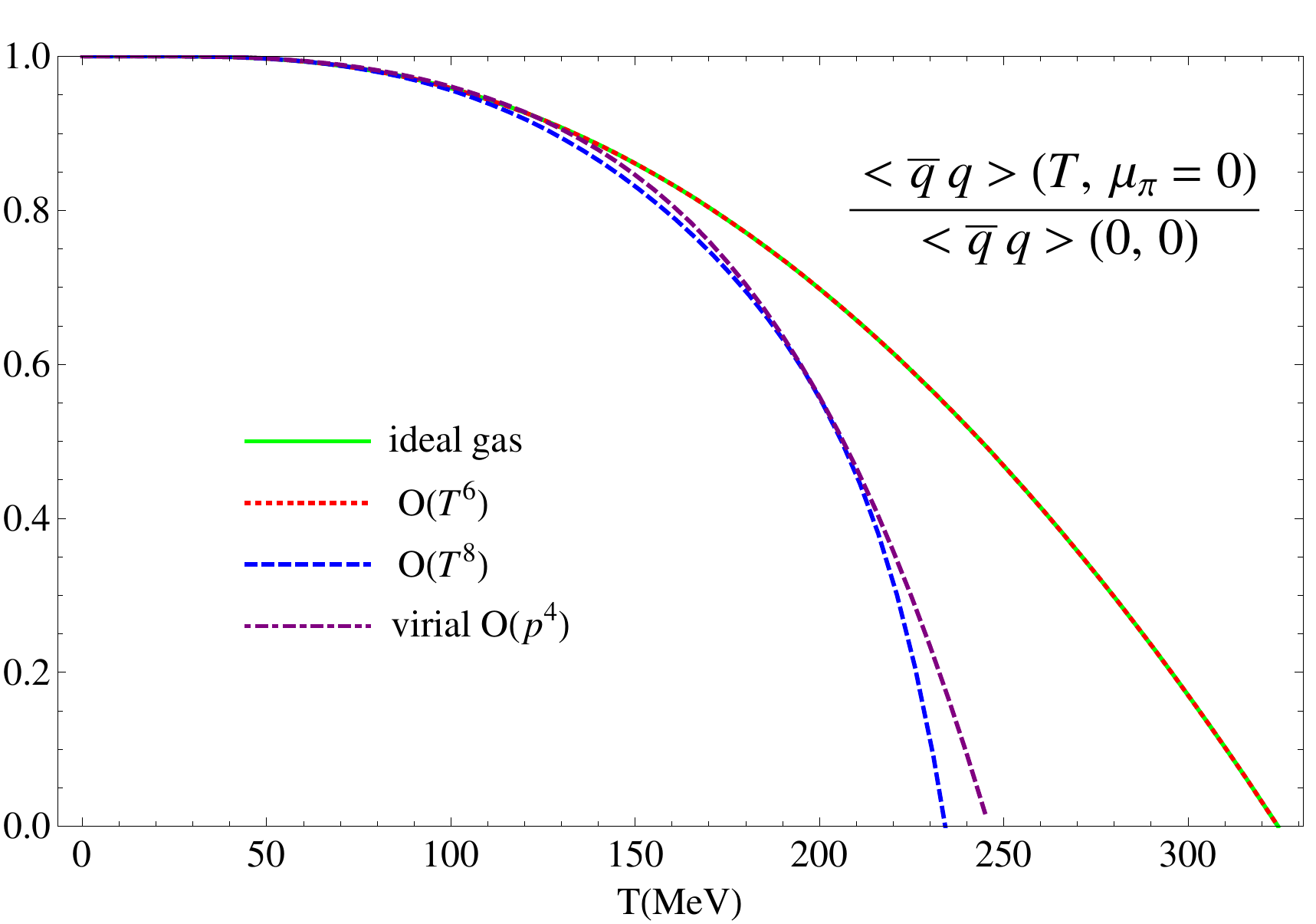}\includegraphics[width=6cm,height=4.2cm]{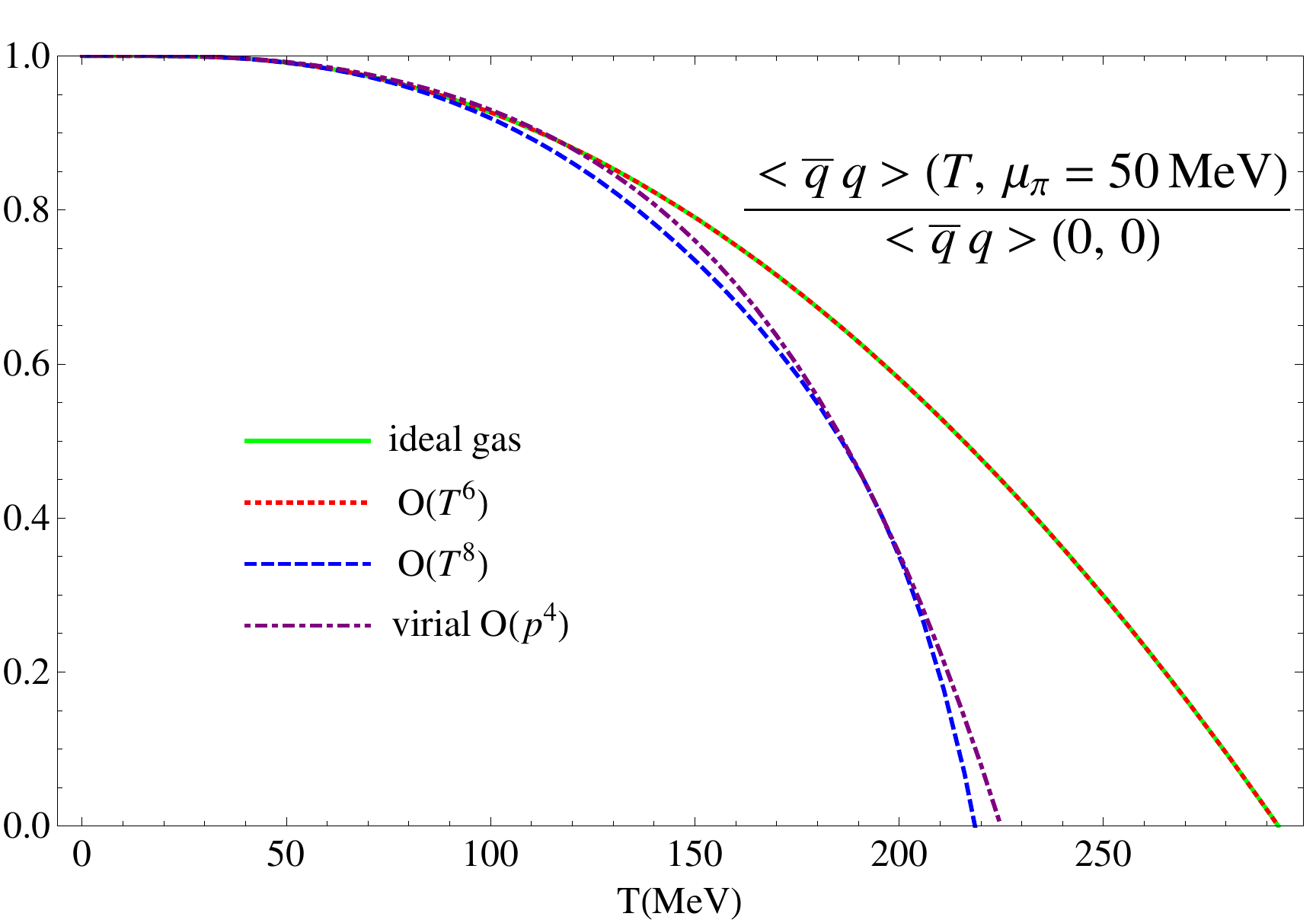}\includegraphics[width=6cm,height=4.2cm]{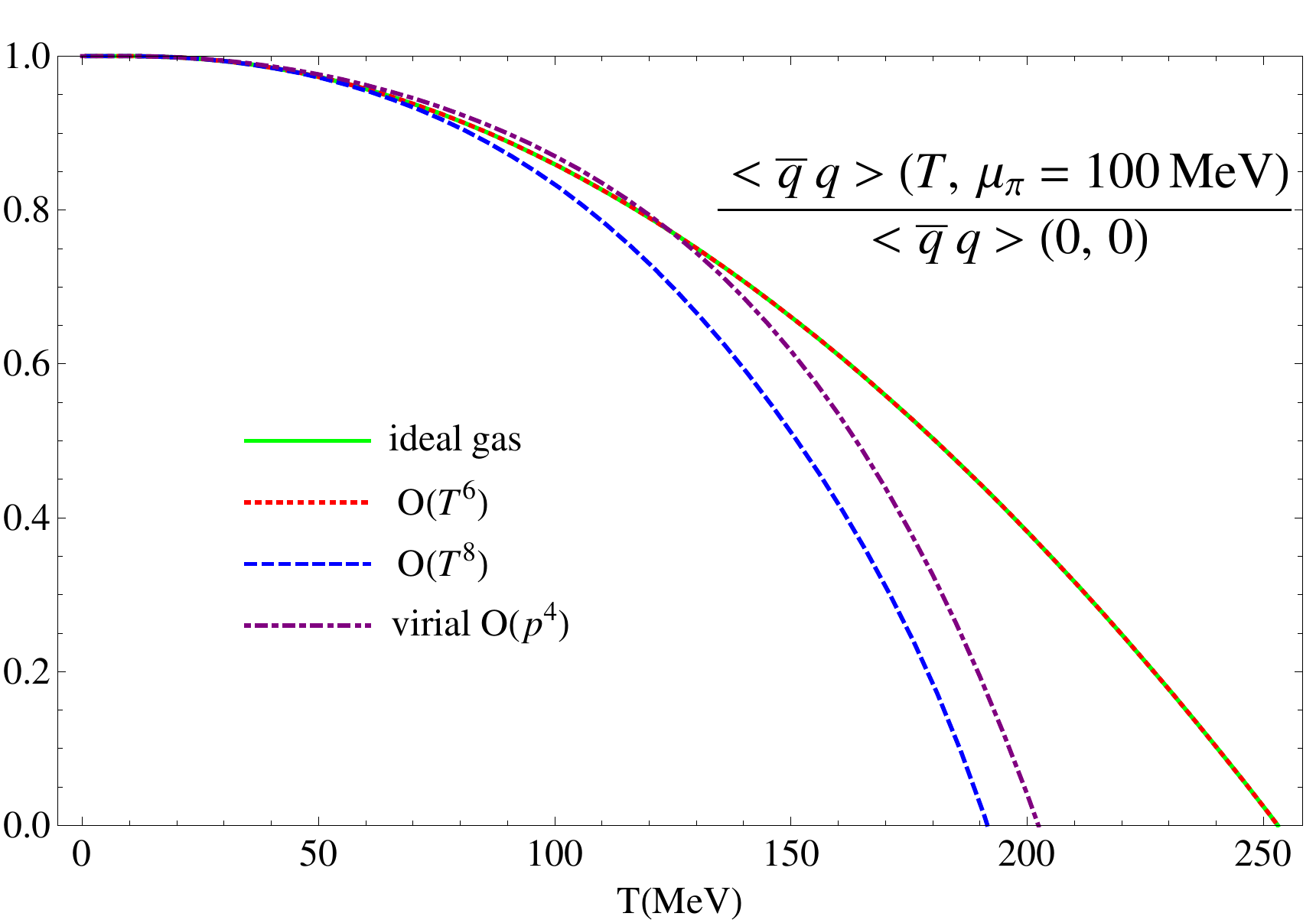}}
\centerline{\includegraphics[width=6cm,height=4.2cm]{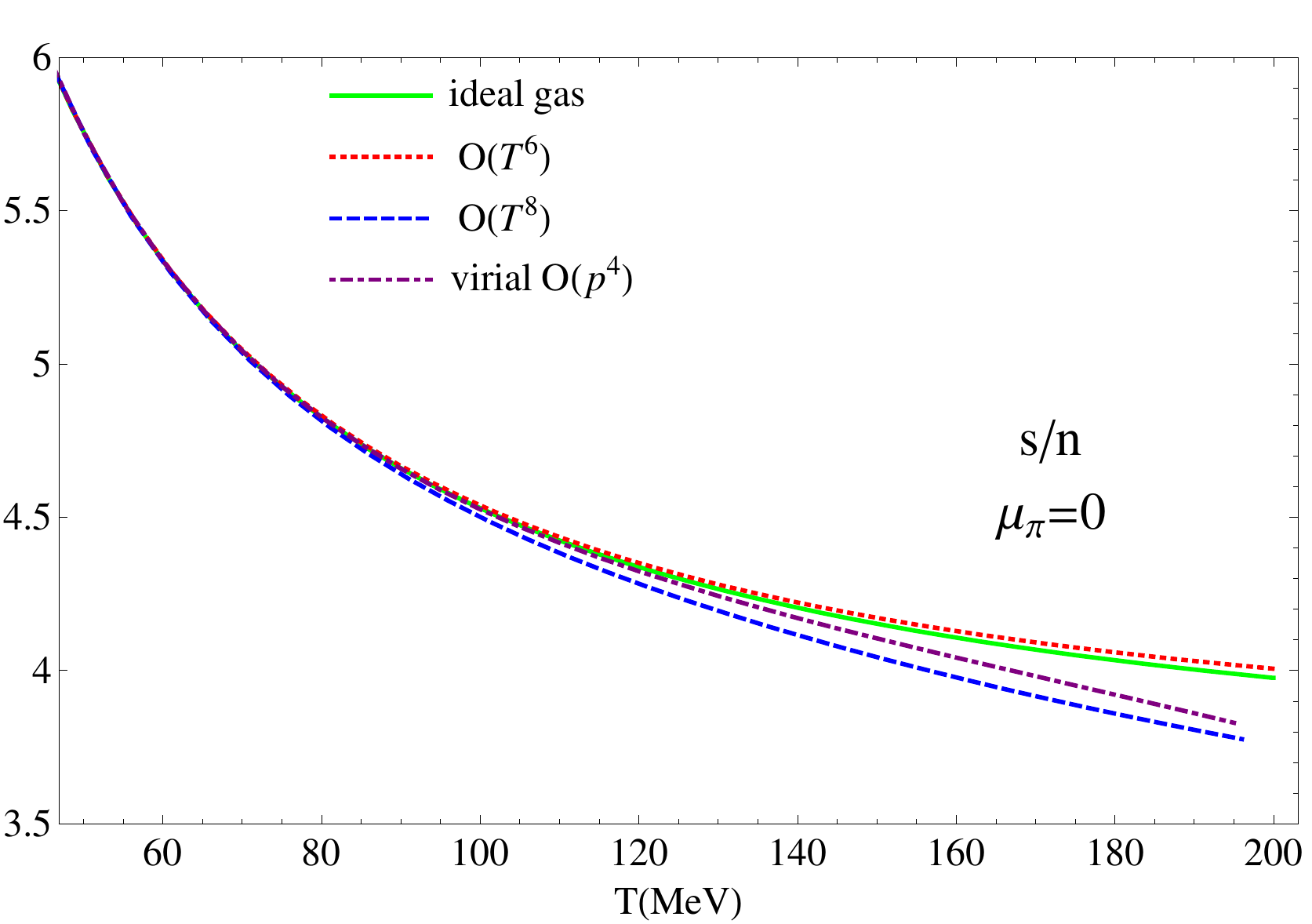}\includegraphics[width=6cm,height=4.2cm]{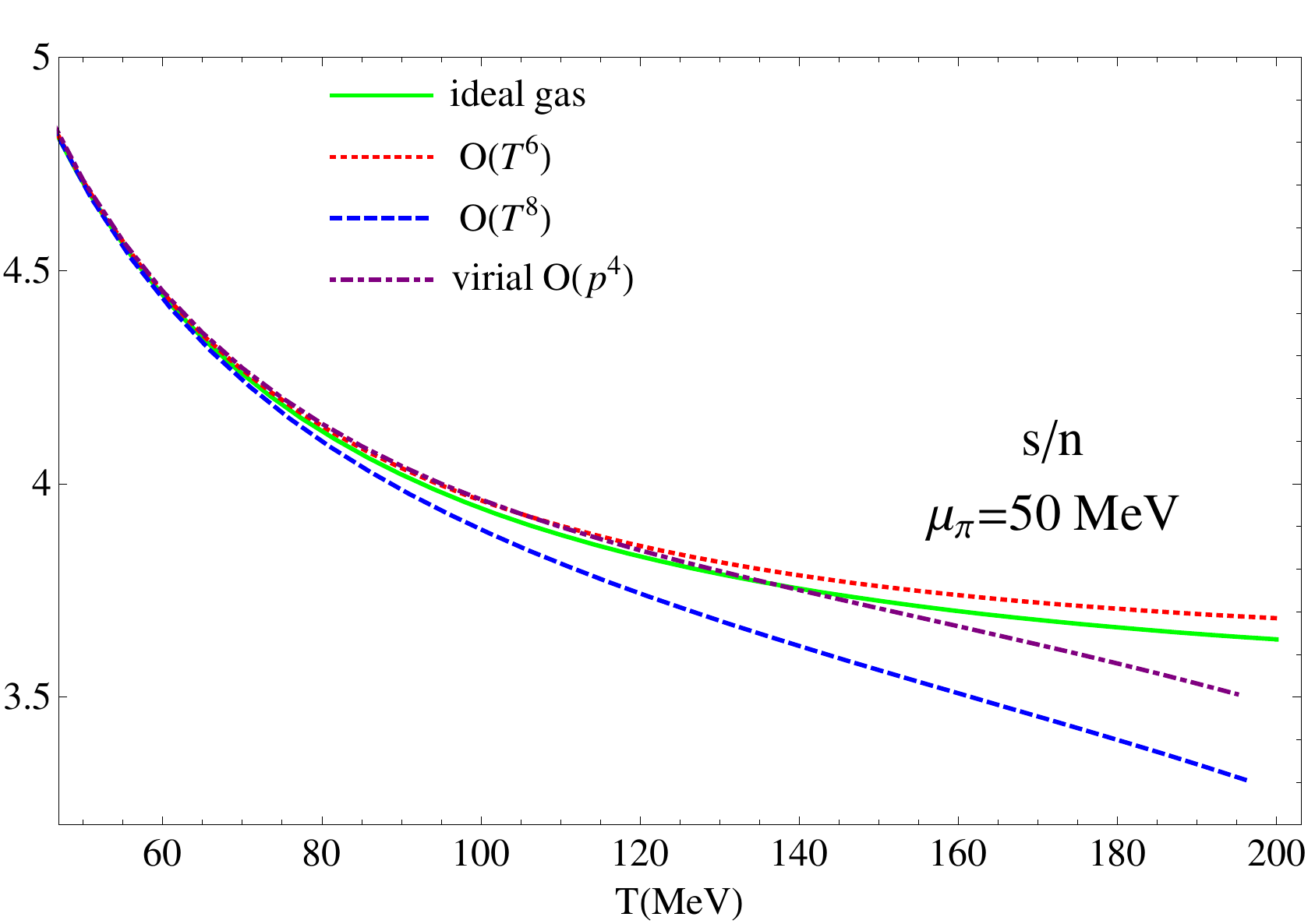}\includegraphics[width=6cm,height=4.2cm]{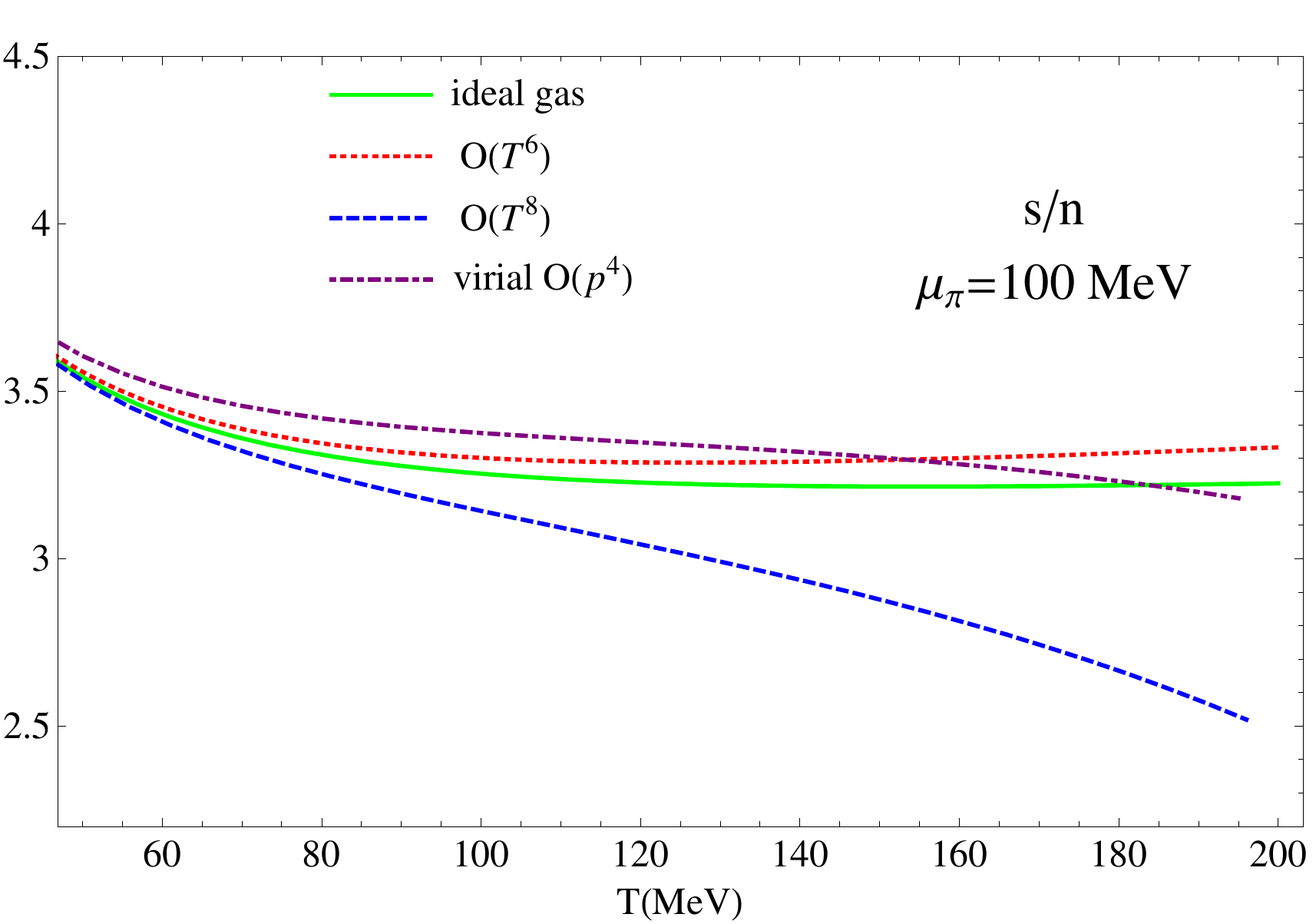}}
% Here is how to import EPS art
%\vspace*{-13cm}
 \caption{\rm \label{fig:resultsther} Results for the pressure, quark condensate and entropy over density ratio  at different chemical potentials and to
  different orders in the ChPT interactions.}
\end{figure*}

We plot our results in Fig.\ref{fig:resultsther}. The first
feature we observe is that the $\Od(T^6)$ and the ideal gas curves
are very close to one another for all the range of temperatures and chemical
potentials showed. Sizable differences due to the interactions
only show up numerically when including the $\Od(T^8)$. This is a
also a feature of the $\mu_\pi=0$ calculation \cite{Gerber:89}.
For instance, in the chiral limit ($m_\pi=0$) and for $\mu_\pi=0$,
the $\Od(T^6)$ in (\ref{z6}) vanishes identically, while the
$\Od(T^8)$ survives, producing conformally anomalous contributions
to the pressure \cite{FernandezFraile:2009mi}. In Fig.\ref{fig:resultsther} we also compare
our results with the virial gas approach \cite{Dobado:1998tv},
where the pressure can be written at low pion density in terms of
the pion scattering phase shifts. In the curves showed in Fig.\ref{fig:resultsther}, the phase shifts have been calculated
perturbatively to $\Od(p^4)$ in ChPT and using the same set of
low-energy constants that for our perturbative  results with the
approach of the present paper. We see that our $\Od(T^8)$ results
with $\mu_\pi\neq 0$ lie reasonably close to the virial result, at
least for not very high $\mu_\pi$. This is a good consistency
check of our present approach.

Another general feature that we observe in the curves is that the
effect of the pion chemical potential is always to increase
thermal effects. Effectively, it acts as similarly as a reduction
of the effective pion mass (this is more accurate for $T\ll m_\pi$
where the typical momenta in the distribution functions are
$p=\Od(\sqrt{T/m_\pi})$) and therefore for  fixed $T$, the
results for increasing $\mu_\pi$ go qualitatively in the same
direction as for increasing $T$ with $\mu_\pi=0$. For instance, we
see that the pressure increases for increasing $\mu_\pi$ and
approaches faster with $T$ the asymptotic limit $P\sim \pi^2 T^4/30$ \cite{Gerber:89} expected in the
chiral limit ($T\gg m_\pi,\mu_\pi$). The effect of interactions is
also to increase the pressure, producing additive contributions in
the ChPT expansion.

The curves for the quark condensate show that the chiral
restoration temperature goes down for $\mu_\pi\neq 0$. This is
also a consequence of the above discussed qualitative behavior,
since the system for $\mu_\pi\neq 0$ is closer to chiral
restoration. With
the numerical values we get, we see that if chemical freeze-out
takes place for temperatures below the chiral phase transition,
then we do not expect to see any change in the value of $T_c$. On
the contrary if $T_{chem}>T_c$ (which is less likely with the
available experimental information) we would expect a reduction in
$T_c$ compared with the estimates taking $\mu_\pi=0$.

\begin{figure*}
%\vspace*{-2cm}
%\hspace*{-2cm}
\includegraphics[scale=.5]{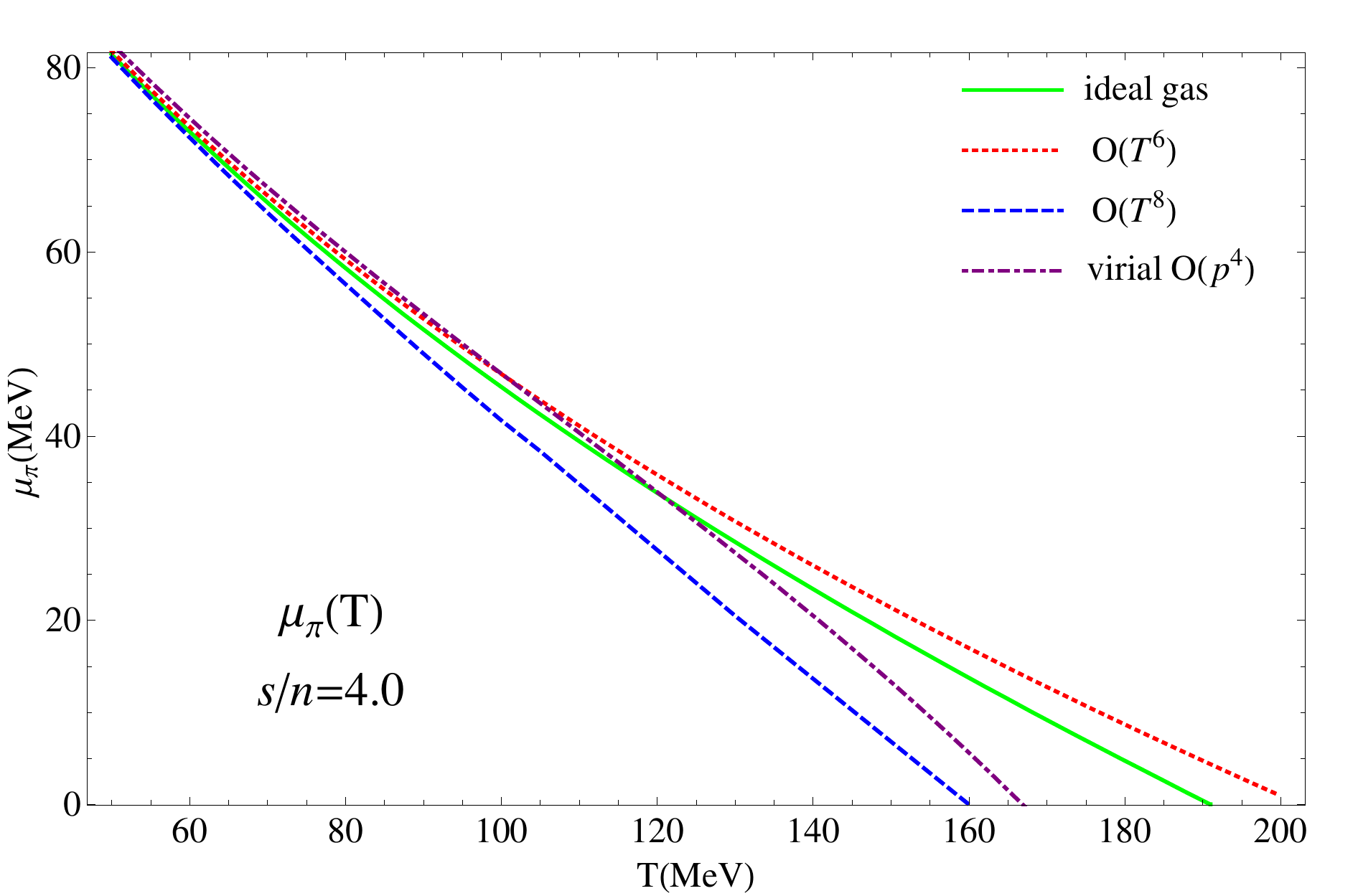}
% Here is how to import EPS art
%\vspace*{-13cm}
 \caption{\rm \label{fig:muT} Dependence of $\mu_\pi(T)$ in the isentropic approximation, with the fixed value
 $s/n=4$.}
\end{figure*}

It becomes clear from our discussion in section \ref{sec:motivation} that incorporating additional physical requirements  allowing to describe $\mu_\pi(T)$   is crucial in our approach, in order to be consistent with the chemical nonequilibrium
evolution. In this sense,
a very interesting observable  is the ratio of entropy density to pion density,
also plotted in Fig.\ref{fig:resultsther}.  It has been  pointed out
\cite{Bebie:1991ij,Hung:1997du,Song:1996ik}  that on general grounds one expects
this ratio to remain almost constant during the expansion. This is
the isentropic expansion approximation, which is exact in the high
$T$ limit $T\gg m_\pi,\mu_\pi$ for the ideal gas. We remark that
we are restricting here to the gas of pions. If  heavier degrees
of freedom are included, such as the $\rho$, one has to account
for the total number of pions $\bar n_\pi=n_\pi+2n_\rho+\dots$
which includes those ``stored" in the $\rho$ if  the channel
$\rho\rightarrow \pi\pi$ is considered as the only source of pion
number changing, and similarly with other resonances (see details
in \cite{Bebie:1991ij,Song:1996ik}). The idea is then that by
fixing $s/n$ to a given value at the chemical freeze out
temperature $T_{chem}$ where $\mu_\pi=0$,  going down in the
temperature scale one can keep $s/n$ fixed by increasing
$\mu_\pi$, as it can be seen in Fig.\ref{fig:resultsther}. This
provides the isentropic dependence $\mu_\pi (T)$, which is given in
\cite{Bebie:1991ij,Song:1996ik} in the ideal gas approximation. We
plot in Fig.\ref{fig:muT} the isentropic curves $\mu_\pi(T)$
with a reference value $s/n=4$, for which $T_{chem}\simeq$ 190 MeV
for the ideal gas. The obtained curves follow a roughly linear behaviour, as expected phenomenologically \cite{Hung:1997du}. The most significant effect we observe is the
reduction of $T_{chem}$ when $\Od(T^8)$ or virial interactions are included.
This is a very natural effect since, as we have discussed in
previous sections, that order in the interaction is the one where
particle-changing processes begin to be relevant and drive the
system back to chemical equilibration. The virial curve  lies
reasonably close to our perturbative $\Od(T^8)$ since the
 two approaches differ significantly only for rather high values of $\mu_\pi$ and $T$, which are not
 reached along the curve $\mu_\pi(T)$. In fact, in the isentropic evolution our $\Od(T^8)$ approach
 is better justified since
 $\mu_\pi (T)\ll T,m_\pi$.    We also remark that the same
effect of faster equilibration is seen when comparing the curves
of the ideal pion gas with that of the ideal pion+resonances gas,
as done in \cite{Bebie:1991ij}. One can check that the curves in
that paper for $s/n$ as a function of $T$ for different $\mu_\pi$
are systematically lower when including resonances,  as in our
case in Fig.\ref{fig:resultsther} when including the $\Od(T^8)$
or in the virial case and therefore the free pion and resonance
gas equilibrates faster, which is the feature that we are able to
reproduce here including higher order pion interactions.

\subsection{Self energy: pion thermal mass and width.}
\label{sec:pimasswidth}

\begin{figure}[h]
\vspace*{-2cm}
%\hspace*{-2cm}\includegraphics[scale=0.6]{tadpole.pdf}
\hspace*{-2cm}\includegraphics[scale=0.7]{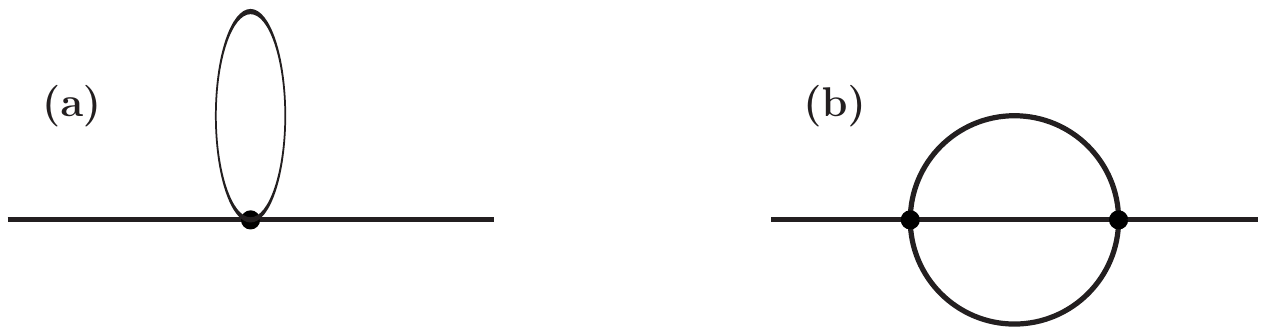}
%\vspace*{-13cm}
\vspace*{-15cm}
 \caption{\rm \label{fig:selfenergy} Diagrams contributing to leading order to the real  (a)
 and imaginary (b) parts of the self-energy.}
\end{figure}

Within the real-time formalism  developed in section
\ref{sec:rtfor}, we can calculate the  pion self-energy for $\mu_\pi\neq 0$, whose leading order corrections to its real
and imaginary parts are given by the diagrams in Fig.\ref{fig:selfenergy}a and \ref{fig:selfenergy}b, respectively, all  vertices coming from the ${\cal
L}_2$ lagrangian.

It is important to remark that when nonequilibrium distributions are considered, as it is our case here, it has been pointed out that additional $\delta^2$-like or pinching-pole ill-defined
contributions arise \cite{Altherr:1994jc}, which should be regularized keeping a nonzero particle width. We will discuss the role of those contributions in the last part of this section.

Consider first diagram \ref{fig:selfenergy}a. It includes a contribution with a constant
vertex proportional to $m^2 \tilde G(0)\partial \tilde
D_{11}(p)/\partial m^2$ which directly renormalizes the pion mass,
following the prescriptions explained in section \ref{sec:rtfor},
and derivative vertices, which contribute either proportional to
$\Box\tilde G(0)=-m^2\tilde G(0)$ (mass renormalization) or as
$\tilde G(0) p^2\partial \tilde D_{11}(p)/\partial m^2=\tilde G(0)\left(\tilde
D_{11} (p)+m^2\partial \tilde D_{11}(p)/\partial m^2\right)$ (mass and
wave function renormalization). One can then follow similar steps
as in the standard derivation of the thermal corrections to the
pion self-energy to this order \cite{Schenk:1993ru,Gasser:1986vb},
the wave function renormalization being directly
related to the thermal $f_\pi$ through the usual
definition in terms of the residue of the axial-axial
current correlator. The ultraviolet divergences arising in the
calculation are absorbed by the renormalization of the low-energy
constants $l_3$ and $l_4$. We finally obtain:

\begin{eqnarray}
m_\pi^2(T,\mu_\pi)&=&m_\pi^2+\frac{m^2}{2f^2}\tilde
g_1(m,T,\mu_\pi)+\Od(m^4) \label{pimassther}\\
f_\pi^2(T,\mu_\pi)&=&f_\pi^2-2\tilde g_1(m,T,\mu_\pi)+\Od(m^4)
\label{fpither}\end{eqnarray} with $m_\pi$ and $f_\pi$ the
$T=\mu_\pi=0$ physical values given in (\ref{pimasszero}) and
(\ref{fpizero}) in terms of $m$ and $f$ to this order.

Taking into account now the corrections to the quark condensate to the same chiral order, i.e.,
 $\Od(T^6)$, which is given from (\ref{conddef}) and (\ref{z6}) using (\ref{derfor}):

\begin{equation}
\langle \bar q q \rangle(T,\mu_\pi)=\langle \bar q q \rangle(0,0)\left[1-\frac{3}{2f^2}\tilde g_1(m,T,\mu_\pi)\right]+\Od(T^8)
\label{condot6}
\end{equation}
we obtain that the Gell-Mann-Oakes-Renner (GOR) relation \cite{gor} holds also for $\mu_\pi\neq 0$ to this order (one-loop ChPT):

 \begin{equation}
 \frac{f_\pi^2(T,\mu_\pi)m_\pi^2(T,\mu_\pi)}{\langle \bar q q \rangle(T,\mu_\pi)}=\frac{f_\pi^2(0,0)m_\pi^2(0,0)}{\langle \bar q q \rangle(0,0)}=-m_q
 \label{gor}
 \end{equation}

 The GOR relation in terms of thermal quantities at $T\neq 0$, $\mu_\pi=0$ had been verified to one-loop in \cite{toublan97}. To this order, the thermal mass varies little, also at $\mu_\pi\neq 0$ (see below) so that the evolution of $f_\pi$ follows that of the quark condensate and both behave as order parameters. However, beyond one-loop, the GOR does not hold for thermal quantities \cite{toublan97}.

Another important observation
is that the shift (\ref{pimassther}) in the mass to this order can be
written, as in the $\mu_\pi=0$ case \cite{Schenk:1993ru} in terms
of the elastic pion-pion forward scattering amplitude, from (\ref{g1tilde}):

\begin{widetext}
\begin{equation}
m_\pi^2 (T,\mu_\pi)-m_\pi^2=-\int
\frac{d^3{\vec{p}}}{(2\pi)^3}\frac{1}{2E_p}\frac{1}{e^{\beta(E_p-\mu_\pi)}-1}
\re \left[T^f_{\pi\pi} (s=(E_p+m_\pi)^2-\vert\vec{p}\vert^2)\right] \label{luschermass}
\end{equation}
\end{widetext}
where $E_p^2=m_\pi^2+\vert\vec{p}\vert^2$ and $T^f_{\pi\pi}(s)$ is the
isospin averaged forward scattering amplitude:

\begin{widetext}
\begin{eqnarray}
T^f_{\pi\pi}(s)&\equiv&T_{\pi\pi}(s,0,u)=\frac{1}{3}\sum_{I=0}^2 (2I+1) T_I(s,0,u)=\frac{32\pi}{3}\sum_{I=0}^2 \sum_{J=0}^\infty (2I+1) (2J+1) t_{IJ}(s)\nonumber\\&\simeq&
\frac{32\pi}{3}\left[t_{00}(s)+9t_{11}(s)+5t_{20}
(s)\right]=-\frac{m^2}{f^2}+\Od(s^2,m^4)
\end{eqnarray}
\end{widetext}
where the last expression is the lowest order $\Od(p^2)$ (tree level diagrams with ${\cal L}_2$ vertices), $T_I(s,t,u)$ are the projections of the scattering
amplitude with definite isospin $I$,  $s,t,u$ are the Mandelstam
variables satisfying $s+t+u=4m_\pi^2$, and $t_{IJ}$ are the
partial waves, defined in the center of mass frame with definite
isospin $I$ and angular momentum $J$. We follow the conventions in
\cite{Gasser:1983yg}. In the previous expression, we have included
only the partial waves with lowest angular momentum $J\leq 1$.
Those with $J>1$ are negligible for $\sqrt{s}$ below inelastic thresholds, such as
the $K\bar K$ one,  and for the temperatures involved here
\cite{Schenk:1993ru}.

The result in (\ref{luschermass}) is the generalization to
$\mu_\pi\neq 0$ of the formula relating  the shift in the
self-energy with the density of states and the scattering
amplitude to lowest order in the density (dilute gas regime)
\cite{Schenk:1993ru}. These type of  relations were first derived by Luscher
\cite{Luscher:1985dn} studying  finite-volume corrections.
 A very interesting point is that it admits a natural
extension \cite{Schenk:1993ru} by considering (in the dilute gas
regime) not only the perturbative tree-level $\Od(p^2)$ amplitude,
but also higher orders, including unitarized amplitudes. In the
latter case, unitarized partial waves $t^U_{IJ}(s)$ for $\pi\pi$
scattering can be constructed to satisfy exactly the unitarity
relation:

\begin{equation}
\im t^U_{IJ} (s)=\sqrt{1-\frac{4m_\pi^2}{s}}\vert t^U_{IJ}\vert^2
\label{unit}
\end{equation}
matching at the same time the perturbative ChPT expansion and
providing  expressions that can be analytically continued to the
complex $s$ plane. All these features are satisfied by the Inverse
Amplitude Method scattering amplitudes \cite{iam} which reproduce
scattering data up to $\sqrt{s}\sim$ 1 GeV and all the low-lying
resonances, which in the pure pion case considered here reduce to
the $\rho(770)$ and the $f_0(600)$ or $\sigma$. Recall that the
 ChPT amplitudes satisfy the unitarity relation (\ref{unit}) only
 perturbatively order by order, violating the unitarity bounds for higher energies and
  thus not being able to reproduce resonances.

\begin{figure*}
%\vspace*{-2cm}
%\hspace*{-2cm}
\centerline{\includegraphics[scale=0.45]{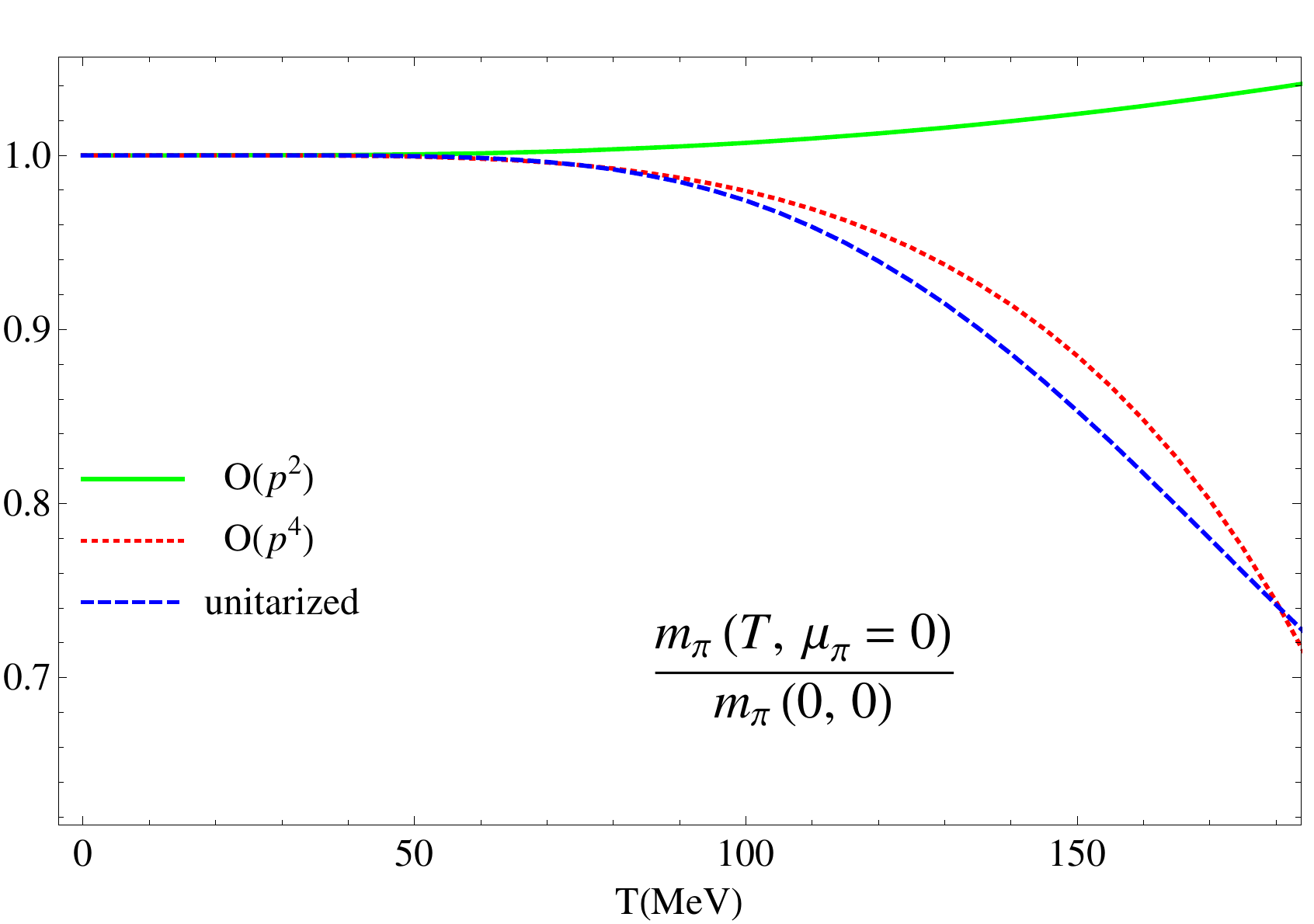}\includegraphics[scale=0.45]{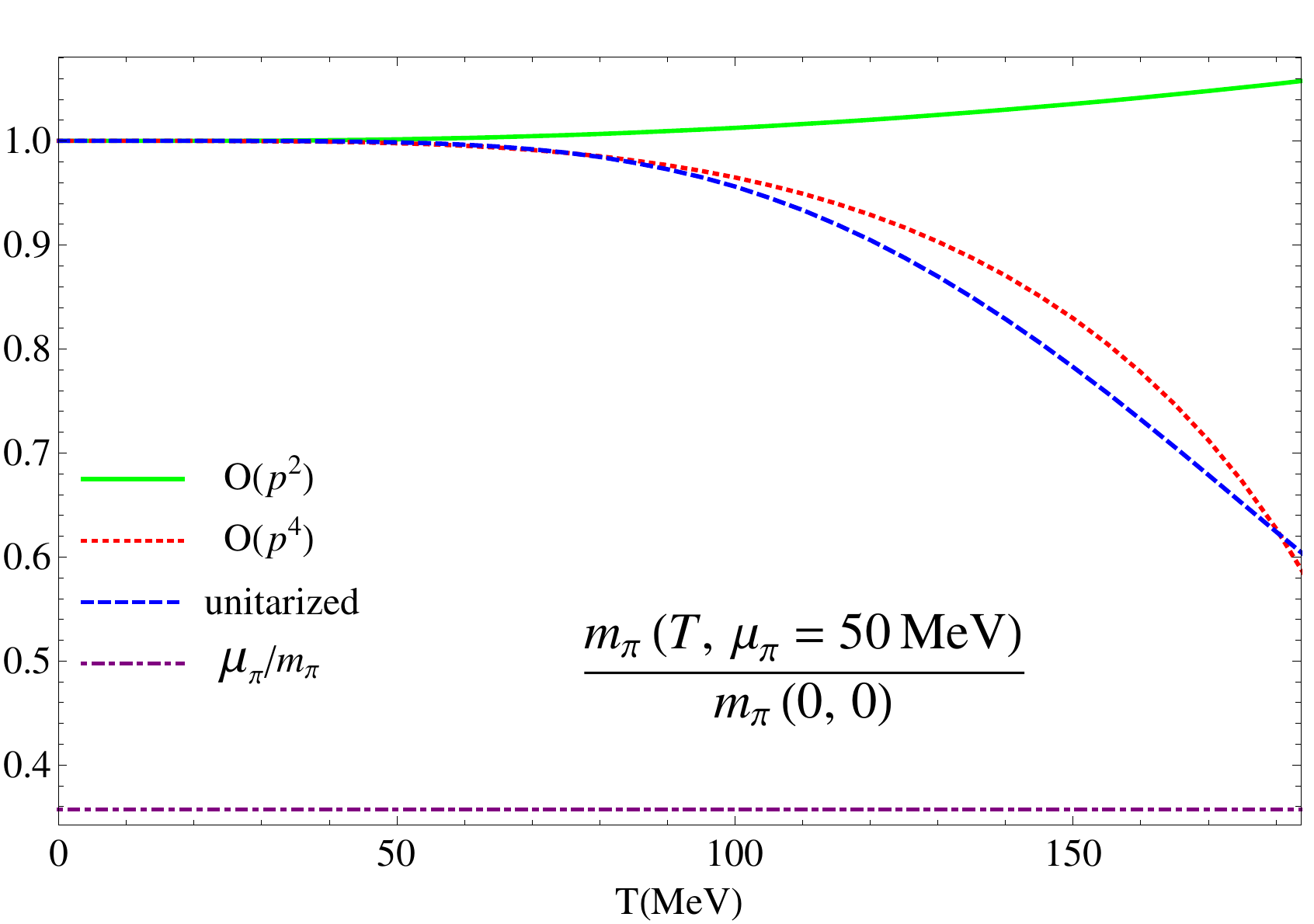}}
\centerline{\includegraphics[scale=0.45]{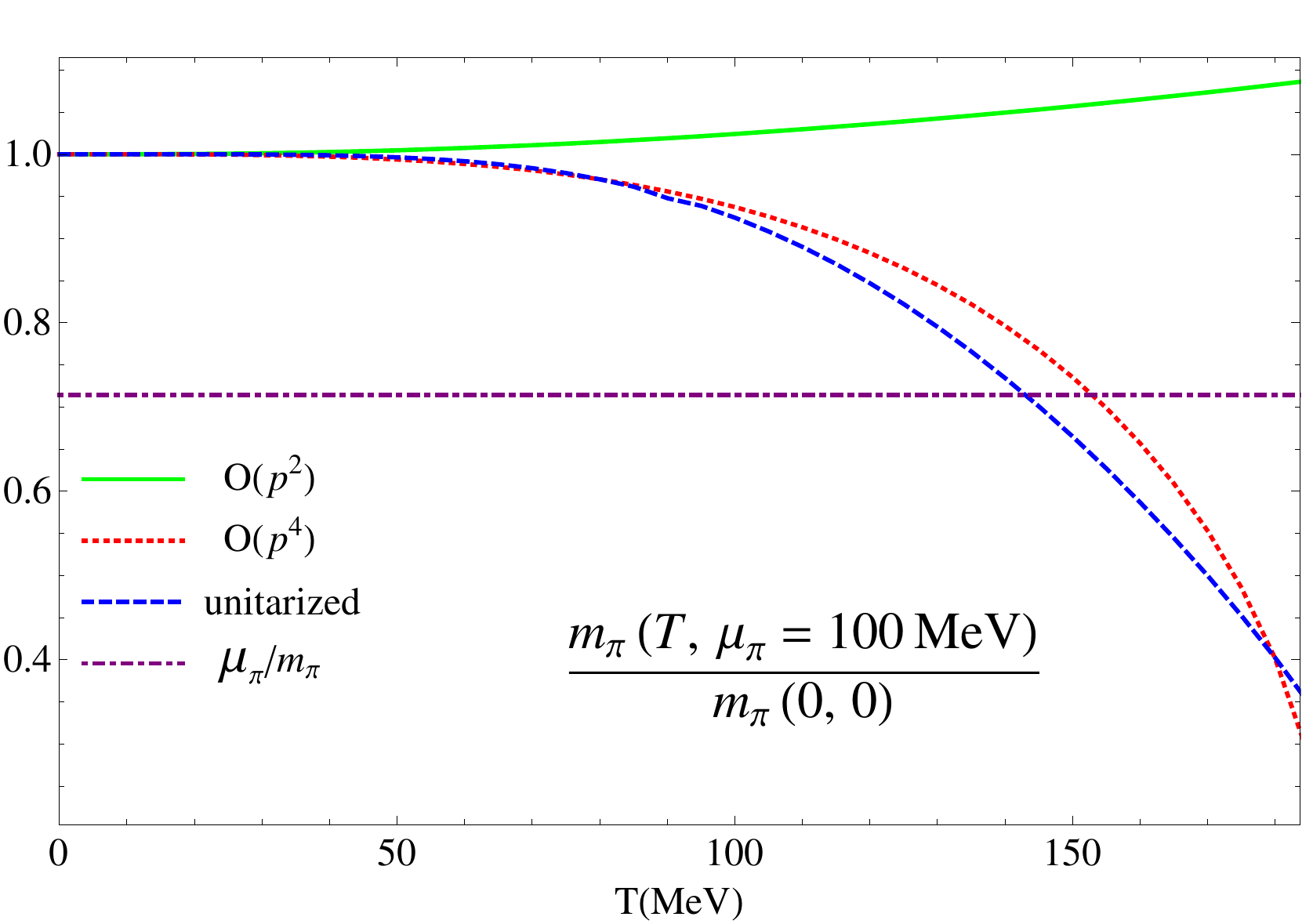}\includegraphics[scale=0.45]{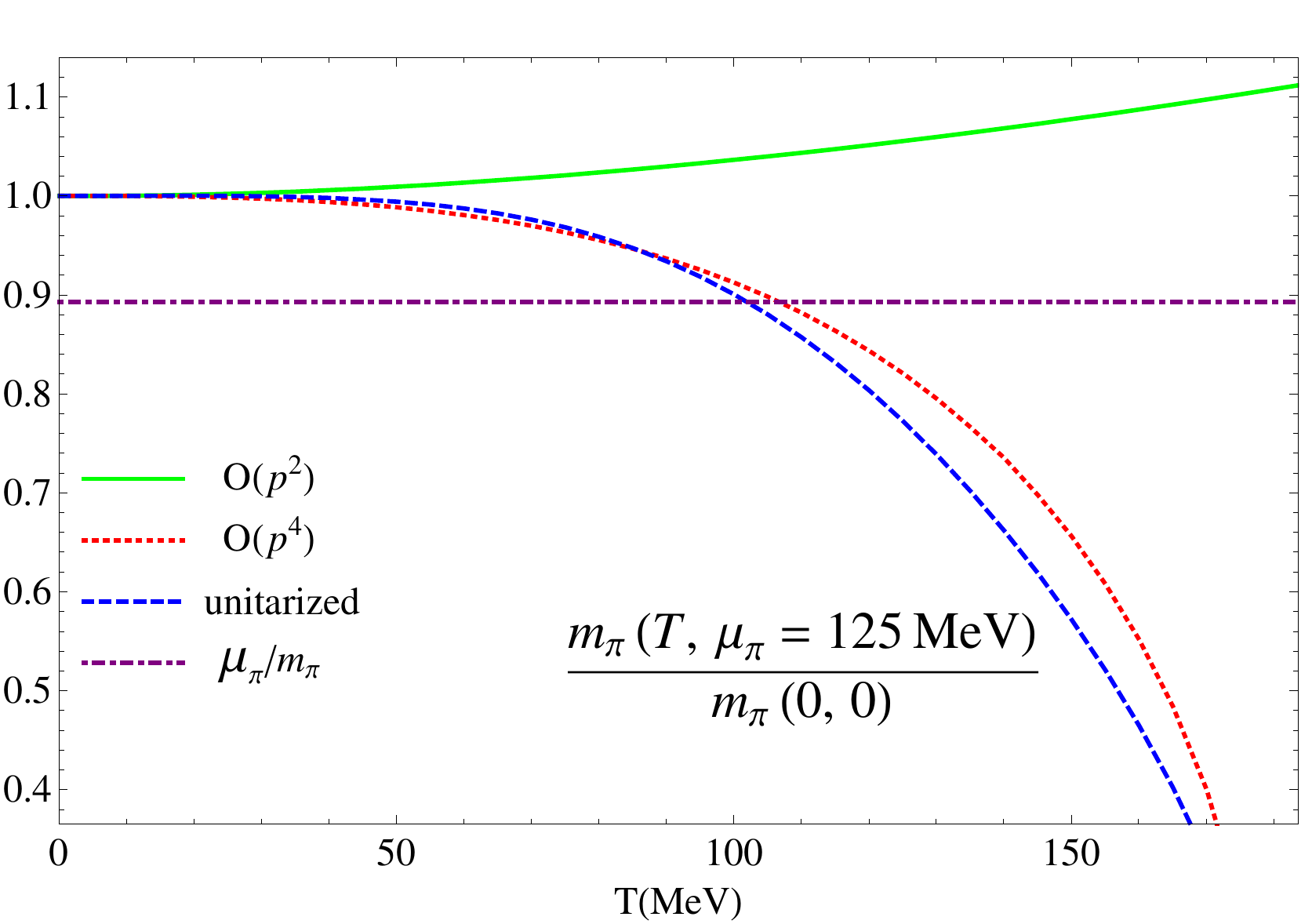}}
% Here is how to import EPS art
%\vspace*{-13cm}
 \caption{\rm \label{fig:resultsmass} Results for the thermal mass dependence on temperature and pion chemical potential, considering different orders in the
  scattering amplitude in (\ref{luschermass}).}
\end{figure*}

In Fig.\ref{fig:resultsmass} we show our results for the
thermal mass, considering $\Od(p^2)$, $\Od(p^4)$ and IAM
unitarized amplitudes in (\ref{luschermass}). We have
   used the same set of low-energy constants as in our previous calculations, i.e, the $\bar l_i$ given at the beginning of section \ref{sec:evalz}. For the case of the
 unitarized amplitudes, this set is adequate to compare with the
perturbative ChPT expressions, although it does not give the best
results for the mass and width of the resonances generated with
the IAM. We have checked that our results do not change
qualitatively by changing for instance to the set given in
\cite{Dobado:2002xf}, which gives better physical values for the
$\rho,f_0(600)$ mass and width.

Our results show that the leading order, the ChPT $\Od(p^2)$ given
by the tadpole diagram in Fig.\ref{fig:selfenergy}a, produces a
thermal mass slightly increasing  with temperature and chemical
potential. However, including the $\Od(p^4)$ or unitarized
corrections to the amplitude, the mass tends to decrease
significantly with $T$ and $\mu_\pi$. Our results at $\mu_\pi=0$
agree with \cite{Schenk:1993ru}. The difference between the
$\Od(p^4)$ and the unitarized curve is not very relevant here. Our $\Od(p^2)$ curve agrees reasonably
 with a linear-sigma model calculation \cite{Ayala:2002qy} which agrees with ChPT to this order at $\mu_\pi=0$ and where  the chemical potential is introduced in
  analogy with the charged scalar field case.

These results suggest an interesting scenario: the pion system
could undergo Bose-Einstein (BE) condensation driven by the
dropping of the thermal mass by interactions. Recall that we are
dealing with BE condensation of both neutral and charged pions,
since we are considering an electrically neutral system with
finite pion density. The physical situation is then different from
the charged pion or kaon condensation taking place in nuclear
matter \cite{sawyerkaplan} or isospin chemical potential
\cite{SonLoewe} scenarios, although the   dropping of the effective mass
takes place also in the former. BE condensation for  pion number and its
possible phenomenological consequences in heavy ion collisions has
been extensively studied in the literature \cite{BE}. Among the
 observable consequences are the anomalous enhancement of
the low-$p_T$ pion spectrum and of number fluctuations in high
multiplicity events.

\begin{figure*}
%\vspace*{-2cm}
%\hspace*{-2cm}
\centerline{\includegraphics[scale=0.5]{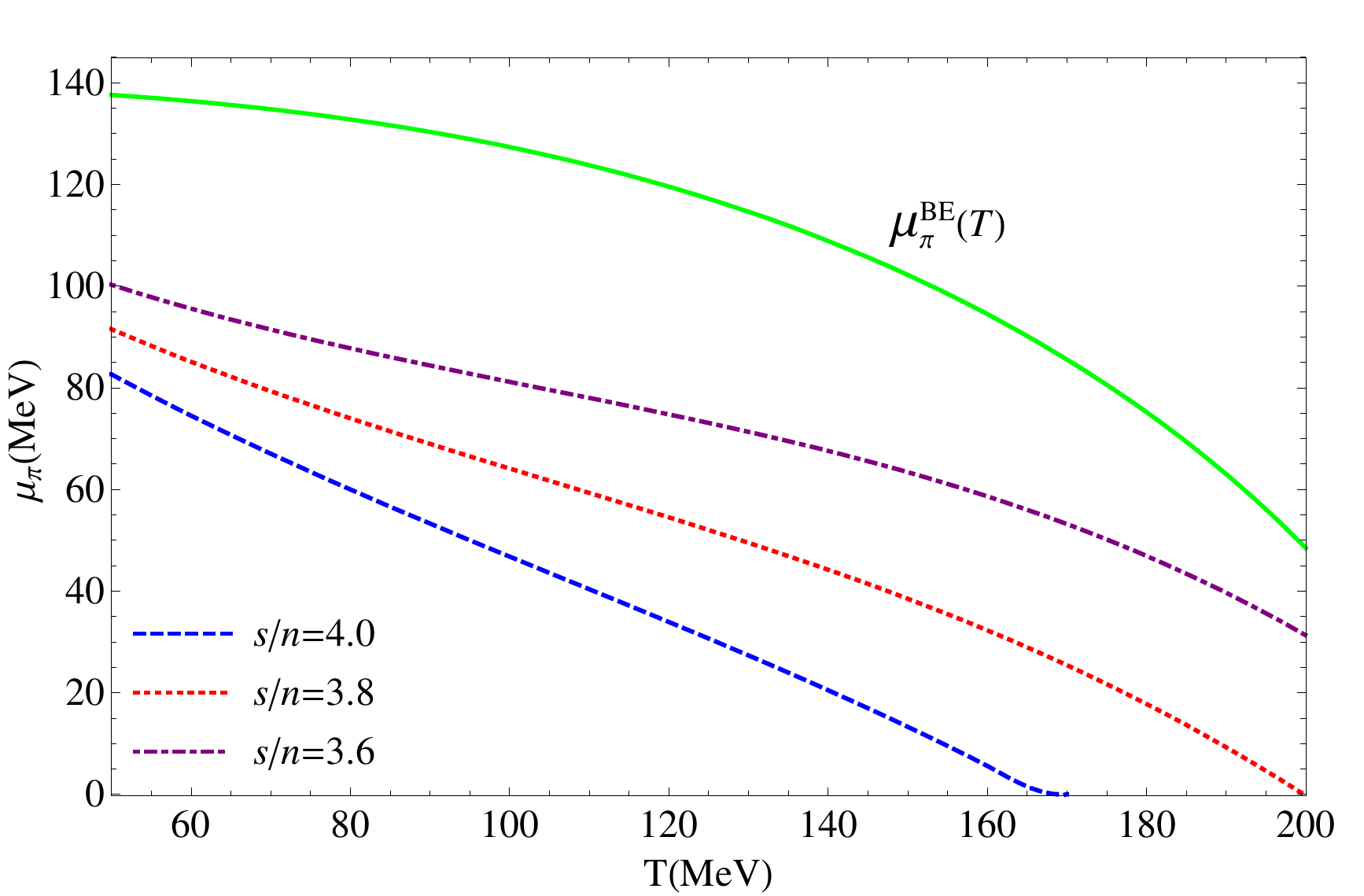}\includegraphics[scale=0.45]{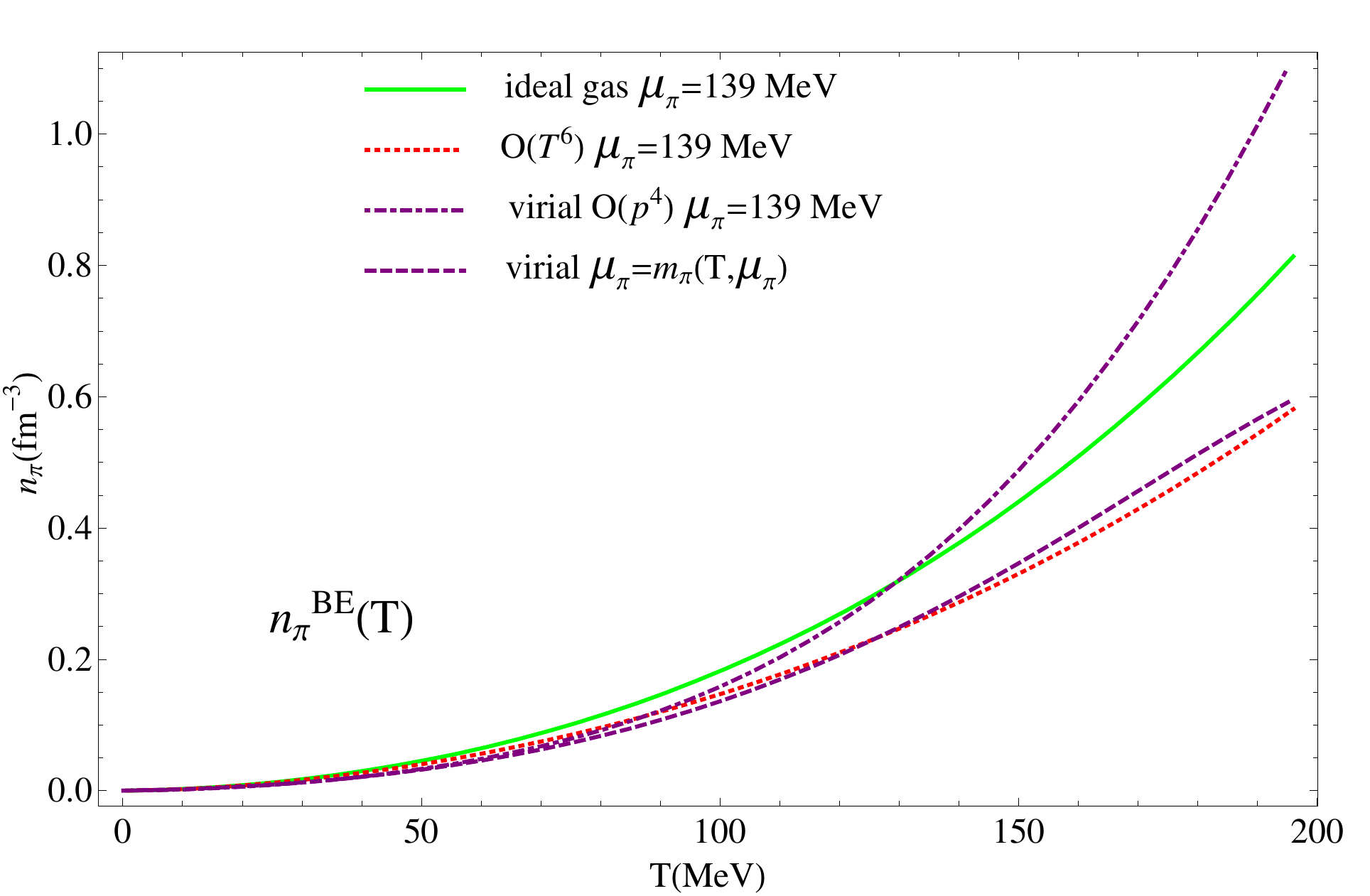}}
% Here is how to import EPS art
%\vspace*{-13cm}
 \caption{\rm \label{fig:BE} Bose-Einstein condensation lines. Left: the curve $\mu^{BE}_\pi=m_\pi(T,\mu_\pi)$
 with the thermal mass from the   $\Od(p^4)$ amplitudes, compared to the isentropic expansion curves
 for the virial case to the same order and for different $s/n$ values. Right:
 pion density versus temperature in the BE limit $\mu_\pi\rightarrow m_\pi^-$ for different orders in the interaction,
 compared to the ideal gas and to the virial case with thermal mass.}
\end{figure*}

In our grand-canonical interacting framework we can describe the
corrections to BE condensation  due to pion interactions. In the
standard free case, the BE limit is reached when
$\mu_\pi\rightarrow m_\pi$ from below (by definition the system is
below the condensed phase). Those values for the pion chemical
potential seem too high compared with those measured in heavy ion
collisions at thermal freeze out $T_{ther}\sim$ 100 MeV
\cite{Bebie:1991ij,Hung:1997du,Kolb:2002ve}. In other words, the
required  pion densities for BE condensation might not be reached.
However, if the effective particle mass $m_\pi(T,\mu_\pi)$ drops by
interactions among the medium constituents, the value
$\mu_\pi=m_\pi(T,\mu_\pi)$ would be  reduced. We show that line in
Fig.\ref{fig:resultsmass}. In Fig.\ref{fig:BE} (left) the
resulting $\mu^{BE}_\pi(T)=m_\pi(T,\mu_\pi^{BE}(T))$ curve is
represented and compared  with the isentropic curves corresponding
to different values of $s/n$. We see that the BE curve thus
defined lies not very far from the isentropic approach and the
expected phenomenological values. Those curves correspond to the
$\Od(p^4)$ amplitudes, both for the thermal mass and for $s$ and
$n$ (in the virial approach). In  Fig.\ref{fig:BE} (right) we
show also the density-temperature curves corresponding to
$\mu_\pi\rightarrow m_\pi^-$ for different orders in the
interaction. The $\Od(T^6)$ allows for lower density values, but
the virial contribution points in the opposite direction. We also
show the curve corresponding to the BE limit by lowering of the
mass, as we have just explained, for the same virial approach,
which  produces a considerable lowering of the required densities.
In any case, the corrections due  to interactions are small near
thermal freeze-out.  We also remark that some of our previous results, including those regarding BE condensation by mass reduction, rely on the validity of the dilute gas regime, for instance when using (\ref{luschermass}), but corrections might be important for temperatures close to the chiral transition or chemical potentials close to $m_\pi$.

Finally, we turn to the calculation of the leading order imaginary
part of the pion self-energy, given in ChPT by the diagram in
Fig.\ref{fig:selfenergy}b. This is the leading order
contribution to the thermal collisional width
$\Gamma_p=-\im \Sigma_R(E_p,\modp)/(2E_p)\ll E_p$ with
$E_p=\sqrt{\modp^2+m_\pi^2}$ and $\Sigma_R$ the retarded
self-energy, which  defines the dispersion relation
$p^2=m_\pi^2+\Sigma_R(p_0,\modp;T)$.

As we have commented in
section \ref{sec:rtfor}, we will evaluate the retarded correlator
in the real-time formalism, following the circling rules in
\cite{Kobes:1990kr}, which also apply to the $\mu_\pi\neq 0$ case.
Applying those rules to diagram \ref{fig:selfenergy}b we have:

\begin{equation}
\im \Sigma_R(p_0,\modp)=-\frac{1}{2}\left[\tilde H^>(p_0,\modp)-\tilde H^<(p_0,\modp)\right]=-\frac{i}{2}\left[\Sigma_{21}(p_0,\modp)-\Sigma_{12}(p_0,\modp)\right]
\end{equation}
where $\tilde H^{>(<)}$ are obtained by using for the three internal lines in the diagram the $\tilde G^{>(<)}$
 RTF propagators. With the usual RT self-energy definition \cite{lebellac} and our convention for the $D_{12}$, $D_{21}$ propagators given in section \ref{sec:rtfor}, we have $H^>=i\Sigma_{21}$ and  $H^<=i\Sigma_{12}$ for  diagram 6b, since, once a particular choice  of $ij$ indices ($i,j=1,2$) has been made for the two vertices in that diagram, the three internal lines carry the same $ij$ combination.

 Now, according to our discussion in section \ref{sec:rtfor} and in Appendix \ref{app:therprop}:

\begin{eqnarray}
\tilde G^> (k)&=&2\pi\delta(p_0^2-E_p^2)\left[\theta(p_0)+n(\vert
p_0 \vert-\mu_\pi)\right]
%\nonumber\\&=&
=e^{\tilde\beta_k k_0}\tilde
G^< (k)=e^{\beta\left[k_0-\mu_\pi {\small \sgn}(k_0) \right]}\tilde G^<
(k)\nonumber\\&&
\end{eqnarray}
 so that we get for the thermal width:
\begin{widetext}
\begin{eqnarray}
\Gamma_p(T,\mu_\pi)&=&\frac{1}{4E_p}\int \prod_{i=1}^3 \frac{d^4 k_i}{(2\pi)^4} \Lambda(k_1,k_2,k_3,p) \tilde
G^> (-k_1)\tilde G^> (k_2)\tilde G^> (k_3)\nonumber\\&\times&\left(1-e^{-\beta\left[E_p-\mu_\pi f(k_1,k_2,k_3)\right]}\right)
 (2\pi)^4\delta(E_p+k_1^0-k_2^0-k_3^0)\delta^{(3)}(\vec{p}+\vec{k}_1-\vec{k}_2-\vec{k}_3)
\label{therwidthpre}
\end{eqnarray}
\end{widetext}
where $k_{1,2,3}$ label the three internal lines,
$\Lambda$ is the squared vertex function coming from the ${\cal
L}_2$ lagrangian and:

\begin{equation}
f(k_1,k_2,k_3)=\sgn(k_2^0)+\sgn(k_3^0)-\sgn(k_1^0)
\end{equation}

Recall that in the $\mu_\pi=0$ case, the $f$-term is absent so that one ends up with a prefactor $e^{-\beta E_p}-1=1/(1+n(E_p))$ in the thermal width. The natural expectation from replacing just the distribution function $n\rightarrow \tilde n_p$ for $\mu_\pi\neq 0$ would be then  $E_p\rightarrow E_p-\mu_\pi$ in that factor, as well as the modifications of the internal distribution functions $n(E_i)\rightarrow n(E_i-\mu_\pi)$ where $E_i$ is short for $E_{k_i}$. This is indeed the result found in \cite{Goity:1989gs} derived from kinetic theory. In our case, it is not obvious that the answer is the same, since the function $f$ above is not  equal to one for the eight possible combinations of signs of the three internal $k_i^0$. We denote them by $s_1s_2s_3$,  with $s_i=\sgn(k_i^0)$. Now,  we take into account that the $\delta$ functions in each of the internal lines put them on-shell, i.e., $k_i^0=\pm E_i$ and   global energy-momentum conservation in the diagram imposed by the $\delta$-functions in (\ref{therwidthpre}). Thus, the combination $+--$ giving $f=-3$ is excluded by energy conservation $E_p+E_1 >0>-E_2-E_3$. On the other hand, from three-momentum conservation and the on-shell conditions we have that for any combination it should hold $A=B$ where we denote $A\equiv E_p^2+E_1^2-E_2^2-E_3^2$ and $B\equiv2(\vec{k}_2\cdot \vec{k}_3-\vec{p}\cdot \vec{k}_1)$ and, in addition  $-C\leq B \leq C$ with  $C=2(p k_1+k_2 k_3)$ and where $k_i$, $p$  are short for $\vert\vec{k}_i\vert$ and $\vert\vec{p}\vert$ respectively. Therefore, the case $++-$ ($f=-1$) is also excluded, since for that combination $E_p+E_1=E_2-E_3$ so that $A=-2(E_p E_1+E_2 E_3)<-C$. The same reason excludes $+-+$ ($f=-1$). Combinations $-++$ ($f=3$) and $---$ ($f=-1$) give $A=2(E_p E_1+E_2 E_3)>C$ and are thus  excluded as well. Therefore, the only combinations remaining are $+++$, $-+-$ and $--+$, the three of them giving $f=1$ and the same contribution from the vertices as for $\mu_\pi=0$, given in \cite{Schenk:1993ru}.

It is not difficult to repeat the above analysis, now with the external energy $p_0=-E_p$. In that case, every combination of relative signs between the $E_i$ is obtained from the previous case by flipping the three $s_i$,  the $A,B,C$ functions being independent of the sign of $p_0$. Thus, applying the same arguments, the only surviving combinations  are now $---$, $+-+$ and $++-$, the three of them giving $f=-1$. Therefore, what we have proven in terms of the $12$ and $21$ components of the self-energy for this diagram is:

\begin{equation}
\Sigma_{21}^{6b} (p_0=\pm E_p,p)=e^{\beta(p_0-\mu_\pi{\small \sgn} (p_0))}\Sigma_{12}^{6b} (p_0=\pm E_p,p)=e^{\tilde\beta_p p_0}\Sigma_{12}^{6b} (p_0=\pm E_p,p)
\label{perio6b}
\end{equation}
which is  the usual equilibrium relation with $\beta\rightarrow\beta_p$. This relation will be of use in the discussion at the end of this section about possible higher order corrections related to pinching poles.

In conclusion, the result (\ref{therwidthpre}) we find with our diagrammatic method is the same  as in kinetic theory \cite{Goity:1989gs}, which, after relabeling $k_1\leftrightarrow -k_3$ in $-+-$ and $k_1\leftrightarrow -k_2$ in $--+$ and performing the three integrals in $k_i^0$  using the on-shell $\delta$-functions, can be written as:

\begin{widetext}
\begin{eqnarray}
\Gamma_p(T,\mu_\pi)&=&\frac{1}{8E_p}\frac{1}{1+n(E_p-\mu_\pi)}\int \prod_{i=1}^3 \frac{d^3 k_i}{(2\pi)^3 2E_i} n(E_1-\mu_\pi)\left[1+n(E_2-\mu_\pi)\right]\left[1+n(E_3-\mu_\pi)\right]
\nonumber\\&\times&\vert T_{\pi\pi}(s,t)\vert^2 (2\pi)^4\delta(E_p+E_1-E_2-E_3)\delta^{(3)}(\vec{p}+\vec{k}_1-\vec{k}_2-\vec{k}_3)
\label{therwidthfull}
\end{eqnarray}
\end{widetext}
where $T_{\pi\pi}$ is the isospin averaged elastic pion scattering amplitude with $s=(E_p+E_1)^2-\vert \vec{p}+\vec{k}_1\vert^2$, $t=(E_p-E_2)^2-\vert \vec{p}-\vec{k}_2\vert^2$.

Taking now the dilute gas regime in the previous expression, which amounts to neglect all the Bose-Einstein functions $n\ll 1$ except $n(E_1-\mu_\pi)$, gives rise to the extension of Luscher's formula in terms of the forward scattering amplitude, as in (\ref{luschermass}) but now for the imaginary part of the self-energy through the pion thermal width (which vanishes at $T=0$):

\begin{widetext}
\begin{equation}
\Gamma^{DG}_p(T,\mu_\pi)=\frac{1}{2E_p}\int
\frac{d^3\vec{k}}{(2\pi)^3}
n(E_k-\mu_\pi)\frac{\sqrt{s(s-4m^2)}}{2 E_k} \sigma_{\pi\pi}
(s)=\frac{1}{2E_p}\int \frac{d^3\vec{k}}{(2\pi)^3 2E_k}
n(E_k-\mu_\pi) \im T^f_{\pi\pi}(s) \label{gammadg}
\end{equation}
\end{widetext}
where we have relabeled $k_1\rightarrow k$ and $\sigma_{\pi\pi}$ is the total $\pi\pi$ cross section

\begin{eqnarray}
\sigma_{\pi\pi}(s)&=&\frac{32\pi}{3s}\sum_{IJ}(2I+1)(2J+1)\vert t_{IJ}(s)\vert^2\nonumber\\
&=&\frac{1}{\sqrt{s(s-4m^2)}}\im T^f_{\pi\pi}(s)
\end{eqnarray}
where the last line is the optical theorem, following from exact unitarity (\ref{unit}).

We remark that our final results both for the real and imaginary parts of the self-energy to this order correspond to the replacement $n\rightarrow\tilde n$ evaluated at positive energies. This is not only natural from the kinetic theory viewpoint but it is also formally obtained by performing such replacement in the analytically continued $\mu_\pi=0$ ITF self-energies.

\begin{figure*}
%\vspace*{-2cm}
%\hspace*{-2cm}
\centerline{\includegraphics[scale=0.35]{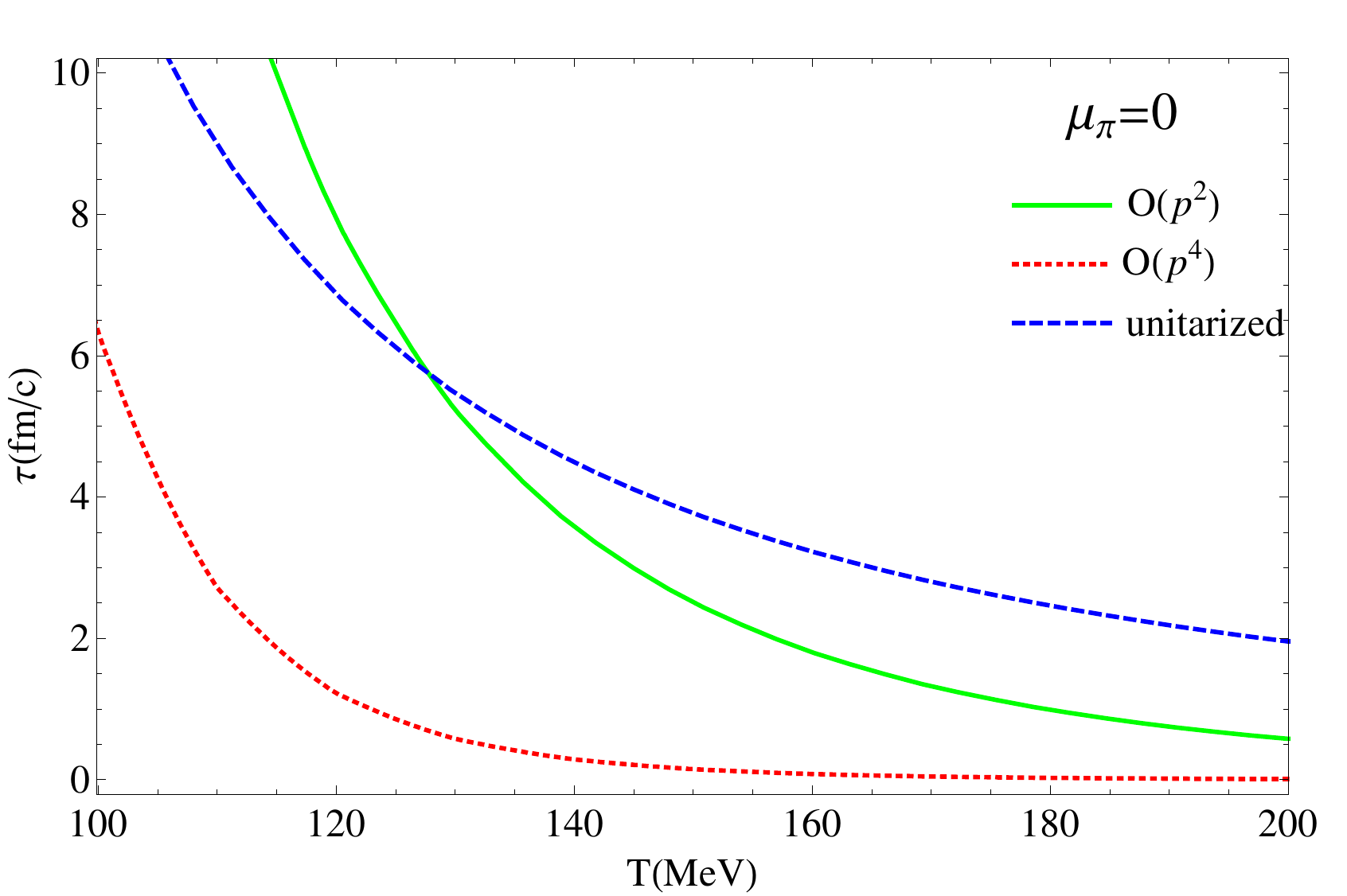}\includegraphics[scale=0.35]{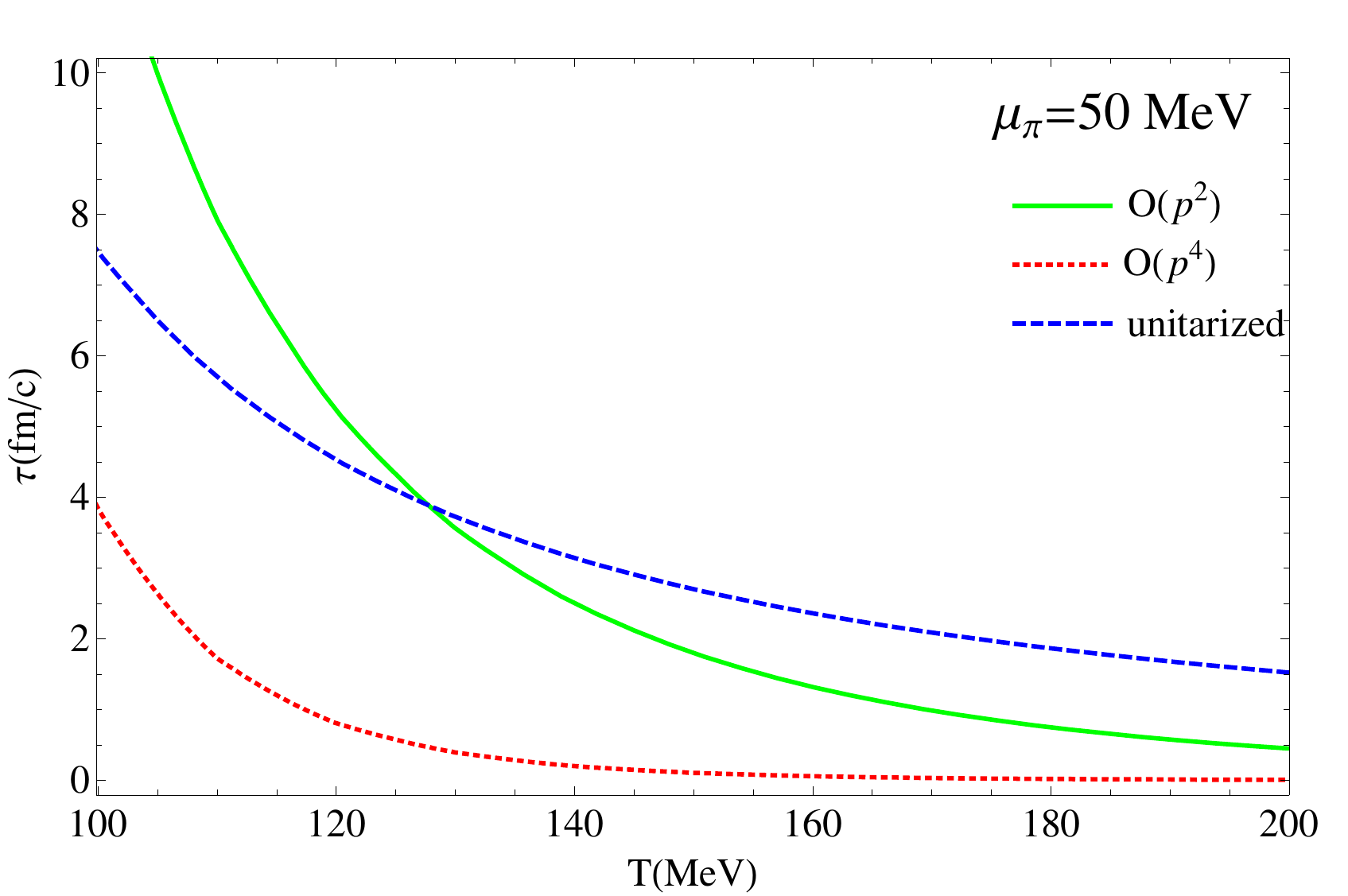}\includegraphics[scale=0.35]{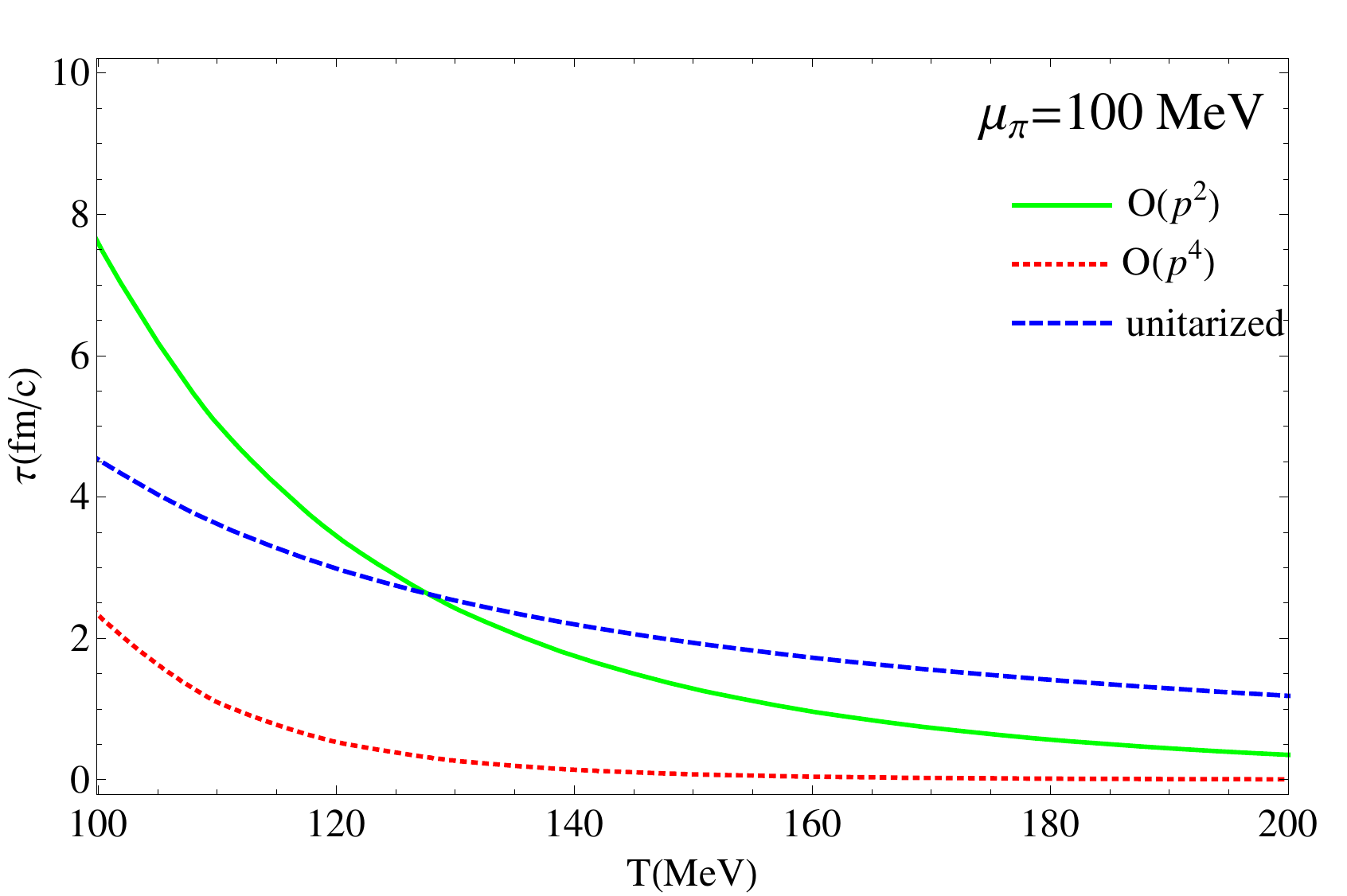}}
% Here is how to import EPS art
%\vspace*{-13cm}
 \caption{\rm \label{fig:tau} Mean collision time in the elastic and dilute limit, considering different orders
 for the pion scattering amplitude and different values for the pion chemical potential.}
\end{figure*}

The thermal width is of phenomenological relevance, since it
enters directly in the calculation of transport coefficients in
the meson gas \cite{FernandezFraile:2009mi,transport}. It is then important to
estimate pion chemical potential effects in the width during the phase of chemical
nonequilibrium where particle number is approximately conserved
and transport phenomena can be described relying on the dominance
of elastic collisions, which is also consistent with the dilute
gas regime. On the other hand, in this regime the mean collision
time defined for ultrarelativistic particles as
$\tau=1/(2\bar\Gamma)$ \cite{Goity:1989gs,Song:1996ik,Hung:1997du} with the averaged width:

\begin{equation}
\bar\Gamma (T,\mu_\pi)=\frac{\int d^3\vec{p}\ \ \Gamma_p(T,\mu_\pi)
n(E_p-\mu_\pi)}{\int d^3\vec{p} \ n(E_p-\mu_\pi)}
\end{equation}
provides direct information about  thermal relaxation. We represent $\tau$
in the dilute approach in Fig.\ref{fig:tau}, using  the
scattering amplitude in (\ref{gammadg}) to different orders,
including the unitarized case. We use the same set of $\bar l_i$
constants as in the rest of the paper.

We see in the figure that the effects of reproducing correctly the energy behavior of the scattering amplitude is important for evaluating the collision time. In particular, the unitarized curve shows important differences with the perturbative ones in the temperature range showed. This was also noticed  in \cite{Schenk:1993ru} at $\mu_\pi=0$ and the importance of including unitarized corrections to the width for transport coefficients in the meson gas has been highlighted in \cite{FernandezFraile:2009mi,transport} for instance regarding violations of AdS/CFT bounds for the shear viscosity over entropy ratio or correlations between the bulk viscosity and the conformal anomaly.

Another clear effect that we observe is the reduction of the mean time with the pion chemical potential, also observed in \cite{Goity:1989gs} with $\Od(p^2)$ amplitudes. Physically, in the temperature regime where $\tau$ is much smaller than the typical plasma lifetime ($\sim$ 10 fm/c) which  at the same time is small compared
to the inelastic collision time driving the system to chemical
equilibrium, the system remains in thermal but not chemical equilibrium. From the estimates of the inelastic collision rates given for instance in \cite{Song:1996ik} and the results in Fig.\ref{fig:tau}, this would happen at $\mu_\pi=0$ typically in the range 120 MeV$<T<$ 180 MeV. However, precisely in that regime, and as we have explained in this paper, $\mu_\pi\neq 0$, its typical values being given by the isentropic curve in Fig.\ref{fig:muT}, which means that the range of thermal equilibrium enlarges from below, from the commented reduction of $\tau$ with $\mu_\pi$. In fact, estimating the thermal freeze out temperature $T_{ther}$ as that where this approximation ceases to be valid, i.e., where $\tau\sim$ 10 fm/c (this type of dynamical condition has been used also in \cite{Kataja:1990tp} to determine the freeze-out conditions) gives a shift in the thermal freeze-out temperature $\Delta T_{ther}\simeq - 20$ MeV with respect to the $\mu_\pi=0$ case,  following approximately the isentropic values in Fig.\ref{fig:muT}. In particular, using the unitarized results in Fig.\ref{fig:tau} we obtain in this way $T_{ther}\simeq$ 95 MeV, close to experimental values.

Finally, let us comment on the pathological nonequilibrium terms found in \cite{Altherr:1994jc} in a $g^2\phi^4$ context. Those terms are of the type of $\delta^2$ functions at the same point, or pinching singularities and therefore have to be regularized by keeping a nonzero particle width in the propagators, i.e., $\Gamma_p\neq 0$ in our case. The first nonvanishing term of this kind in the $g^2\phi^4$ theory is the three-loop diagram given in Figure 2b of \cite{Altherr:1994jc}. Note that formally this is an $\Od(g^6)$ correction, while the diagrams we have considered here in Fig.\ref{fig:selfenergy}, for which there are no such pathologies, would be $\Od(g^2)$ and $\Od(g^4)$ respectively in that counting. Nevertheless, the argument in \cite{Altherr:1994jc} is that those contributions are proportional to the inverse width $1/\Gamma$ with $\Gamma=\Od(g^4)$ and therefore could become of the same order as the leading ones. The form of such leading pinching-pole term \cite{Altherr:1994jc} in our case is:

\begin{equation}
\int dP \Delta_R(p)\Delta_A(p)\left\{\left[(1+\tilde n_p(p_0)\right]\Sigma_{12}^{6b}-\Sigma_{21}^{6b}\tilde n_p(p_0)\right\}
\label{patho1st}
\end{equation}
where the   self-energy components of diagram 6b that we have analyzed above enter directly to this order and $\Delta_{R,A}$ denote the retarded/advanced propagators. Let us isolate the leading $\Gamma_p\rightarrow 0^+$ behaviour of the previous expression. The product $\Delta_R\Delta_A$ is the characteristic pinching-pole contribution appearing typically in diagrammatic  calculations of transport coefficients \cite{FernandezFraile:2009mi,transport} and in the $\Gamma_p\rightarrow 0^+$ behaves as:

\begin{equation}
\Delta_R(p)\Delta_A(p) \stackrel{\Gamma_p\rightarrow 0^+}{\approx}  \frac{\pi}{2 E_p\Gamma_p}\delta(p_0^2-E_p^2)
\label{ralo}
\end{equation}
which puts on-shell ($p_0=\pm E_p$) the integrand of (\ref{patho1st}). Now, we recall the relation (\ref{perio6b}) we have derived for diagram 6b, together with the properties of the ``modified" distribution function, in particular $1+\tilde n_p(p_0)=e^{\tilde\beta_p} \tilde n_p(p_0)$. Altogether, this means that the leading order pathological contribution (\ref{patho1st}) vanishes in our case in the $\Gamma_p\rightarrow 0^+$ limit. There may be higher order corrections of this kind, but in accordance with the power counting in \cite{Altherr:1994jc}, those would be subleading with respect to the contributions in Fig.\ref{fig:selfenergy} that we have analyzed in this section.

The previous analysis showing the absence of pathological terms is only valid to lowest order in those terms. Further conclusions can only be reached with a complete ChPT analysis of higher order pinching-pole contributions, extending that in \cite{Altherr:1994jc}, which is beyond the scope of this work. Nevertheless, it is worth pointing out that the low-$T$ ChPT counting does not involve any coupling constant, which implies  important differences with respect to the $g^2\phi^4$ one in the perturbative behaviour of pinching diagrams at low $T$ \cite{transport}. In addition, the analysis of \cite{Altherr:1994jc} shows that the nonequilibrium pathological terms are always proportional to $\delta n$, the deviation from the Bose-Einstein distribution function. Thus, in our case, we can use an argument similar to the one we invoked in section \ref{sec:evalz} when dealing with the $\Od(T^8)$ terms in the partition function. Namely, that in the ChPT counting those diagrams are expected to be important for temperatures for which $\beta\mu(T)\sim\delta n \ll 1$, giving a further suppression. We finally remark that in the $\mu_\pi\neq 0$ scenario, the presence of nonvanishing pinching-pole terms have also their origin in the presence of particle-changing processes (see also our discussion in section \ref{sec:evalz}). In fact, self-energy combinations of the form (\ref{patho1st}) enter directly in the Boltzmann equation describing the rate of particle number change \cite{Altherr:1994jc}. The fact that the leading correction of that type (\ref{patho1st}) vanishes in our case seems to be related to the fact that the leading self-energy corrections arising from the diagrams in Fig.\ref{fig:selfenergy} can always be expressed in terms only of the elastic $\pi\pi$ scattering amplitude, as we have  discussed extensively throughout this section.

\section{Conclusions}

In this work we have developed a path-integral diagrammatic
formalism in order to deal with chemical nonequilibrium effects in
interacting scalar field theories, in the regime where particle number is approximately conserved. Within the theoretical framework of holomorphic
path integrals and thermal field theory, we have derived the
relevant  Feynman rules for nonzero particle number chemical potential $\mu$, whose validity is
restricted to the temperature regimes where one can neglect
particle-changing processes. This derivation in the interacting case is original of this paper, to the best of our knowledge.

We have addressed some subtleties related to the choice of contour in complex times, leading to the
 extension of  real and imaginary time formalisms at $\mu\neq0$. We have shown that the consistent formulation is the real-time one, in agreement with
  other nonequilibrium formulations. The imaginary-time formalism can lead to spurious contributions, related to the loss of periodicity or global KMS
   conditions. These problems are not present in the real-time formalism, once a proper energy representation for the propagators is chosen, in accordance with
    the standard $\mu=0$ choice. In addition, following  previous studies in the literature at $\mu=0$, we have been able to construct the combinations of real-time diagrams
    leading to  retarded correlators and to closed diagrams contributing to the free energy.

    We have applied this formalism to the case of a pion gas, relevant for Relativistic Heavy Ion Collisions between thermal and chemical freeze-out with nonzero pion number chemical potential $\mu_\pi(T)$. Our description is consistent if the $T$-dependence of $\mu_\pi$  encodes the time evolution of the plasma between those phases.    The relevant diagrammatic scheme for temperatures below chiral
     restoration is Chiral Perturbation Theory.  In this framework, we have calculated the leading corrections to the ideal gas coming from chiral interactions. To leading order $\Od(T^6)$ the corrections to the pressure can be expressed in terms of tadpole diagrams and are numerically rather small up to  $T_c$. To next to leading order $\Od(T^8)$, closed diagrams contributing to the free energy can be obtained from particle-changing processes, which signals the onset of the number conservation approximation break-up. Nevertheless, since $\mu_\pi$ is small for temperatures where those ChPT corrections become important, they can be reliably calculated and produce sizable deviations from
      the free gas. The results to that order agree reasonably well with a virial expansion analysis.  Our results for thermodynamical observables show that both chiral interactions and $\mu_\pi$ tend to increase the pressure. The chiral restoration critical temperature decreases with increasing $\mu_\pi$, which would be of relevance only if chiral restoration takes place for lower temperatures than chemical freeze-out. We have also calculated the isentropic $\mu_\pi(T)$ curve for different orders in the interactions. The corrections to the ideal gas show a significant reduction of the chemical freeze-out temperature, which is the expected effect of interactions, since they increase the probability of producing  inelastic processes. The same effect had been observed previously in a free gas of pions and resonances.

     Our approach allows to derive  thermal corrections to the pion self-energy at $\mu_\pi\neq 0$, from the leading-order ChPT diagrams, both for the real and imaginary parts of the retarded correlator. The imaginary part comes from a two-loop diagram, for which the use of RTF rules for the construction of the retarded function is crucial. After a detailed evaluation, our diagrammatic result is shown to coincide with the expected expressions from kinetic theory arguments.
      We have also discussed the role of  higher-order pinching pole contributions to self-energies, providing different arguments which support that those corrections are subleading in our approach.
     The real part of the self-energy gives the thermal mass, which together with the condensate and the pion decay constant to the same order, satisfy the $\mu_\pi\neq 0$ extension of the Gell-Mann-Oakes-Renner relation. In addition, both the real and imaginary parts satisfy a Luscher-like relation in terms of the forward pion scattering amplitude. This relation allows  to calculate in the dilute regime the self-energy corrections for higher orders in the ChPT amplitudes, including unitarized expressions which have the physically expected energy behavior and reproduce the lightest resonance states. The results for the thermal mass show a clear decreasing both with $T$ and $\mu_\pi$ for $\Od(p^4)$ and unitarized amplitudes. This suggests the interesting possibility of reaching Bose-Einstein condensation when the effective thermal mass approaches the chemical potential. This mechanism would require lower pion densities to reach BE condensation. We have discussed this possibility, which is a purely interacting effect, within the isentropic values and comparing the pion densities with those in the standard approach of considering the ideal gas BE limit $\mu\rightarrow m_\pi^-$ with $m_\pi$ the vacuum mass. Our mass-dropping BE curve is not far, but still above the isentropic ones for reasonable values of chemical freeze-out. Finally, using also the  scattering amplitudes, we have evaluated the corrections to the mean collision time at $\mu_\pi\neq 0$. The mean time decreases with $T$ and $\mu_\pi$ for all orders in the interaction, which implies a sizable reduction, compared to the $\mu_\pi=0$ case, of the thermal freeze-out temperature, estimated as that where $\tau$ equals the typical plasma lifetime.

     Summarizing, the diagrammatic field-theory scheme developed in the present work provides, in our opinion, useful results regarding the chemically non equilibrated phase  of the meson gas resulting from a Relativistic Heavy Ion Collision. In future works we plan to generalize the analysis presented here to include also the strange sector (kaons and eta) as well as to extend  previous studies of transport coefficients by including  pion chemical potentials along the lines presented here.

\begin{acknowledgments}
We  acknowledge financial support from the Spanish research
projects   FPA2007-29115-E,
PR34-1856-BSCH, CCG07-UCM/ESP-2628,  FPA2008-00592, FIS2008-01323 and from the
 FPI programme (BES-2005-6726).
\end{acknowledgments}

\appendix

\section{Holomorphic path integrals} \label{app:holo}

We review here some of the key aspects of the holomorphic path
integral representation which are used in the main text. We will
follow the discussion in \cite{ZJ02}, to which we refer for more
details.

We consider the space $\mathcal{S}$ of complex analytic functions
of one complex variable and define the following scalar product:
\begin{equation}\label{scalarprod}
\langle f|g\rangle\equiv\int\frac{\mathrm{d}\bar{z}\
\mathrm{d}z}{2\pi i }\ \mathrm{e}^{-\bar{z}z}\
\overline{f(z)}g(z)
\end{equation}
where the bar denotes complex conjugation ($z$ and $\bar{z}$
are treated as independent variables), and the notation for the
measure means
\begin{equation}
\int\frac{\mathrm{d}\bar{z}\
\mathrm{d}z}{2\pi i }\equiv\int\limits_{-\infty}^\infty\frac{\mathrm{d}x\
\mathrm{d}y}{\pi}
\end{equation}
with $z\equiv x+ i y$. We also define the states $\langle
z|$ in the dual space $\mathcal{S}^\ast$  such that $\langle
z|f\rangle\equiv f(z)$, with $|f\rangle\in\mathcal{S}$. Then, the
set $\{f_n\}_0^\infty$, with
\begin{equation}\label{basis}
f_n(z)\equiv\frac{z^n}{\sqrt{n!}}
\end{equation}
constitutes an orthonormal basis for $\mathcal{S}$ with the inner
product (\ref{scalarprod}). This implies in particular:

\begin{equation}\label{delta}
\int\frac{\mathrm{d}z'\ \mathrm{d}\bar{z}'}{2\pi i }\
\mathrm{e}^{-z'\bar{z}'}\ \mathrm{e}^{\bar{z}'z}\ f(z')=f(z)
\end{equation}

We can also calculate the scalar product:
\begin{equation}
\langle z|\bar{z}'\rangle=\sum\limits_{n=0}^\infty\
f_n(z)\overline{f_n(z')}=\sum\limits_{n=0}^\infty\
\frac{1}{n!}(z\bar{z}')^n=\mathrm{e}^{z\bar{z}'}
\end{equation}
where we will denote the dual of $\langle z|$ by
$|\bar{z}\rangle$. Now, from the definition (\ref{scalarprod}),
the identity operator can be written as:
\begin{equation}
\hat{1}=\int\frac{\mathrm{d}\bar{z}\ \mathrm{d}z}{2\pi i }\
\mathrm{e}^{-\bar{z}z}\ |\bar{z}\rangle\langle z| \label{idop}
\end{equation}
Since the functions (\ref{basis}) constitute an orthonormal basis,
we can calculate the trace of an operator as follows:
\begin{equation}
\mathrm{Tr}\{\cdot\}=\sum\limits_{n=0}^\infty\ \langle
f_n|\cdot|f_n\rangle=\int\frac{\mathrm{d}\bar{z}\
\mathrm{d}z}{2\pi i }\ \mathrm{e}^{-\bar{z} z}\ \langle
z|\cdot|\bar{z}\rangle \label{tracehol}
\end{equation}
The prescription (\ref{aaplusholo}) defines a representation of the creation and annihilation
operators on $\mathcal{S}$. Therefore,

\begin{equation}\label{actcrean}
\langle z|\hat{a}^\dagger|\bar{z}'\rangle=z\langle
z|\bar{z}'\rangle=z\mathrm{e}^{z\bar{z}'}\ ,\quad\langle
z|\hat{a}|\bar{z}'\rangle=\frac{\partial}{\partial z}\langle
z|\bar{z}'\rangle=\bar{z}'\mathrm{e}^{z\bar{z}'}
\end{equation}
For the purpose of obtaining a path integral, we need to know how
to calculate the matrix elements (kernels) of the kind
$\mathcal{O}(z,\bar{z}')\equiv\langle
z|\hat{O}(\hat{a}^\dagger,\hat{a})|\bar{z}'\rangle$, where
$\hat{O}$ is an operator expressed in terms of creation and
annihilation operators. If the operator is expressed in
normal-order form (which we denote by the subscript $\mathrm{N}$)
i.e, arranging creation operators to the left and annihilation to
the right, the kernels can be written in a particularly useful
way, from (\ref{actcrean}), as:

\begin{equation}
\langle
 z|\hat{O}_\mathrm{N}|\bar{z}'\rangle=O_\mathrm{N}(z,\partial/\partial
z)\mathrm{e}^{z\bar{z}'}=O_\mathrm{N}(z,\bar{z}')\mathrm{e}^{z\bar{z}'}
\label{opform}
\end{equation}

In particular, we will need the kernel corresponding to
``time-evolution":

\begin{equation}
\mathcal{U}(z,\bar{z}';t_f-t_i)\equiv\langle
z|\mathrm{e}^{- i (t_f-t_i)\hat{H}_\mathrm{N}}|\bar{z}'\rangle
\label{unitop}
\end{equation}
with $\hat{H}_\mathrm{N}$ the normal-ordered hamiltonian of the
system.

For that purpose, as customarily, we divide the interval $t_f-t_i$
into $n$ subintervals of infinitesimal length $\varepsilon$ and we
will take the $n\rightarrow\infty$ limit in the end. Having in
mind the application to thermal field theory, we will take complex
times $t\in C$ where $C$ is the  contour starting at $t_i$ and
ending at $t_f=t_i-i\beta$ showed in Fig.\ref{fig:realcontour}.

Now, from (\ref{opform}), for an infinitesimal time interval:

\begin{equation}
\mathcal{U}(z_1,\bar{z}_2;\varepsilon)\simeq\mathrm{e}^{- i \varepsilon
H_\mathrm{N}(z_1,\bar{z}_2)}\mathrm{e}^{z_1\bar{z}_2}
\end{equation}
so that, inserting the identity operator (\ref{idop}) $n-1$ times in
(\ref{unitop})  one gets:

\begin{eqnarray}
\mathcal{U}(z,\bar{z}';t_f-t_i)=
\int\prod\limits_{k=1}^{n-1}\frac{\mathrm{d}z_k\
\mathrm{d}\bar{z}_k}{2\pi i }\mbox{exp}\left\{z_1\bar{z}'
%\right.\nonumber\\+\left.
+\sum_{k=1}^{n-1}\left[(z_{k+1}-z_k)\bar{z}_k
+\varepsilon\hat{H}_N(z_{k+1},\bar{z}_k)\right]\right\}
\end{eqnarray}
with $z_n=z$.

For the Hamiltonian (\ref{hamil}) one has
$\hat{H}=\hat{H}_N+\omega/2$ and the previous integral can be
explicitly calculated by using the standard formula \cite{ZJ02} :

\begin{equation}
\label{multipleint} \int\prod\limits_{k=1}^n\frac{\mathrm{d}z_k\
\mathrm{d}\bar{z}_k}{2\pi i }\
\mathrm{e}^{-\bar{z}Az+\bar{u}z+u\bar{z}}=(\det
A)^{-1}\mathrm{e}^{\bar{u}A^{-1}u}
\end{equation}
which, taking the $n\rightarrow\infty$ limit, yields:

\begin{equation}\label{evol}
\mathcal{U}_0(z,\bar{z}';t_f-t_i)=\exp\left(z\bar{z}'\mathrm{e}^{-i\omega(t_f-t_i)}
+\mathnormal{\Sigma}[j]\right)
\end{equation}
where the subscript ``0" distinguishes the particular case of the
hamiltonian (\ref{hamil}) and
\begin{eqnarray}\label{Sigma}
\Sigma[j]&=& i \int\limits_C\mathrm{d}t
\left[z\frac{\mathrm{e}^{- i \omega(t_f-t)}}{\sqrt{2\omega}}
j(t)+\bar{z}'\frac{\mathrm{e}^{ i \omega(t_i-t)}}{\sqrt{2\omega}}j(t)\right]
%\nonumber\\&-&
-\int\limits_C\mathrm{d}t\ \mathrm{d}t'\
j(t)\theta(t-t')\frac{\mathrm{e}^{- i \omega(t-t')}}{2\omega}j(t')
\end{eqnarray}

\section{Free Thermal propagators and partition function at $\mu\neq0$}
\label{app:therprop}

In this appendix  we review some important aspects regarding the canonical description of the free theory and the different representations for free  propagators in
 Thermal Field Theory at nonzero chemical potential, paying special attention to the differences between the case of particle number chemical potential  and that of exactly conserved charges such as the electric charge for  complex scalar fields.

Let us consider first  the case of a free neutral scalar field
$\phi(x)$. In that case, one
can evaluate the partition function and the propagator (two-point
function) directly in the complete set $|N_1,N_2,\ldots\rangle$,
corresponding to eigenstates of the hamiltonian operator with $N_1$
particles in the state 1, $N_2$ particles in the state 2, and so on:
\begin{align}
&\langle N_1,N_2,\ldots|N_1',N_2',\ldots\rangle=\delta_{N_1
N_1'}\cdot\delta_{N_2 N_2'}\cdot\ldots\notag\\
&\hat{H}_0|N_1,N_2,\ldots\rangle = \left(N_1 E_1+\ldots+\sum\limits_{i=1}^\infty\frac{E_i}{2}\right)|N_1,N_2,\ldots\rangle\notag\\
&\hat{N}|N_1,N_2,\ldots\rangle =
(N_1+N_2+\ldots)|N_1,N_2,\ldots\rangle
\end{align}
with $\sum N_i=N$. For non-interacting bosons of mass $m$,
$E_i\equiv\sqrt{m^2+|\vec{p}_i|^2}$, and the (infinite) term
$\sum_i E_i/2$ is the vacuum energy. As customary, we
consider first the system in a finite volume $V=L^3$, which we will
later take to infinity, so that spatial momenta are discretized as
$|p_i|=\frac{\pi n_i}{L}$ with integers $n_i$ and energy levels are
labeled by $\vec{n}\equiv(n_x,n_y,n_z)$. The free partition function
reads then:
\[
\tilde{Z}_\beta^0=\prod\limits_{\vec{n}}\sum\limits_{N=0}^\infty\mathrm{e}^{-\beta
N(E_{\vec{n}}-\mu)}\mathrm{e}^{-\beta
E_{\vec{n}}/2}=\prod\limits_{\vec{n}}\frac{\mathrm{e}^{-\beta
E_{\vec{n}}/2}}{1-\mathrm{e}^{-\beta(E_{\vec{n}}-\mu)}}
\]
where the condition $\mu<E_{\vec{n}}$ must be satisfied for all $\vec{n}$. Thus,
in the $V\rightarrow\infty$ limit:

\begin{equation}\label{parti}
\log
\tilde{Z}_\beta^0=-V\int\frac{\mathrm{d}^3\vec{p}}{(2\pi)^3}\
\left[\frac{\beta
E_p}{2}+\log\left(1-\mathrm{e}^{-\beta(E_p-\mu)}\right)\right]
\end{equation}
where $E_p^2=\vert\vec{p}\vert^2+m^2$. Therefore, in the following we must restrict to a
chemical potential $\mu\leq m$ (below the Bose-Einstein condensation limit)
to ensure the convergence of the previous expressions.

In order to obtain the free particle propagator in the canonical
formalism, defined as the two-point function:

\begin{equation}\label{propdef}
\tilde{G}(x)\equiv\langle
\hat{T}\hat{\phi}(x)\hat{\phi}(0)\rangle_{\beta,\mu}\equiv
\tilde{Z}^{-1}_\beta\
{\rm Tr}\left\{\mathrm{e}^{-\beta(\hat{H}-\mu\hat{N})}\hat{T}\hat{\phi}(x)\hat{\phi}(0)\right\}
\end{equation}
where $\hat{T}$ is the time-ordering operator, we expand the field
as customarily in terms of creation and annihilation operators:

\begin{equation}\label{freefield}
\hat{\phi}(\vec{x})=\frac{1}{V}\sum\limits_{n}\frac{1}{\sqrt{2E_{\vec{n}}}}
\left(\hat{a}_{\vec{n}}\mathrm{e}^{ i 2\pi\vec{n}\cdot\vec{x}/L}+
\hat{a}^\dagger_{\vec{n}}\mathrm{e}^{- i 2\pi\vec{n}\cdot\vec{x}/L}\right)
\end{equation}

with  commutation relation
\begin{equation}\label{commrel}
[\hat{a}_{\vec{n}},\hat{a}^\dagger_{\vec{n}'}]=V\delta_{\vec{n},\vec{n'}}
\end{equation}

The free hamiltonian and the number operator are given in terms of
creation and annihilation operators as:
\begin{align}
&\hat{H}_0=\sum\limits_{\vec{n}}\frac{1}{V}E_{\vec{n}}(\hat{a}_{\vec{n}}^\dagger \hat{a}_{\vec{n}}+\frac{1}{2}V)\ ,\notag\\
&\hat{N} =
\sum\limits_{\vec{n}}\frac{1}{V}\hat{a}_{\vec{n}}^\dagger
\hat{a}_{\vec{n}}
\end{align}

 Now, the {\em real} time
evolution of the field is given by
$\hat{\phi}(t,\vec{x})\equiv\mathrm{e}^{ i \hat{H}
t}\hat{\phi}(\vec{x})\mathrm{e}^{- i \hat{H} t}$ with $t\in
\mathbb{R}$. We will calculate the trace in (\ref{propdef}) using

\begin{equation}\label{bosedisfun}
\langle
\frac{1}{V}\hat{a}^\dagger_{\vec{n}}\hat{a}_{\vec{n}}\rangle_{\beta,\mu}=
\frac{1}{\mathrm{e}^{\beta(E_{\vec{n}}-\mu)}-1}\equiv
 n(E_{\vec{n}}-\mu)
\end{equation}

so that we get for the free propagator, after taking the
$V\rightarrow\infty$ limit:

\begin{equation}\label{freeprop}
\tilde{G}(x)=\theta(t) \tilde{G}^{>}(x)+\theta(-t)
\tilde{G}^{<}(x)
\end{equation}
with:
\begin{align}
\tilde{G}^{> (<)}(x)=&\int\frac{\mathrm{d}^3\vec{p}}{(2\pi)^3}
\mathrm{e}^{ i {\vec p}\cdot {\vec x}}
\tilde{G}^{>(<)}(t,p)\label{Gmixdef}
\\
 \tilde{G}^>(t,p)=&\frac{1}{2E_p}
 \left[\mathrm{e}^{- i E_p t}
 (1+n(E_p-\mu))+\mathrm{e}^{ i E_p t}n(E_p-\mu)
 \right]\label{G>}\\
 \tilde{G}^<(t,p)=&\frac{1}{2E_p}
 \left[\mathrm{e}^{ i E_p t}(1+n(E_p-\mu))
 +\mathrm{e}^{- i E_p t}n(E_p-\mu)\right]\label{G<}
 \end{align}

Note that the above propagators are obtained from the $\mu=0$ ones by the following replacement in the distribution function:

\begin{equation} \label{ntildedef}
n(x)\rightarrow{\tilde n}_p(x)\equiv\frac{1}{e^{{\tilde\beta}_p x}-1}
\end{equation}
with:

\begin{equation}\label{betatilde}
\tilde{\beta}_p\equiv\beta\left(1-\frac{\mu}{E_p}\right)
\end{equation}

Therefore, we have for instance ${\tilde n}_p(E_p)=n(E_p-\mu)$ and
the ${\tilde n}_p$ function satisfies:

\begin{equation} 1+{\tilde n}_p(x)+{\tilde n}_p(-x)=0
\label{disfunprop}
\end{equation}

Thus, the free propagator satisfies the following KMS-like
periodicity condition in the  mixed representation:

\begin{equation}\label{periodc}
\tilde{G}^>(t,p)=\tilde{G}^<(t+ i \tilde{\beta}_p,p)
\end{equation}
and in Fourier space we can write a spectral representation:

\begin{align}
\tilde{G}^>(p_0,p)=&\left[1+\tilde{n}_p(p_0)\right]\rho
(p_0,p)\nonumber\\
\tilde{G}^<(p_0,p)=&e^{-\tilde{\beta}_p
p_0}\tilde{G}^>(p_0,p)=\tilde{n}_p(p_0)\rho (p_0,p)
\label{spectralrep}\end{align}
where \begin{equation} \rho
(p_0,p)=2\pi\sgn(p_0)\delta\left[(p_0)^2-E_p^2\right]\end{equation}
is the free spectral function, which is independent of temperature
and chemical potential.

Now, using:

\begin{equation}
\theta (t)=i\int_{-\infty}^\infty \frac{dk_0}{2\pi} \frac{e^{-ik_0 t}}{k_0+i\epsilon}
\end{equation}
with $\epsilon\rightarrow 0^+$, we can write for the propagator in (\ref{freeprop}) in momentum space:

\begin{equation}
\tilde G(p_0,p)=\frac{i}{p_0^2-E_p^2+i\epsilon}+2\pi\delta(p_0^2-E_p^2)
n(\vert p_0 \vert-\mu) \label{tildeGp0}
\end{equation}

Note that we have used $\tilde n_p(E_p)\delta(p_0^2-E_p^2)=\tilde
n_p(\vert p_0 \vert)\delta(p_0^2-E_p^2)=n(\vert p_0
\vert-\mu)\delta(p_0^2-E_p^2)$ and we have chosen the ``$\vert p_0
\vert$-prescription" which, as explained in the main text,
guarantees the decoupling of the imaginary-leg contribution to
real-time Green functions.

\begin{figure}
\hspace*{-1.5cm}
\includegraphics[scale=.45]{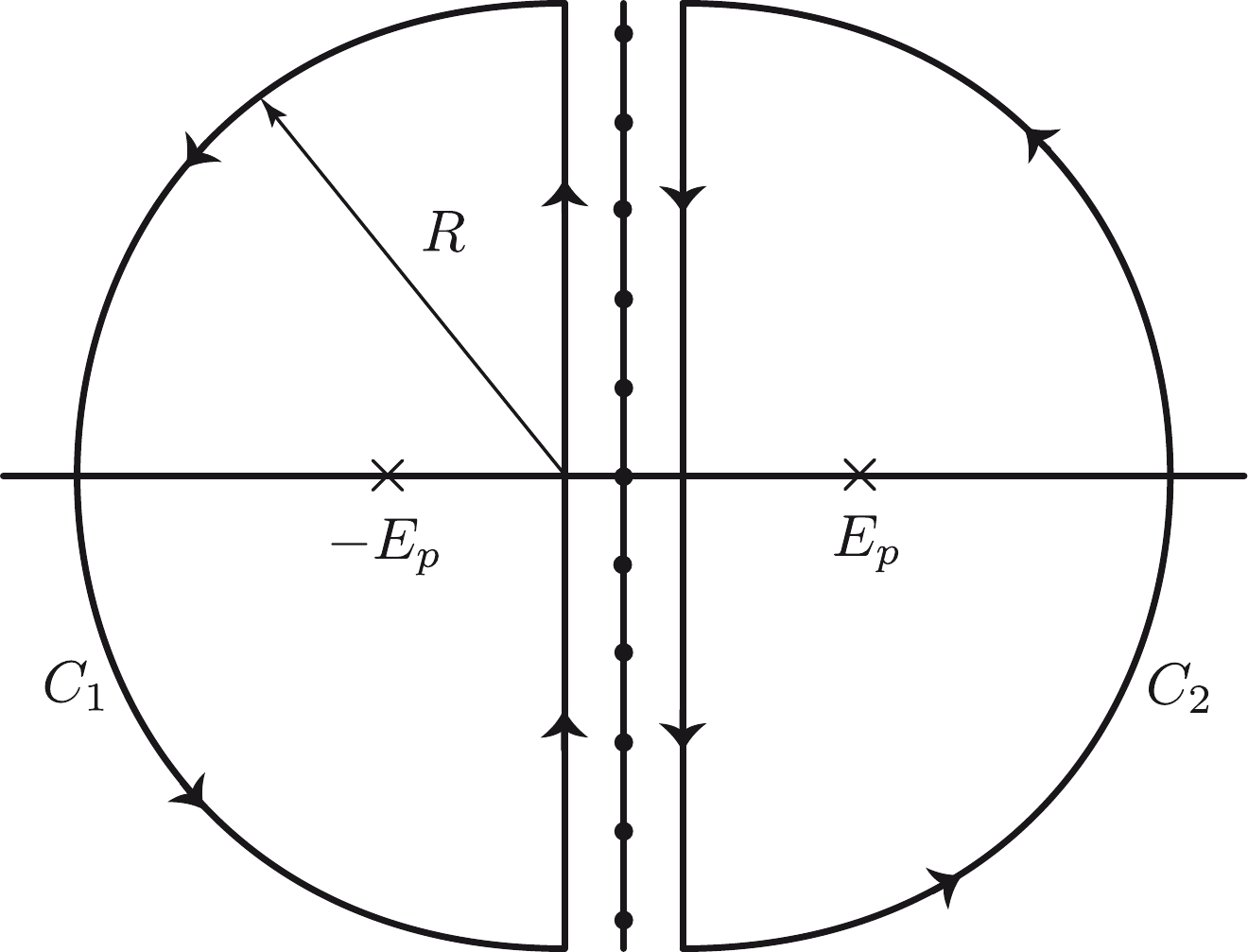}
%\vspace*{-9.5cm}
%\vspace*{-11cm}
 \caption{\rm \label{fig:matsubaracont} Contours used to derive the Matsubara representation of the free imaginary-time propagators. The black dots on the imaginary axis denote the
 Matsubara frequencies and $R\rightarrow\infty$.}
\end{figure}

The free propagators in (\ref{G>})-(\ref{G<}) can be extended to
imaginary times $t=-i\tau$ corresponding time differences along  the imaginary-time leg
$C_4$ in Fig.\ref{fig:realcontour}. Thus, we define $\tilde\Delta_T(\tau,p)=\tilde G^>(-i\tau,p)$ for $\tau\geq 0$ and $\tilde\Delta_T(\tau,p)=\tilde G^<
(-i\tau,p)$ for $\tau\leq 0$. Now, if we try  to construct a Matsubara frequency representation in this case, we have, from the mixed representation (\ref{G>}) and using (\ref{ntildedef})-(\ref{betatilde}):

\begin{equation}\label{matsu}
\tilde\Delta_T(\tau\geq 0,p)=\frac{1}{2\pi i} \oint_{\small C_1\cup C_2} \frac{e^{z\tau}}{e^{\betatp z}-1} \ \frac{1}{z^2-E_p^2}\stackrel{0\leq\tau\leq\betatp}{\large =} \frac{1}{\betatp}\sum_n
\frac{e^{i(\tilde\omega_n)}}{\tilde\omega_n^2+E_p^2}
\end{equation}
where the $C_{1,2}$ contours are showed in Fig.\ref{fig:matsubaracont}, the black dots on the imaginary axis being the ``modified" Matsubara frequencies $\tilde\omega_n=2\pi n/\betatp$. A very important point here is that the last step in the above equation is only valid for $\tau\in[0,\betatp]$, otherwise the integrals along the circular arcs with $R\rightarrow\infty$ do not vanish. Thus, the Matsubara Fourier representation is only valid in that interval, which is smaller than $[0,\beta]$.

Carrying out the same procedure with $\tilde\Delta_T(\tau\leq 0,p)$ using (\ref{G<}) leads to the same ``modified" Matsubara representation for $\tau\in[-\betatp,0]$. In fact, we see that the KMS-like condition (\ref{periodc})
translates into the imaginary-time propagator as:

\begin{equation}\label{itpropbc}
\tilde\Delta_T(\tau+\betatp,p)=\tilde\Delta_T(\tau,p)
\end{equation}
so that this propagator does not satisfy the usual equilibrium KMS condition $\tilde\Delta_T(\tau+\beta,p)=\tilde\Delta_T(\tau,p)$.

At this point, it is instructive to compare the above free propagators for chemical nonequilibrium with those obtained when an exact conserved charge is present. As we are going to see, there are crucial differences between the two cases. For definiteness, we consider the electric charge for the case of a complex scalar field and denote the corresponding chemical potential by $\mu_Q$. In that case, the counterparts of (\ref{G>})-(\ref{G<}) for the free propagator $G_Q(x)=\langle
\hat{T}\hat{\phi^\dagger}(x)\hat{\phi}(0)\rangle$ are \cite{Landsman:1986uw}:

\begin{align}
G_Q^>(t,p)=&\frac{1}{2E_p}
 \left[\mathrm{e}^{- i E_p t}
 (1+n(E_p-\mu_Q))+\mathrm{e}^{ i E_p t}n (E_p+\mu_Q)
 \right]\\
 G_Q^<(t,p)=&\frac{1}{2E_p}
 \left[\mathrm{e}^{ i E_p t}(1+n(E_p+\mu_Q))
 +\mathrm{e}^{- i E_p t}n(E_p-\mu_Q)\right]
 \end{align}

 Note that, unlike our previous case in (\ref{G>})-(\ref{G<}), the  chemical potential enters now with {\em opposite} sign for the positive and negative frequencies, which comes from the opposite charge of particles and antiparticles, necessary to maintain the chemical equilibrium imposed by charge conservation. Due to this sign difference,  the above propagator satisfies now the following condition:

\begin{equation}
G_Q^>(t,p)=\mathrm{e}^{-\beta\mu_Q}G_Q^<(t+i\beta,p)
\end{equation}
i.e, in this case the KMS symmetry realizes simply as a modification of the $\mu_Q=0$ KMS boundary condition  by a constant $\mu_Q$-dependent  factor, which is a result completely different  from the previous case, c.f., eq.(\ref{periodc}), where the loss of KMS involves the $p$-dependent  $\betatp$, which {\em cannot} be rewritten as a multiplicative factor:

\begin{align}
\tilde G^>(t,p)
= &\frac{1}{2E_p} \left[\mathrm{e}^{-\beta\mu_\pi}\mathrm{e}^{- i E_p (t+i\beta)}n(E_p-\mu_\pi)
 +\mathrm{e}^{\beta\mu_\pi}\mathrm{e}^{ i E_p (t+i\beta)}(1+n(E_p-\mu_\pi))\right]
 \neq  \mathrm{e}^{-\beta\mu_\pi} \tilde G^<(t+i\beta,p)
 \end{align}

Another way to arrive to the same conclusion is to derive directly the periodicity relation from the thermal averages. In the charged scalar case, it is crucial to use that the field is a charge eigenstate, i.e. $[\hat Q,\hat\phi]=-\hat\phi$, $[\hat Q,\hat\phi^\dagger]=\hat\phi^\dagger$ \cite{Landsman:1986uw}. This, together with charge conservation $[\hat Q,\hat H]=0$ leads to:

\begin{equation}
{\rm Tr}\left[\hat\phi^\dagger (t-i\beta)\hat\phi (0)\mathrm{e}^{-\beta(\hat H-\mu_Q \hat Q)}\right]=\mathrm{e}^{-\beta\mu_Q} {\rm Tr}\left[\hat\phi (t) \hat\phi^\dagger (0) \mathrm{e}^{-\beta(\hat H-\mu_Q\hat Q)}\right]
\end{equation}

However, in the  case  of a real field and the number operator, even though $[\hat N,\hat H]=0$ in the free case, $[\hat N,\hat\phi]\neq \pm\hat\phi$ which prevents the previous relation to hold.

Defining now the imaginary-time propagators as above, we  get the same factor in the $\mu_Q$ case:

\begin{equation}
\Delta^Q_T(\tau+\beta,p)=\mathrm{e}^{-\beta\mu_Q}\Delta^Q_T(\tau,p)
\label{kmscharg}
\end{equation}

In fact, it is not difficult to see that in the $\mu_Q$ case, this simple form of KMS symmetry still allows for a well-defined Matsubara IT frequency representation:

\begin{equation}
\tilde\Delta_T^Q(\tau\geq 0,p)=\frac{1}{2\pi i} \oint_{\small C'_1\cup C'_2} \frac{e^{z\tau}}{e^{\beta (z+\mu_Q)}-1} \ \frac{1}{z^2-E_p^2}\stackrel{0\leq\tau\leq\beta}{\large =}\frac{1}{\beta}\sum_n
\frac{e^{i(\omega_n+i\mu_Q)}}{(\omega_n+i\mu_Q)^2+E_p^2}
\label{chargedmats}
\end{equation}
where $C'_{12}$ correspond to the contours in Fig.\ref{fig:matsubaracont} but with the vertical line displaced to $z=-\mu_Q$ ($\mu_Q<m$) and the dots in that line being now the standard Matsubara frequencies $\omega_n=2\pi n/\beta$. Therefore, in this case the ordinary IT formalism is recovered for $\tau\in[-\beta,\beta]$ simply by changing in the Feynman rules $\omega_n\rightarrow \omega_n+i\mu_Q$.

Most of the results showed in the main text can be written in
terms of the above thermal propagators evaluated at the origin in
position space and functions related to them. From
(\ref{Gmixdef})-(\ref{G<}) we have (for $\mu\leq m$) at
$\tau=t=\vec{x}=0$:

\begin{align}
{\tilde G}^>(0)&={\tilde G}^<(0)={\tilde \Delta}_T (0)=\tilde G(0)
%\nonumber\\&=
=\left[\tilde G(0)\right]^{T=\mu=0} +{\tilde g}_1(m,T,\mu)
\label{tadpole}
\end{align}
where the $T=\mu=0$ contribution is ultraviolet divergent. In
dimensional regularization it is given by:

\begin{equation}
\left[\tilde
G(0)\right]^{T=\mu=0}
=\int\frac{d^{D-1}p}{(2\pi)^{D-1}}\frac{1}{2E_p}=\frac{\Gamma[1-\frac{D}{2}]m^{D-2}}{(4\pi)^{D/2}}
\label{tadpolezero}
\end{equation}
while the $T,\mu$-dependent contribution $\tilde g_1$ is finite.
We are following the same notation as in \cite{Gerber:89} so that
$\tilde g_1$ is the $\mu\neq 0$ extension of their function
$g_1(T)$, to which it reduces for $\mu=0$. We have:

\begin{equation}
{\tilde g}_1(m,T,\mu)=\frac{1}{2\pi^2}\int_0^\infty dp \frac{p^2}{E_p} \frac{1}{e^{\beta(E_p-\mu)}-1}
\label{g1tilde}
\end{equation}

Note that in dimensional regularization one has, as in the $\mu=0$
case:

\begin{eqnarray}
\left.\left[\partial_{\tau}^2-\nabla^2\right]\tilde\Delta_T(\tau,\vec{x})\right\vert_{\tau=\vec{x}=0}&=&m^2{\tilde
\Delta}_T (0)\nonumber\\
\left.\Box\tilde G(t,\vec{x})\right\vert_{t=\vec{x}=0}&=&-m^2
\tilde G(0) \label{secder}
\end{eqnarray}
and $\partial_\mu\tilde G(0)=\partial_\mu\Delta_T(0)=0$.

Let us also define, following again the notation in
\cite{Gerber:89}:

\begin{equation}
{\tilde g}_0(m,T,\mu)=-\frac{T}{\pi^2}\int_0^\infty dp \ p^2 \log\left[1-e^{-\beta(E_p-\mu)}\right]
\label{g0tilde}
\end{equation}
so that, taking into account that $\partial E_p/\partial m^2=1/(2E_p)$, we can write the free partition function (\ref{parti}) separating its divergent contribution in dimensional regularization as:

\begin{equation}
\log
\tilde{Z}_\beta^0=\frac{\beta V}{2} \left[\frac{\Gamma\left[-\frac{D}{2}\right]m^D}{(4\pi)^{D/2}}+{\tilde g}_0(m,T,\mu)\right]
\label{freeparfung0}
\end{equation}

Finally, note that the functions $\tilde g_0$ and $\tilde g_1$ satisfy a similar relation as in \cite{Gerber:89}:

\begin{equation}
\tilde g_1(m,T,\mu)=-\frac{\partial}{\partial m^2}g_0(m,T,\mu)
\label{derfor}
\end{equation}

\end{document}